\documentclass[useAMS,usenatbib,usegraphicx]{mn2e}
\usepackage{natbib}
\usepackage{color}
\usepackage{graphicx}
\usepackage{bm}
\usepackage{aas-macros}

\title{Constraints on the dynamical evolution of the galaxy group M81}

\author[W. Oehm, I. Thies and P. Kroupa]
{W. Oehm$^{1}$\thanks{physik@wolfgang-oehm.com},
I. Thies$^{2}$\thanks{ithies@astro.uni-bonn.de, Postal address: Auf dem H\"ugel 71, D-53121 Bonn, Germany} and
P. Kroupa$^{2,3}$\thanks{pavel@astro.uni-bonn.de, Postal address:
  Argelander Institute for Astronomy (AIfA), Auf dem H\"ugel 71,
  D-53121 Bonn, Germany} \\
   $^{1}$ scdsoft AG, Albert-Nestler-Str. 10, D-76131 Karlsruhe, Germany \\
   $^{2}$ Helmholtz-Institut f\"ur Strahlen und Kernphysik (HISKP),
   Nussallee 14-16, D-53115 Bonn, Germany \\
  $^{3}$ Charles University in Prague, Faculty of Mathematics and Physics,
Astronomical Institute, V  Hole\v{s}ovi\v{c}k\'ach 2, CZ-180 00 Praha 8, \\Czech Republic }

\begin{document}
\maketitle


\begin{abstract}
  According to the standard model of cosmology, galaxies are embedded
  in dark matter halos which are made of particles beyond the standard
  model of particle physics, thus extending the mass and the size of the
  visible baryonic matter by typically two orders of magnitude. 
The observed gas distribution throughout the nearby M81
    group of galaxies shows evidence for past significant
    galaxy--galaxy interactions but without a merger having occurred.
    This group is here studied for possible dynamical solutions within
    the dark-matter standard model.
In order to cover a comprehensive set of
    initial conditions, the inner three core members M81, M82 and
    NGC~3077 are treated as a three-body model based on Navarro-Frenk-White
    profiles. The possible orbits of these galaxies are examined
    statistically taking into account dynamical friction.  
    Long living, non-merging initial constellations which allow
    multiple galaxy--galaxy encounters comprise unbound galaxies only,
    which are arriving from a far distance and happen to
    simultaneously encounter each other within the recent
    500~Myr. 
  Our results are derived by the employment
  of two separate and independent statistical methods, namely a Markov
  chain Monte Carlo method and the genetic algorithm using the SAP
  system environment. The conclusions reached are
  confirmed by high-resolution simulations of live self-consistent
  systems (N-body calculations). Given the observed positions of
    the three galaxies the solutions found comprise predictions for
    their proper motions. 
 \end{abstract}


\maketitle

\section{Introduction}

The law of gravitation by Newton and its reformulation as a
geometrical space-time distortion through matter by Einstein has been
empirically derived within the Solar system.  It is commonly
extrapolated, by orders of magnitude, to the very strong and very weak
curvature regimes \citep{Koyama2015}. However, in the very weak
curvature regime the extrapolation fails to account for the
observations, such as the flat shape of the rotation curves of
galaxies. One attempt at solving this extrapolation is to hypothesize
that the observed deficit comes about due to the presence of particle
dark matter, therewith proposing an extension of the standard
model of particle physics. In the standard dark-energy plus
  cold-dark-matter model of cosmology
 (LCDM) the new particles must only interact gravitationally and at
most weakly with each other and with the 
variety of known particles of the standard model of particle physics.

The question whether the existence of dark matter halos can be
inferred observationally has a long history. While the approximately
flat rotation curves to large radii observed in disk galaxies
unambiguously imply either particle dark matter halos or non-Newtonian
gravitation \citep{Famaey2012}, the potentials around elliptical
galaxies are much less constrained because of the lack of observable
tracers and because velocity anisotropies confuse the dynamical
evidence (e.g. \citealt{Samurovic2005}, \citealt{Samurovic2006},
\citealt{Samurovic2008}, \citealt{Samurovic2014} and \citealt{Richtler2011}).  

One important testable implication of the particle dark matter halos
is that they imply significant dynamical friction when a galaxy
encounters another galaxy \citep{Kroupa2015}. Indeed, ``Interacting
galaxies are well-understood in terms of the effects of gravity on
stars and dark matter'' \citep{Barnes1998}, and interacting major
galaxies merge within about 1-1.5~Gyr after infall \citep{Privon2013}.
This is the sole origin of the hierarchical, merger-driven formation
of structure in the LCDM model, which is popularly assumed to be
the main physical process driving the formation and evolution of the
galaxy population observed at the present cosmological epoch.  Thus,
if observed galactic systems show either evidence for dynamical
friction or absence of this important process then this would
test the presence of dark matter halos \citep{Kroupa2015}.

Located about 3.6 Mpc from the Local Group of galaxies, the M81 group
of galaxies is the closest similar constellation of galaxies. It
therefore offers opportunities for the investigation of the dynamical
behavior of such loose but typical groups of galaxies, based on
thorough available observational data. Thus, if information exists
which constrains the past encounters between group members then the
process of dynamical friction can be tested for. For example, the
Local Group of galaxies is constrained by the two major galaxies not
allowed to have merged, such that Andromeda and the Milky Way never
had a close encounter in the past if the LCDM is valid \citep{Zhao2013}.

The M81 group of galaxies is comprised of a major Milky-Way type
galaxy (M81, baryonic mass $M_{\rm
  M81}$~$\approx$~3~$\cdot~10^{10}M_\odot$) and four companion
galaxies, two of which are themselves dwarf but nevertheless
substantial star-forming disk galaxies (M82, $M_{\rm
  M82}$~$\approx$~1~$\cdot~10^{10}M_\odot$ and NGC~3077, $M_{\rm
  N3077}$~$\approx$~2~$\cdot~10^{9}M_\odot$).  The interesting
observational finding is that these galaxies are enshrouded by
HI-emitting gas spanning the dimensions of the inner group which is
about 50-100~kpc.  This requires the galaxy pairs M81/M82 and
M81/NGC~3077 to have interacted closely at least once, in order to
strip the outer gaseous disks and rearrange the matter in the long
tidal features (\citealt{Cottrell1977}, \citealt{Gottesman1977},
\citealt{Hulst1979}, \citealt{Yun1993} and \citealt{Yun1994}). This
constrains the distribution of dark matter in the group by the process
of dynamical friction, because the time span between the encounters
and the merging due to dynamical friction is constrained to be at
least about a crossing time, i.e. about 0.5--1~Gyr (for a typical
velocity of $200\,$km/s and over a spatial range of 50--100$\,$kpc).

The M81 group must have assembled through the infall of the individual
galaxies according to the merger tree which is a necessary part of the
hierarchical structure formation in the LCDM model. The
currently observed configuration of the M81 group constrains the
pre-infall initial conditions since for example tightly bound initial
conditions are most likely ruled out given that the M81 group has not
merged yet to an early-type galaxy. Full-scale simulations of live
self-gravitating disk galaxies, each with their inter-stellar media
are too prohibitive, given that only four dimensional observational
information is available in six-dimensional phase space, to allow a
wide scan of parameters to seek those pre-infall configurations which
lead to the presently observed group structure. Therefore, as a first
step simplified but very rapid simulations are performed here taking
into account the essential physical process, namely dynamical friction
on the putative dark matter halos. For this purpose the three-particle
integration is combined in a novel approach with the genetic algorithm
and the Markov chain Monte Carlo method to yield robust estimates of
the allowed pre-infall configurations. 

Thus, as a first step, we perform numerical simulations of the
possible orbits of the three core members M81, M82 and NGC~3077,
treated as a three-body problem with rigid Navarro-Frenk-White
profiles (NFW, \citealt{Navarro1995}) for the DM halos. Based on the
knowledge of today's plane of sky (POS) coordinates, the distances and
the radial velocities, as well as the knowledge about the North Tidal
Bridge (NTB) and South Tidal Bridge (STB) obtained by radio
astronomical observations, two different methods (Markov chain Monte
Carlo (MCMC) and genetic algorithm(GA)) are utilized for the delivery
of statistical populations for the unknown, or just roughly known
physical entities under the constraint that known conditions be
fulfilled.  For a comprehensive statistical evaluation, the simplified
three-body model serves as a first basis since numerical N-body
simulations wouldn't allow a comprehensive scan of initial parameters
within an appropriate project time.  By then comparing individual
solutions with high-resolution simulations of live self-consistent
systems the results of these estimates are verified.

In Section~\ref{sec:DMHDF} we briefly discuss the impact of dark
matter (DM) halos on the orbits of bodies upon intruding their
interior and present the model for the Coulomb logarithm examined in
this publication.  Section~\ref{sec:IG} presents facts about the M81
group important for our investigations, and our approach how to
achieve meaningful statistical results concerning their dynamical
behaviour.  Sections~\ref{sec:MCMC} and~\ref{sec:GA} deal with the
details of our application of the mentioned statistical methods, MCMC
and GA, to the inner M81 group and the results obtained there, followed 
by a general discussion comparing those methods (Section~\ref{sec:comp}). In
Section~\ref{sec:N-body} we present the results of individual N-body
calculations compared to the orbits of our three-body calculations,
randomly extracted from the statistical population.  Finally, in
Section~\ref{sec:conc} we discuss the results obtained.  For easy
reading, Appendix~\ref{app:abr} provides a list of acronyms used in
this publication.

\section{Dark Matter Halos and Dynamical Friction}
\label{sec:DMHDF}

Exploring the dynamics of bodies travelling along paths in the
interior of DM~halos implies that the effects of dynamical friction
have to be taken into account in an appropriate manner \citep{Chandra1942}.
For isotropic distribution functions the decelaration of an
intruding body due to dynamical friction is described by
Chandrasekhar's formula \citep{Chandra1943}, which reads for a
Maxwellian velocity distribution with dispersion $\sigma$ (for
details see \citet{Binney}, chap. 8.1):

\begin{equation}
\label{eq:chandra}
\frac{d\vec{\textbf{v}}_M}{dt} = -\frac{4{\pi}G^2M{\rho}}{{\textnormal{v}_M}^3}\ \textnormal{ln}{\Lambda} 
\left[\textnormal{erf}(X) - \frac{2X}{\sqrt{{\pi}}}\textnormal{e}^{-X^2} \right] \vec{\textbf{v}}_M\ ,
\end{equation}
with \emph{X} = v$_M$/($\sqrt{2}\sigma$). The intruder of mass \emph{M} and relative velocity $\vec{\textbf{v}}_M$ is decelareted by $d\vec{\textbf{v}}_M$/\emph{dt} in the background density $\rho$ of the DM~halo.

Simulating galaxy-galaxy encounters \citet{Petsch2008} showed that a
modified model for the Coulomb logarithm ln$\Lambda$, originally
proposed by \citet{Jiang2008}, describes the effects of dynamical
friction in a realistic manner. This mass- and distance-dependent
model reads:

\begin{equation}
\label{eq:cl}
\textnormal{ln}\Lambda = \textnormal{ln}\left[1 + \frac{M_{halo}(r)}{M}\right]\  ,
\end{equation} 
where $M_{halo}(r)$ is the mass of the host dark matter halo within
radial distance $r$.

Chandrasekhar's formula only gives an estimate at hand. However,
high-resolution simulations of live self-consistent systems presented
in Section~\ref{sec:N-body} confirm our approach of employing this
semi-analytical formula in our three-body calculations.

\section{The inner Group}
\label{sec:IG}

As already indicated, the three core members M81, M82 and NGC~3077 are
treated as a three-body system, based on rigid NFW-profiles for the DM
halos. In our notation the indices 1, 2 and 3 represent M81, M82 and
NGC~3077, respectively. M81 appears to be the most massive object
accompanied by M82 and NGC~3077.

\subsection{Observational Data}

The observational data for the plane-of-sky (POS) coordinates and the
radial velocities are taken from the NASA/IPAC Extragalactic Database
(NED)\footnote{http://ned.ipac.caltech.edu/}, query submitted on 8 Feb
2014, and are presented in Table~\ref{coordinates}. The distances,
however, are only roughly known: even for M81 an uncertainty of about
10\% due to the variation of recent results published over the last decade can
be extracted from NED. In our context we are interested in the
relative distances of the three core members related to the M81 frame
at present, rather than absolute distances. Therefore we take the
average value for M81 as granted and cater for the mentioned
uncertainty by means of appropriate ranges for the distances of the
companions M82 and NGC~3077, as shown in Table~\ref{coordinates}. As a
final remark regarding the phase space, the POS velocity components
are not known at all.

Apart from optical observations, radio astronomical data are at our
disposal. M81, M82 and NGC~3077 are known to be embedded in a large
cloud of atomic hydrogen as observations of the 21cm HI emission line
have shown \citep{Appleton1981}. \citet{Cottrell1977} and
\citet{Gottesman1977} identified the NTB connecting M81 and M82, as
\citet{Hulst1979} did for the STB between M81 and NGC~3077. The
results of those observations were confirmed by \citet{Yun1993} and
\citet{Yun1994} by investigating the inner M81 group with the Very
Large Array Telescope. Since then, the fact that both companions must
have experienced close encounters with M81 in the recent cosmological
past has been commonly agreed to be an established feature of the
inner group. A review of the entire situation is presented by
\citet{Yun1999}.

\begin{table*}           
\centering                          
\begin{tabular}{c c c c c c c c}        
\hline\hline                 
Object      & RA                   & DEC               & Heliocentric & Distance                        & $L_\textnormal{v}$            & Baryonic Mass                     & DM Halo Mass\\ 
                 & (EquJ2000)     & (EquJ2000)   & Velocity       &                                       &   (visual)                              & ($M_{\odot}$)                     & ($M_{\odot}$) \\ 
\hline                                   
M81          &  09h55m33.2s & +69d03m55s & -34 km/s    & $3.63 \ Mpc$                  &  2.04~$\cdot~10^{10}$    &   3.06~$\cdot~10^{10}$   &   1.17~$\cdot~10^{12}$ \\
M82          &  09h55m52.7s & +69d40m46s & 203 km/s   & $(3.53 \pm 0.6) \ Mpc$  &  8.70~$\cdot~10^{9}$      &   1.305~$\cdot~10^{10}$  &   5.54~$\cdot~10^{11}$   \\
NGC 3077 & 10h03m19.1s  & +68d44m02s &   14 km/s   & $(3.83 \pm 0.6) \ Mpc$  &  1.95~$\cdot~10^{9}$      &   2.925~$\cdot~10^{9}$     &   2.43~$\cdot~10^{11}$    \\
\hline                                   
\end{tabular}
\caption{Observational and derived data for the inner M81 group members.}
\label{coordinates}      
\end{table*}              

\subsection{Previous Results}
\label{PrevRes}

\citet{Thomasson1993} simulated the tidal interaction between M81
and NGC~3077, reproducing the STB and the spiral structure of M81. The
orbit generating the best fit shows a pericentre passage with a
distance of 22~kpc about 400~Myr ago (i.e. at time $-400\,$Myr).

\citet{Yun1999} additionally considered the tidal interaction between
M81 and M82, too, reproducing the NTB as well as the STB. The
pericentre passages occur at -220~Myr with a separation of 25~kpc and
at -280~Myr with a separation of 16~kpc for M82 and NGC~3077,
respectively. (A GIF-movie is still available at
http://www.aoc.nrao.edu/$\sim$myun/movie.gif).

Full N-body simulations by \citet{Thomson1999} for the three core
members did not yield satisfactory results because the galaxies merge.
Indeed, due to the effect of dynamical friction within the DM halos,
the solutions found by \citet{Yun1999} and \citet{Thomson1999} would
result in merged galaxies\footnote{\emph{"...this simulation resulted
    in the mergers of the companions onto M81, probably because the
    dynamical dissipation associated with the truncated isothermal
    model halo used was too efficient."}(\citealt{Yun1999},
  p. 88).}. Obviously, this begs the question how likely it is to
catch the M81 system, which is also the next and nearest Local Group
equivalent, just in this special evolution phase after the close
encounters between its inner three members and before any one of them
has merged.
Further investigations of the behaviour of the inner M81 group have
not been published since~\citet{Yun1999} and \citet{Thomson1999}.

\subsection{DM halo masses}
\label{sec:masses}

Following \citet{Behroozi2013} we derived the baryonic and DM Halo
masses from the bolometric luminosities, as extracted fom NED (query
submitted on 3 Nov 2015). The results are shown in
Table~\ref{coordinates}.

The masses of the DM halos listed in Table \ref{coordinates} may appear very
high compared to the baryonic masses. However, following fig. 7 and
8 in \citet{Behroozi2013} DM-to-baryonic mass ratios of the order of
100:1 are expected for the baryonic masses listed in Table~\ref{coordinates}. 
The lowest ratio is about 40:1 at a DM halo mass of
$10^{12}\,M_\odot$, as can be seen for M81.

This is in line with \citet{Binney} (p. 18, table 1.2), who quote for the M81 comparable 
Milky Way a baryonic mass of $5 \cdot 10^{10}\,M_\odot$, and a DM halo mass 
of $2 \cdot 10^{12}\,M_\odot$.

\subsection{The Model}
\label{sec:model}

The DM halo of either galaxy is treated as a rigid halo with a density
profile according to \citet{Navarro1995} (NFW-profile), truncated at
the the radius $R_{200}$:

\begin{equation}
\label{eq:NFW}
\rho (r)=\frac{\rho_0}{r/R_s\left(1+r/R_s\right)^2}\ ,
\end{equation}
with $R_s=R_{200}/c$, $R_{200}$ denoting the radius yielding an
average density of the halo of 200~times the cosmological critical
density

\begin{equation}
\rho_{crit}=\frac{3 H^2}{8 \pi G}\  ,
\end{equation}
and the concentration parameter~$c$

\begin{equation}
\log_{10}c=1.02-0.109\left(\log_{10}\frac{M_\mathrm{vir}}{10^{12}M_{\odot}}\right)\ 
\end{equation}
(see \citet{Maccio2007}). The parameters for the three NFW-profiles are presented in Table~\ref{NFW}.

\begin{table}          
\centering                          
\begin{tabular}{c c c c }        
\hline\hline                 
Object      & $R_{200} (kpc)$    & $\rho_0$ ($M_{\odot}/pc^3$)    & $c$         \\ 
\hline                                   
M81          &  210.4                   & 7.210~$\cdot~10^{-3}$             & 10.29  \\
M82          &  164.0                   & 8.810~$\cdot~10^{-3}$             & 11.17  \\
NGC 3077 &  124.6                   & 1.100~$\cdot~10^{-2}$             & 12.22  \\
\hline                                   
\end{tabular}
\caption{NFW profile parameters for the inner M81 group members.}
\label{NFW}       
\end{table}               

\begin{figure}           
\centering
\includegraphics[width=6cm]{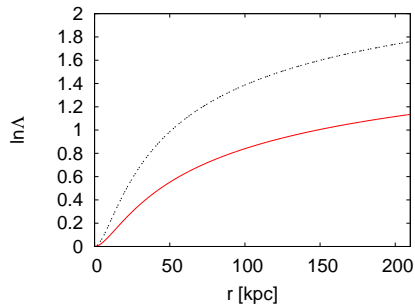}
\caption{The Coulomb logarithm ln$\Lambda$ for M82 (lower curve) and NGC~3077 (upper curve) penetrating the DM halo of  M81, shown as a function of the distance from the DM~halo centre of M81.}
\label{CouLog}            
\end{figure}

As explained in Section~\ref{sec:DMHDF} in case of overlapping DM
halos, not only the gravitational force derived from the NFW profiles
but also the effect of dynamical dissipation has to be taken into
account. As an example, the Coulomb logarithm (Eq.~\ref{eq:cl}) to be
incorporated in Eq.~\ref{eq:chandra} is displayed in
Figure~\ref{CouLog} for either companion M82 and NGC~3077 penetrating
M81's DM halo.

In the M81 reference frame, at present, with $x1 = y1 = z1 = 0$ and
$v_{1_x} = v_{1_y} = v_{1_z} = 0$, our three-body problem is described
by the POS coordinates and radial velocities of the two companions M82
and NGC~3077 relative to M81, deducted from Table~\ref{coordinates},
their unknown POS velocity components, and their roughly known
relative distances also derived from Table~\ref{coordinates}.
Table~\ref{parameters} shows the full set of open parameters.

The details regarding the numerical solution for the calculation of
the forces between the galaxies, as well as the details concerning the
numerical solution of the dynamical equations thereafter are presented
in Appendix~\ref{app:num}. Figure~\ref{forces} shows the forces
between the galaxies due to gravity and dynamical friction, dependent
on the relative velocities.

\begin{table}         
\centering                          
\begin{tabular}{c c c c c c}        
\hline\hline                 
$P_1$ & $P_2$ & $P_3$ & $P_4$ & $P_5$ & $P_6$ \\ 
\hline                     
$v_{2_x}$ & $v_{2_y}$ & $z_2$ & $v_{3_x}$ & $v_{3_y}$ & $z_3$  \\
\hline                                   
\end{tabular}
\caption{Open parameters for the inner M81 group.}
\label{parameters}      
\end{table}               

\begin{figure*}
\centering
\begin{tabular}{ccc}
\includegraphics[width=5.5cm]{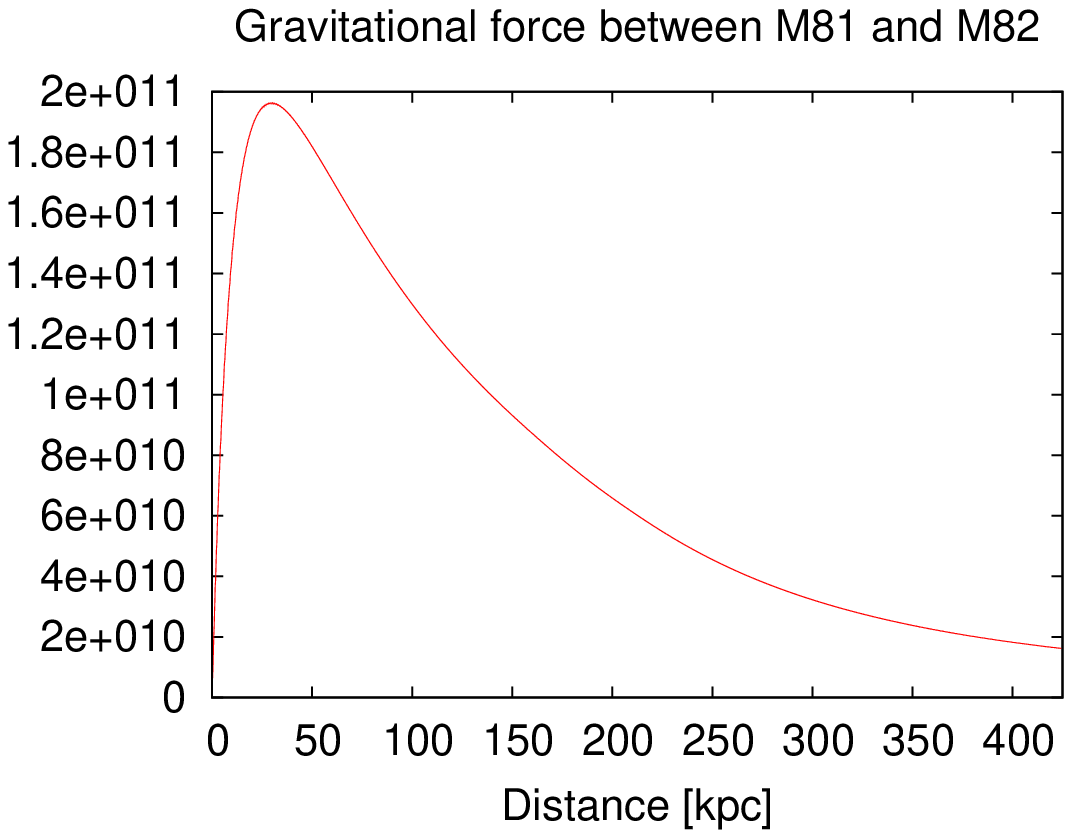} & \includegraphics[width=5.5cm]{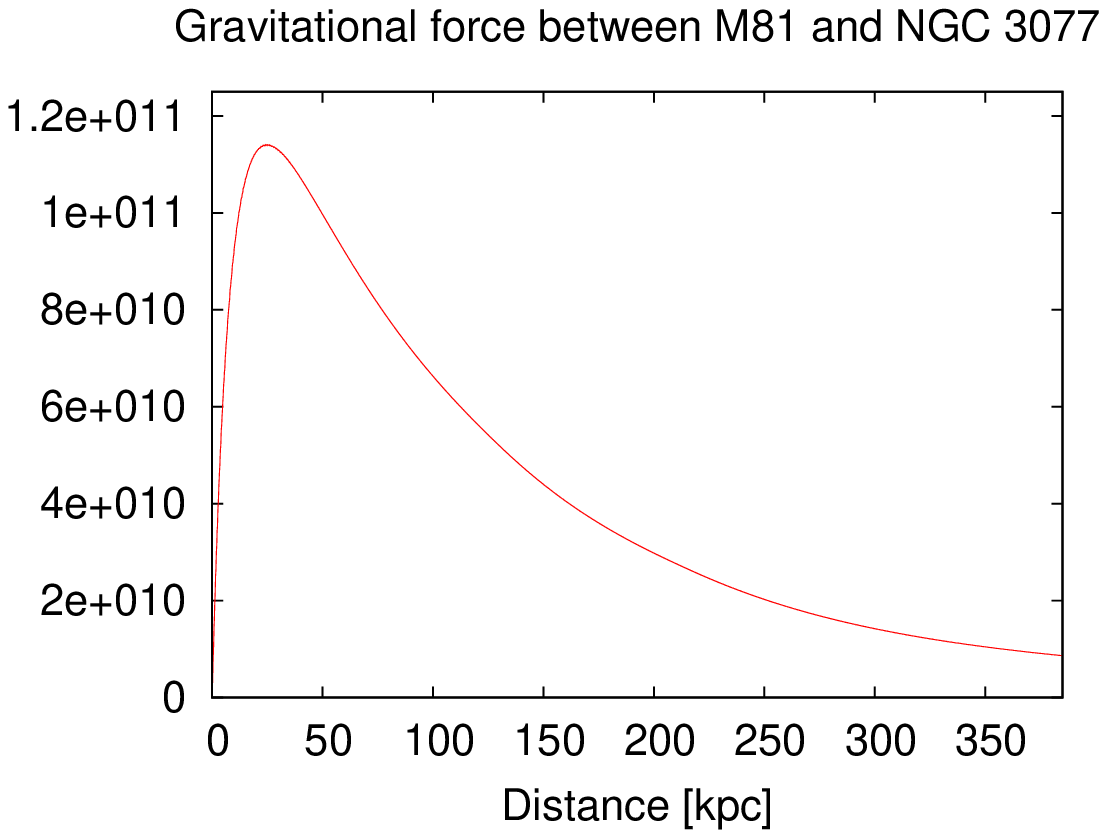} & 
\includegraphics[width=5.5cm]{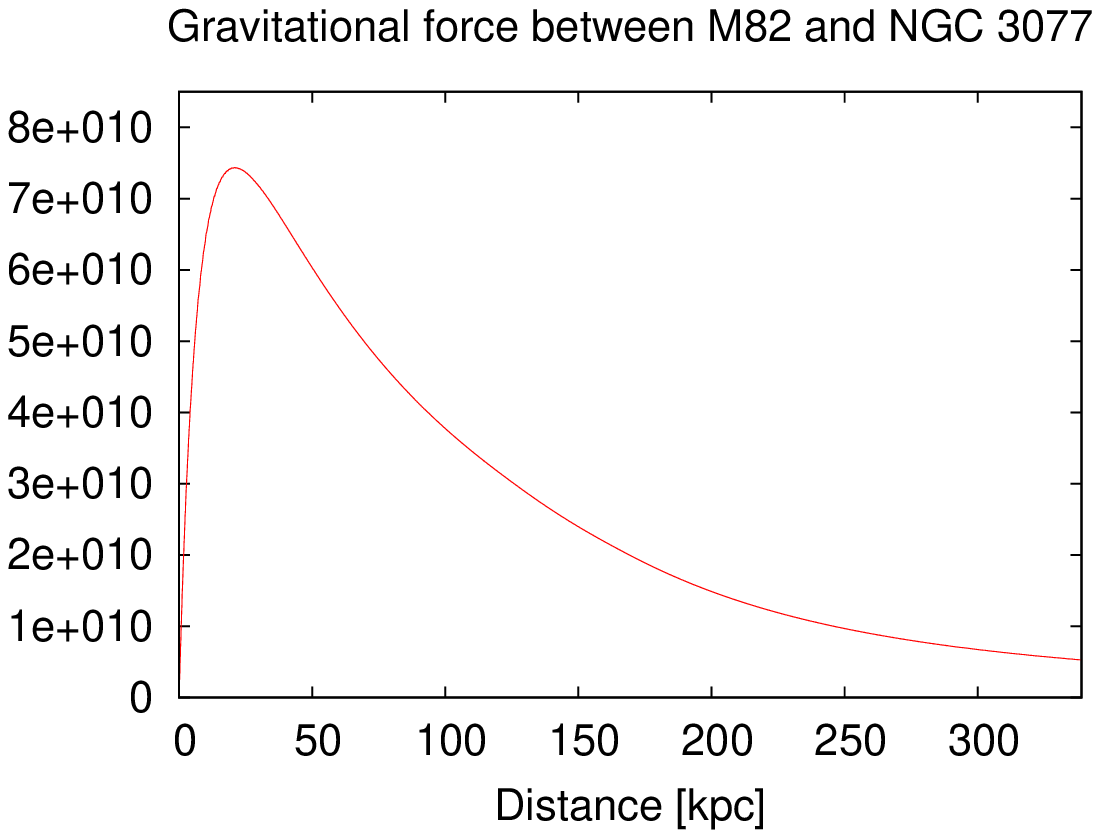}\\
\includegraphics[width=5.5cm]{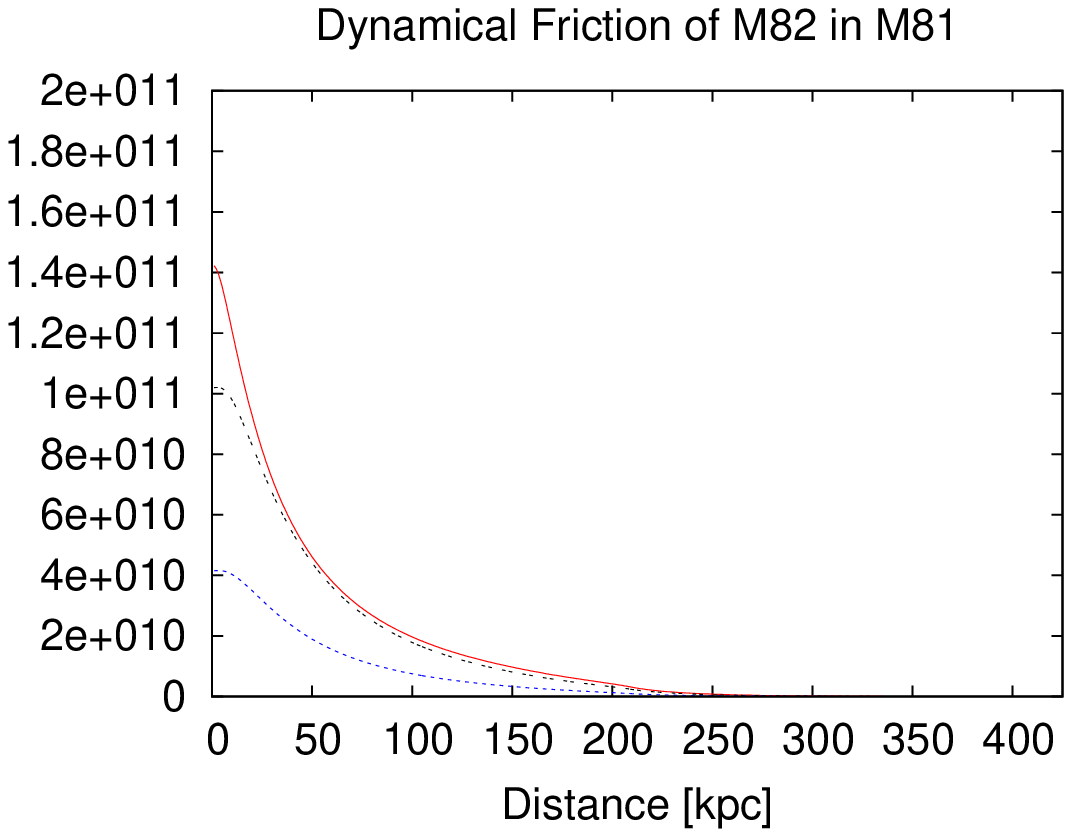} & \includegraphics[width=5.5cm]{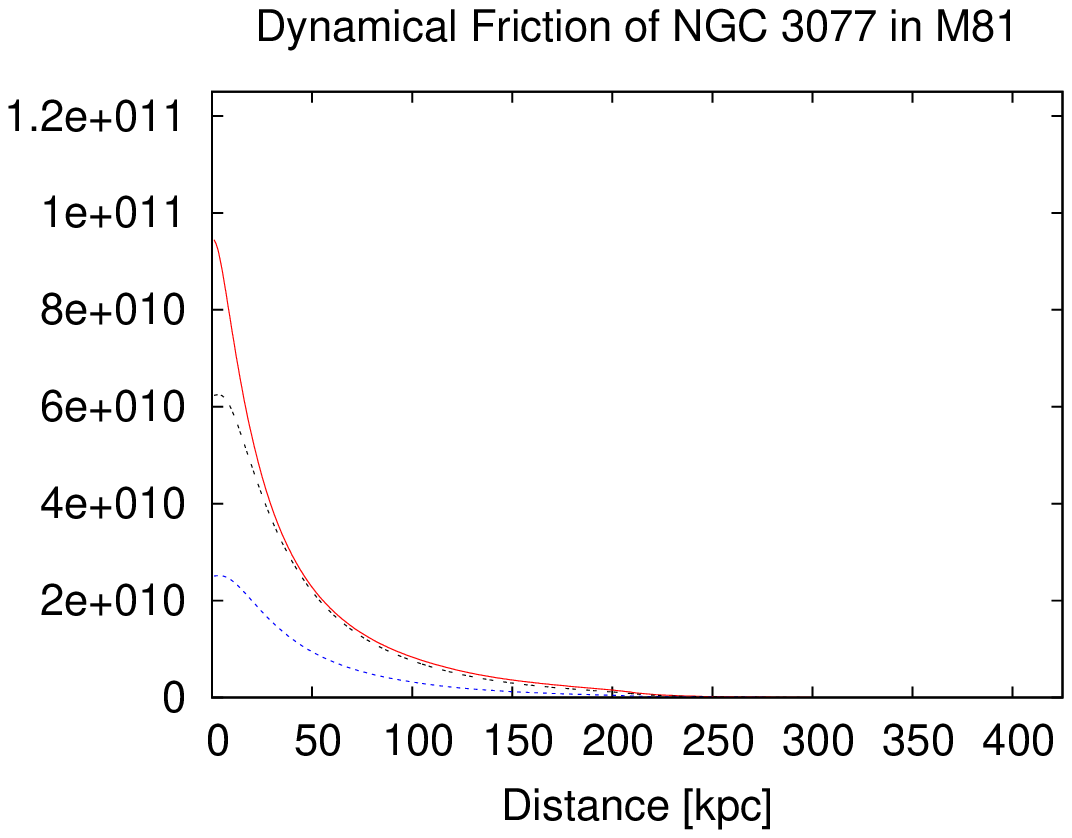} & 
\includegraphics[width=5.5cm]{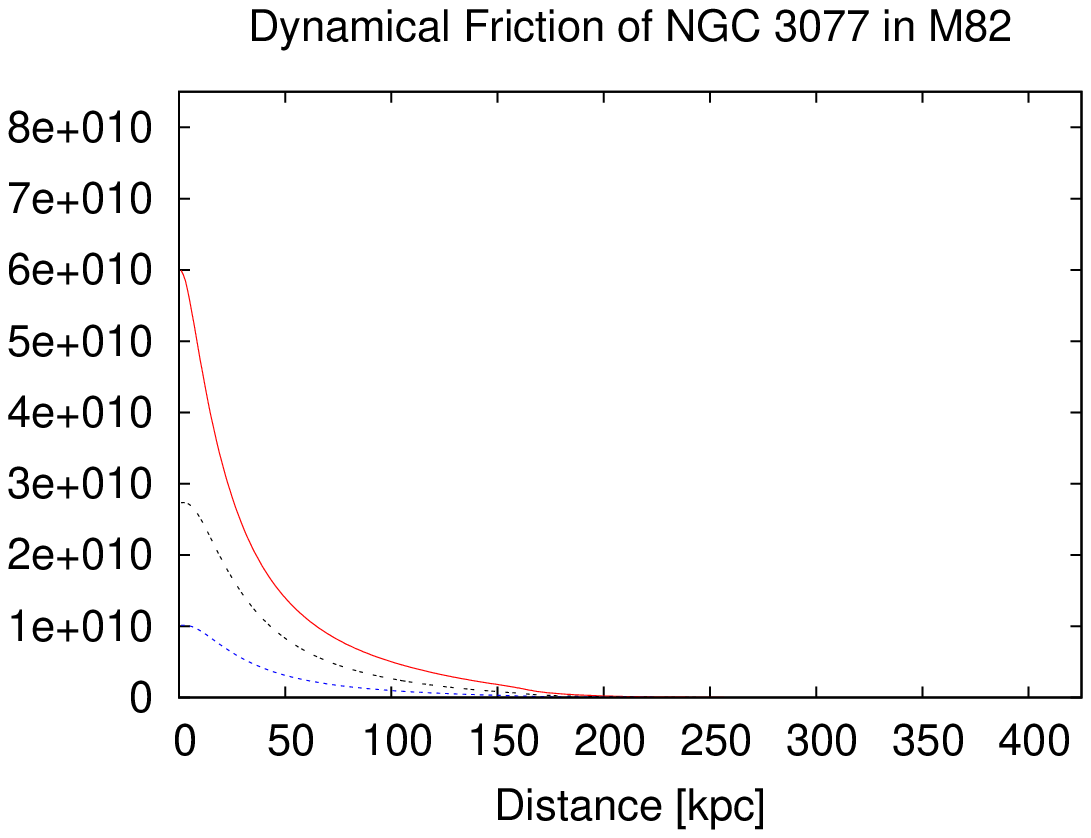}\\
\includegraphics[width=5.5cm]{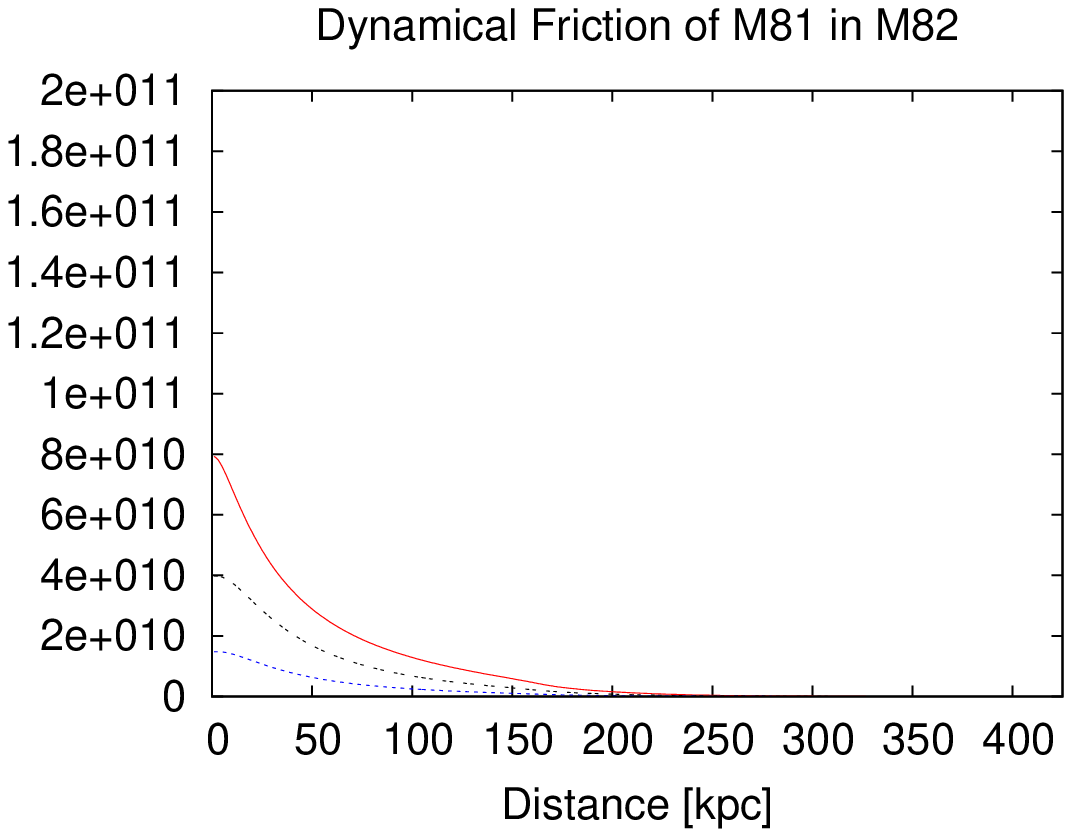} & \includegraphics[width=5.5cm]{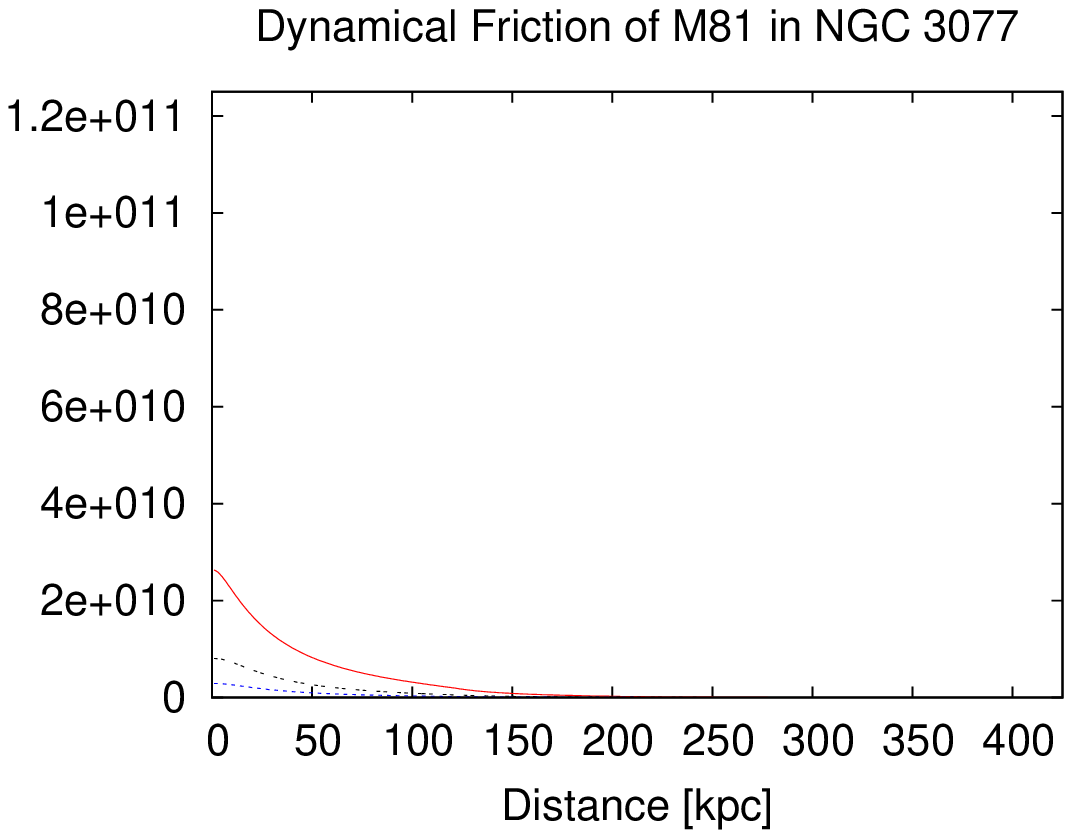} & 
\includegraphics[width=5.5cm]{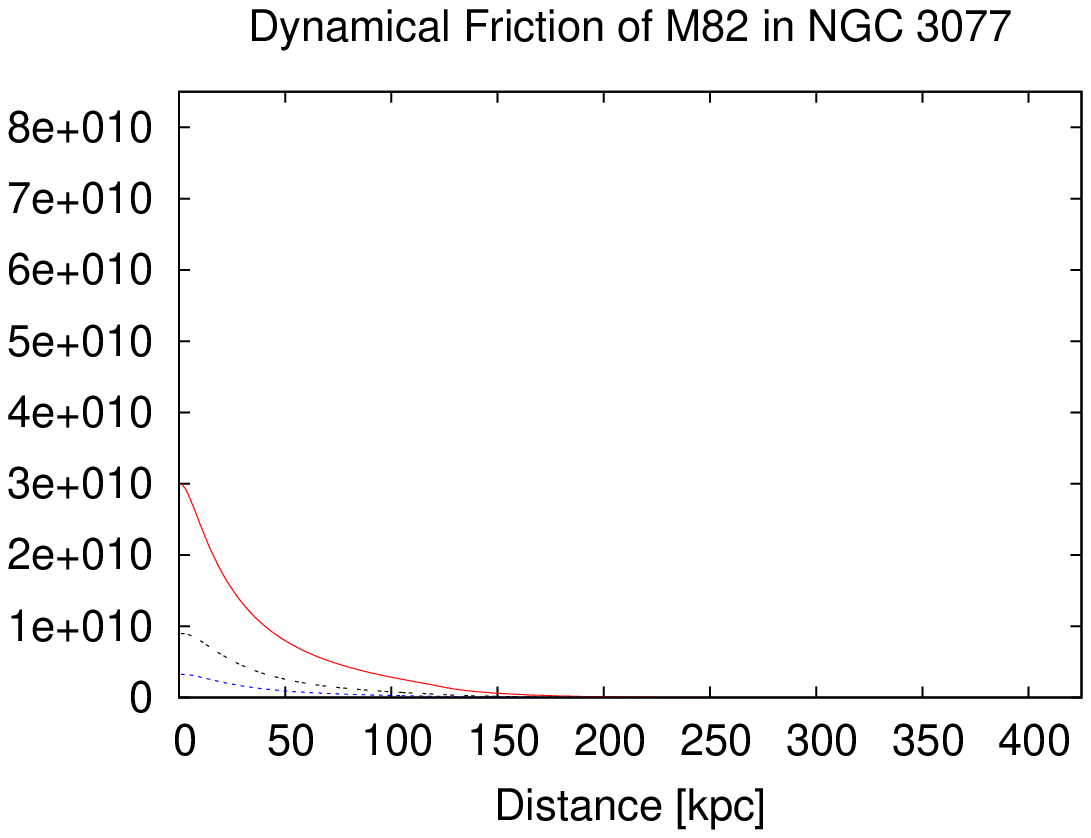}\\
\includegraphics[width=5.5cm]{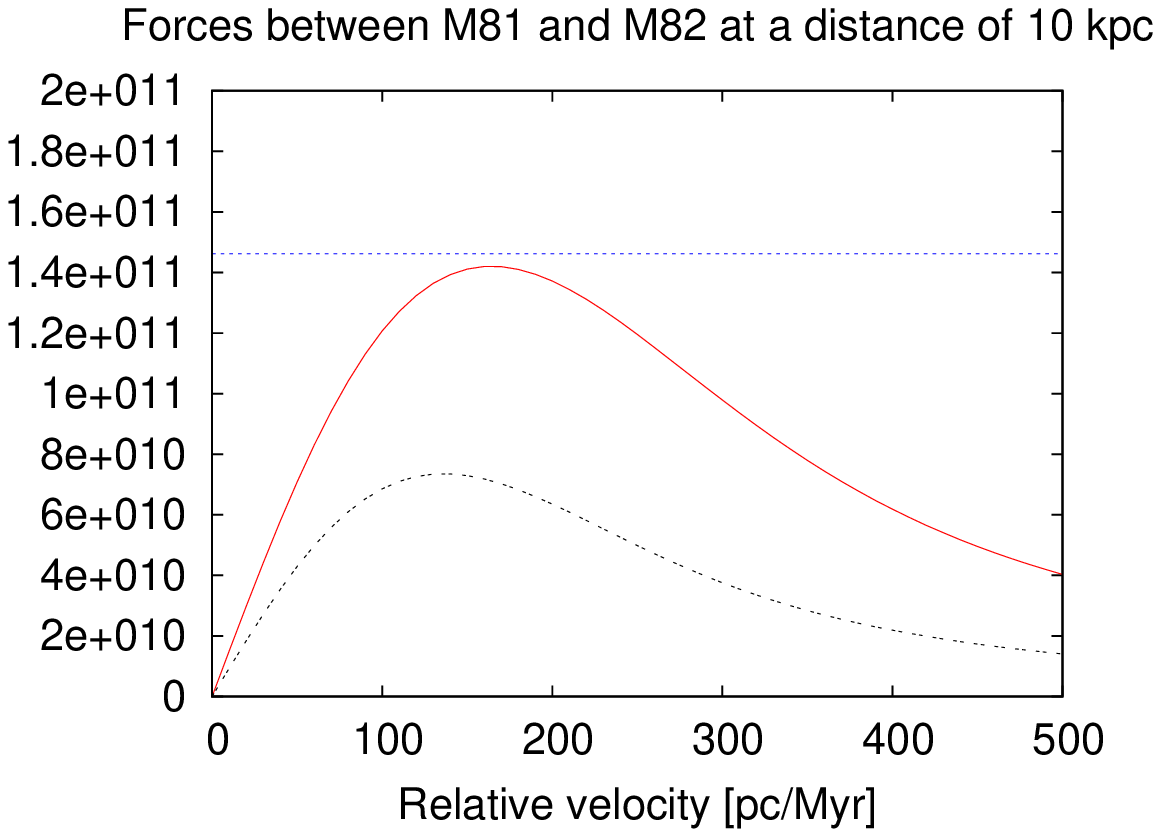} & \includegraphics[width=5.5cm]{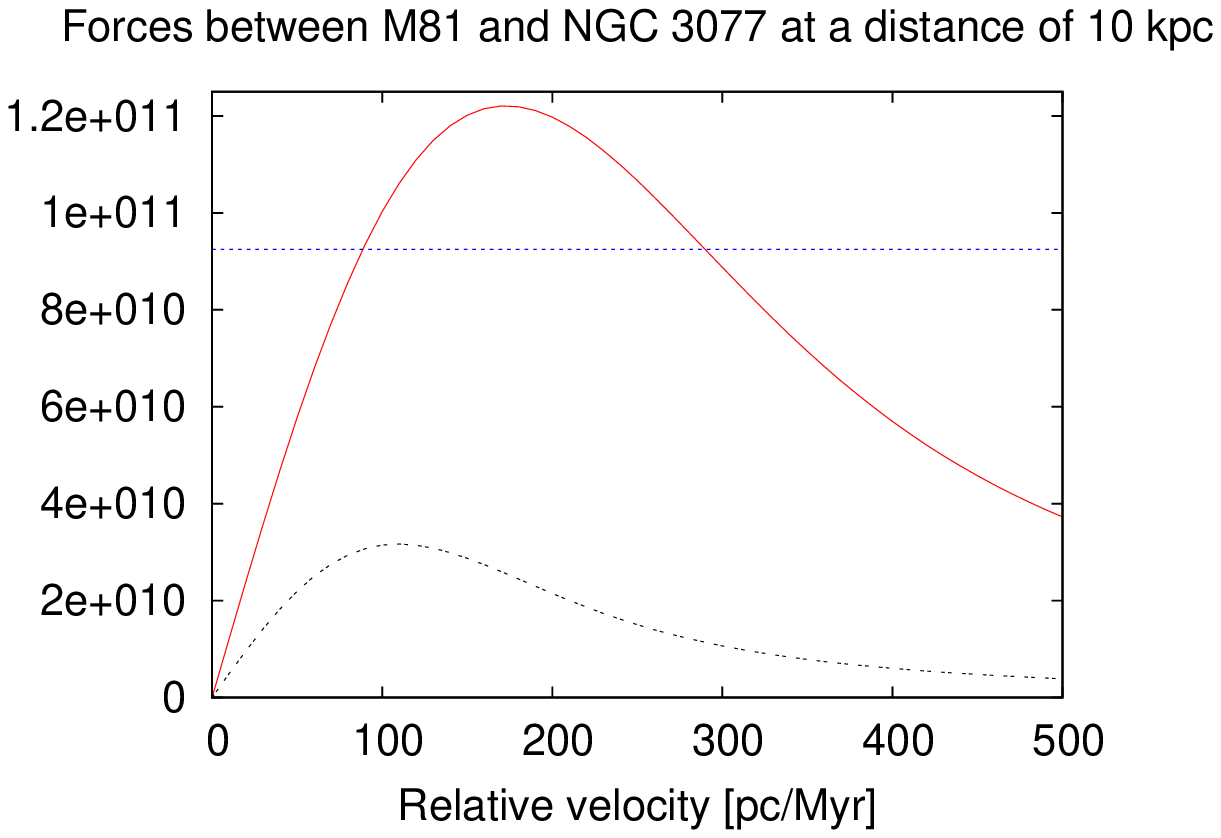} & \includegraphics[width=5.5cm]{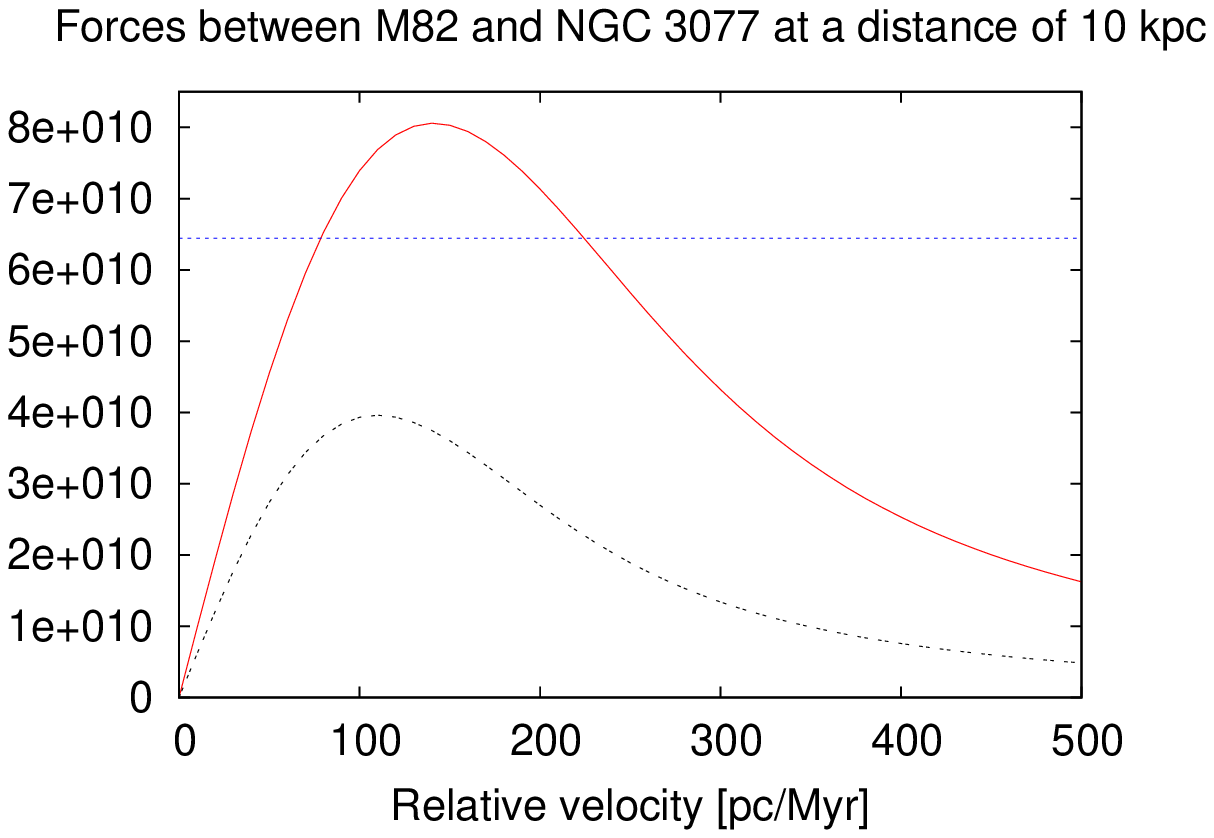}\\
\end{tabular}
\caption{The variety of forces between the galaxies presented in
  galactic units ($M_{\odot}$, pc, Myr). {\bf First row}: The
  gravitational forces between M81 and M82 (left), M81 and NGC~3077
  (centre) and M82 and NGC~3077 (right). The distances between the
  centres of the galaxies are extended by 50~kpc beyond the separation
  distance of the DM halos, thus visualizing the continous transition
  to Newton's law of gravitation for point masses
  upon separation. {\bf Second row}:
  Forces, caused by dynamical friction, on M82 moving in the DM halo
  of M81 (left), NGC~3077 in M81 (centre) and NGC~3077 in M82 (right),
  for relative velocities of 100~pc/Myr (top curves), 300~pc/Myr
  (middle curves) and 500~pc/Myr (bottom curves). {\bf Third row}:
  Same as second row, but vice versa. {\bf Fourth row}: Forces due
  to gravitation (constant values) and dynamical friction at a
  separation distance of 10~kpc between the centres of the DM halos,
  shown in dependence on the relative velocity. Left: M82 in M81 (top
  curve) and M81 in M82 (bottom curve). Centre: NGC~3077 in M81 (top
  curve) and M81 in NGC~3077 (bottom curve). Right: NGC~3077 in M82
  (top curve) and M82 in NGC~3077 (bottom curve).}
\label{forces}
\end{figure*}

\subsection{Approach}
\label{sec:approach}

At first, calculating three-body orbits backwards up to $-7$~Gyr,
statistical populations for the open parameters
(Table~\ref{parameters}) are generated by means of the methods of
Sections~\ref{sec:MCMC}~(MCMC) and \ref{sec:GA}~(GA). Following the
results of the previous publications mentioned in
Section~\ref{PrevRes} we added, additionally to the known initial
conditions at present, the rather general condition (later referred to
as COND) that \emph{both companions M82 and NGC~3077 encountered M81
  within the recent 500~Myr at a pericentre distance below 30~kpc}.

Each three-body orbit of those populations is fully determined by all
the known and open parameters provided by either MCMC or GA.
Starting at time $-7$~Gyr and
calculating the corresponding three-body orbits forward in time up to
$+7$~Gyr, the behaviour of the inner group is investigated
with respect to the question of possibly ocurring mergers in the
future. 

\section{Statistical Methods I: Markov Chain Monte Carlo}
\label{sec:MCMC}

We follow a methodology proposed by \citet{Goodman2010} employing an
affine-invariant ensemble sampler for the Markov chain Monte Carlo
method (MCMC). A solid instruction "how to go about implementing this
algorithm" can be found in \citet{Foreman2013}, who also provide an
online link to a Python implementation. However, we decided to realize
this algorithm on our own utilizing the ABAP development workbench
(see Appendix~\ref{app:pd}). The basics of applying this formalism to
our situation are outlined in Appendix~\ref{app:MCMC}.

\subsection{Definition of the Posterior Probability Density} 

First of all, concerning the open parameters, we need to account for
the distance ranges specified in Table~\ref{coordinates} as well as
the fact that realistic velocities should be considered. This is
ensured by an appropriate definition of the prior distribution
$p(\vec{X})$.

Regarding the distance ranges for M82 and NGC~3077 we define ($i = 2, 3$)

\begin{equation}
\label{eq:prio-z}              
p_{z}(\vec{X})_i \propto \left\{
\begin{array}{ll}
\exp\left({-{\displaystyle\frac{\left( {z_i}-{z_{min}} \right)^2}{2\cdot {z_0}^2}}}\right), & z_i < z_{min}\ ,\\
\exp\left({-{\displaystyle\frac{\left( {z_i}-{z_{max}} \right)^2}{2\cdot {z_0}^2}}}\right), & z_i > z_{max}\ ,\\
1, &  \textrm{otherwise}\ ,\\
\end{array} \right. 
\end{equation}
with the appropriate values for $z_{min}$ and $z_{max}$ for either
companion derived from Table~\ref{coordinates}, and a value of 100~kpc
for~$z_0$. To account for realistic velocities $v_i = \
\mid\vec{v}_i\mid$ (about $\le$~500~km/s) in the centre-of-mass frame
we implemented ($i = 2, 3$)

\begin{equation}
\label{eq:prio-v}              
p_{v}(\vec{X})_i \propto \left\{
\begin{array}{ll}
1, &  v_i \le v_{max}\ ,\\
\exp\left({-{\displaystyle\frac{\left( {v_i }-{v_{max}} \right)^2}{2\cdot {v_0}^2}}}\right), & v_i > v_{max}\ ,
\end{array} \right. 
\end{equation}
with $v_{max}$~=~400~pc/Myr and $v_0$~=~100~pc/Myr. All together our prior distribution is given by

\begin{equation}
\label{eq:prio}
p(\vec{X}) \ \propto \ \prod_{i=2}^{3} \  p_{z}(\vec{X})_i  \ \cdot \  p_{v}(\vec{X})_i \ .
\end{equation}

Exploiting the minimal distances $d_{12}$ and $d_{13}$ within the
recent 500~Myr between M81/M82 and M81/NGC~3077, respectively, the
condition COND specified in Section~\ref{sec:approach} is incorporated
by the following contribution to the likelihood function ($i = 2, 3$):

\begin{equation}
\label{eq:like-dist}              
P_{C}(\vec{X} \mid D)_i \propto \left\{
\begin{array}{ll}
1, &  d_{1i} \le d_{per}\ ,\\
\exp\left({-{\displaystyle\frac{\left( {d_{1i}}-{d_{per}} \right)^2}{2\cdot {d_0}^2}}}\right), &  d_{1i} > d_{per}\ ,
\end{array} \right. 
\end{equation}
with $d_{per}$~=~25~kpc and ${d_0}$~=~5~kpc. Denoting the initial
velocities at $-7$~Gyr (prior to the encounters) in the centre-of-mass
frame by $u_i=\mid\vec{u}_i\mid$, we generate realistic values for
those by

\begin{equation}
\label{eq:like-v}              
P_{u}(\vec{X} \mid D)_i \propto \left\{
\begin{array}{ll}
1, & u_i \le v_{max}\ ,\\
\exp\left({-{\displaystyle\frac{\left( {u_i}-{v_{max}} \right)^2}{2\cdot {v_0}^2}}}\right), &  u_i  > v_{max}\ ,
\end{array} \right. 
\end{equation}
taking the same values for $v_{max}$ and $v_0$ as in
Eq.~\ref{eq:prio-v}. Assembling both aspects we arrive at the
likelihood function

\begin{equation}
\label{eq:like}
P(\vec{X} \mid D) \ \propto \ \prod_{i=2}^{3} \ P_{C}(\vec{X} \mid D)_i \ \cdot \ P_{u}(\vec{X} \mid D)_i \ .
\end{equation}

\subsection{The First Ensemble}
\label{sec:first}

Generating random numbers $n\in\{0,...,1000\}$ separately for each
walker and open parameter, the first ensembles are created according
to ($i\in\{1,...,6\}$)

\begin{displaymath}
P_i = (P_i)_{min} + n \cdot \left[  (P_i)_{max} - (P_i)_{min} \right] / 1000\ .
\end{displaymath}
The minimum and maximum values for the distances are derived from
Table~\ref{coordinates}, and for the POS velocity components chosen to
be $\pm$~500~pc/Myr.

\subsection{Results}
\label{sec:MCMC-results}

\begin{figure}
\centering
\includegraphics[width=8cm]{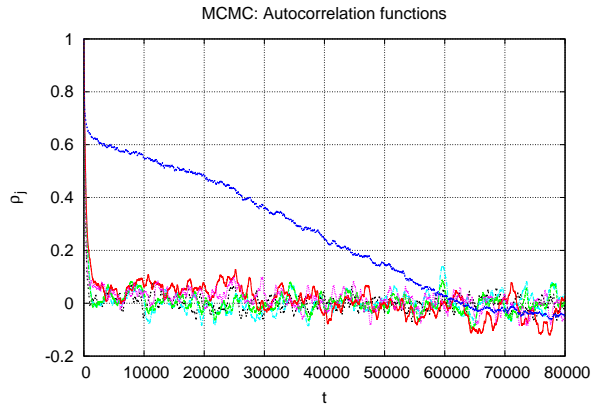}
\caption{Autocorrelation functions $\rho_j(t)$ ($j$~=~1,...,6), calculated for $300\,000$ ensembles of 100 walkers and displayed for the first $80\,000$ ensembles.}  
\label{ACF}
\end{figure}

The comparison of the autocorrelation functions (Eq.~\ref{eq:ac2}) for
ensembles of either 100 or 1000 walkers ($N_L=100,~1000$), creating a
chain of $50\,000$ ensembles in either case, did not show a significant
advantage for the choice of the larger ensemble size over the smaller
one. Therefore the evaluations were performed with $N_L=100$ creating
a chain of $300\,000$ ensembles. Figure~\ref{ACF} shows the behaviour
of the corresponding autocorrelation functions.  The acceptance rates
for the succeeding walkers, generated by the stretch moves, appeared
to be $\approx0.3$, in accordance with the fact that a range of
$(0.2,0.5)$ is considered to be a good value in that regard.

Examining the autocorrelation functions, one realizes that a fairly
good convergence towards the requested value of~0 is already achieved
after a couple of thousands of ensembles, except for parameter 6 which
represents the distance of NGC~3077, where it takes about $60\,000$
ensembles. However, afterwards the autocorrelation functions keep
varying, mostly within the range of $(-0.1,0.1)$. A detailed discussion of this 
behaviour is left for Section~\ref{sec:comp}.

As explained in Section~\ref{sec:approach} the ensemble chains serve
as input for three-body integration runs into the future. Taking each
of the ensembles t~=~$75\,000$, $100\,000$, ..., $300\,000$ as a
starting point for continuing the MCMC procedure, we extracted the
first 1000 occurences of walkers fulfilling condition COND in each
case.  This produces ten independent follow-up sets of walkers. As
soon as one of the companions is seperated less than 15~kpc (the
baryonic radius of M81's disc) from M81 and does not leave this
spatial extension thereafter anymore, we regard this time as the
time of the merger process.

\begin{figure}
\centering
\begin{tabular}{cc}
\includegraphics[width=4cm]{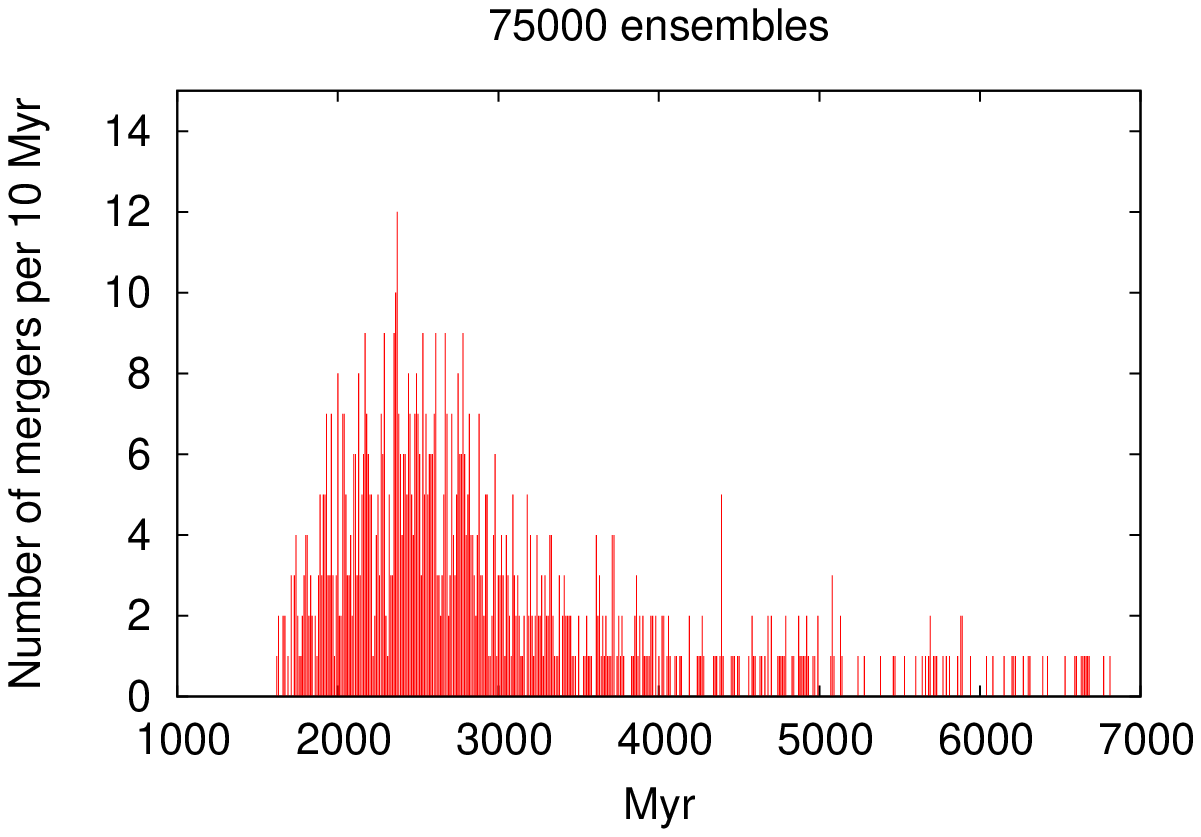} & \includegraphics[width=4cm]{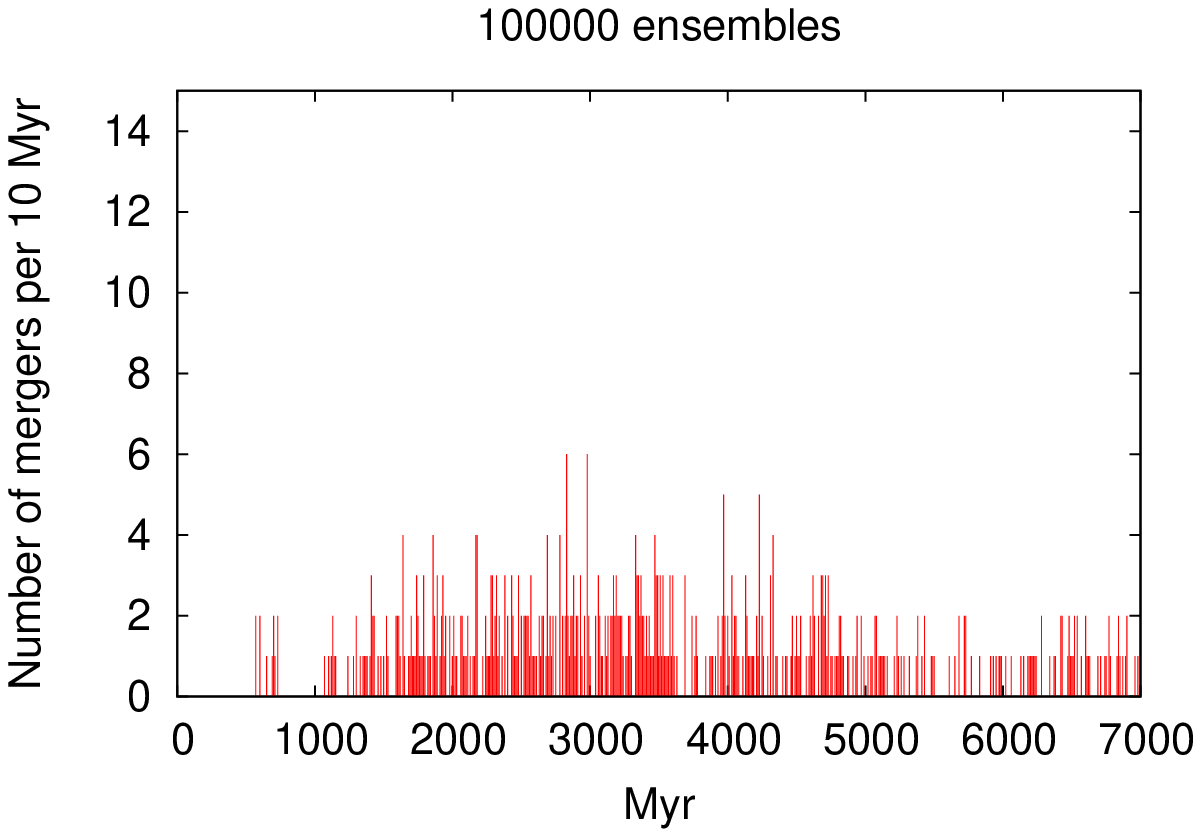}\\
\includegraphics[width=4cm]{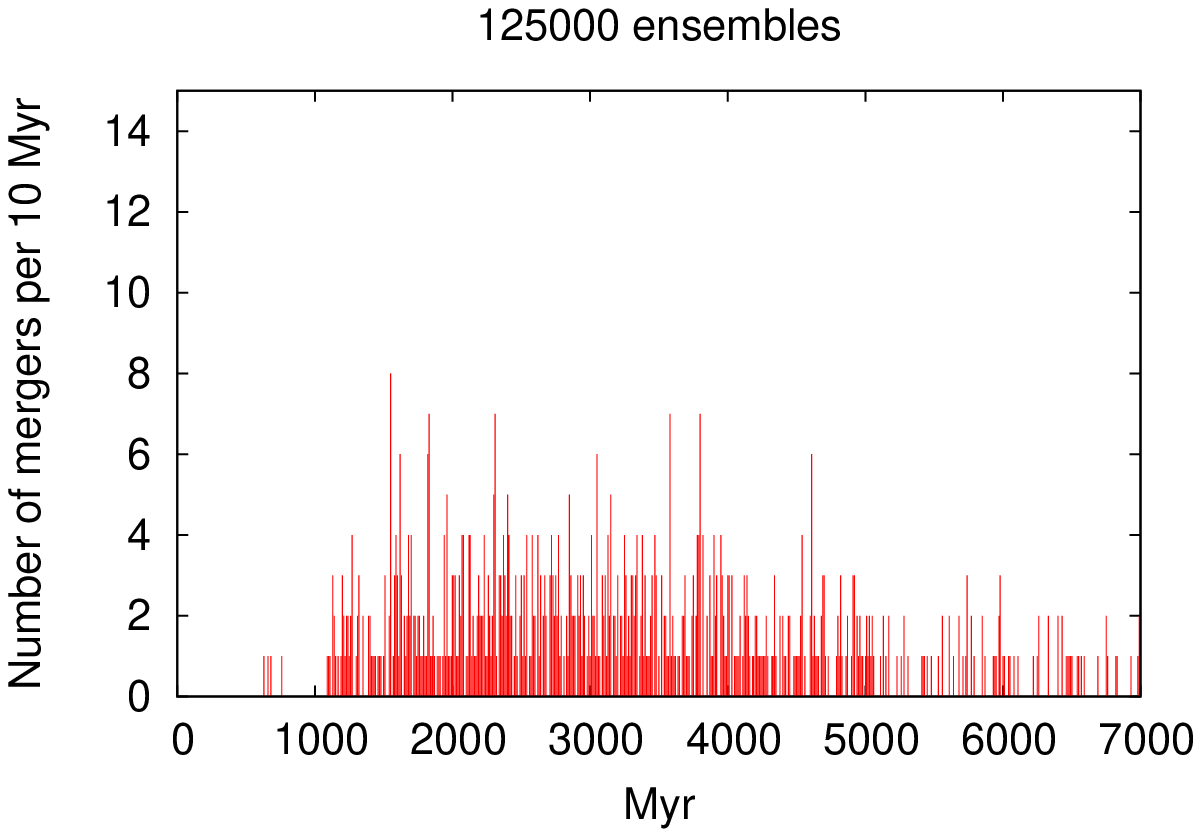} & \includegraphics[width=4cm]{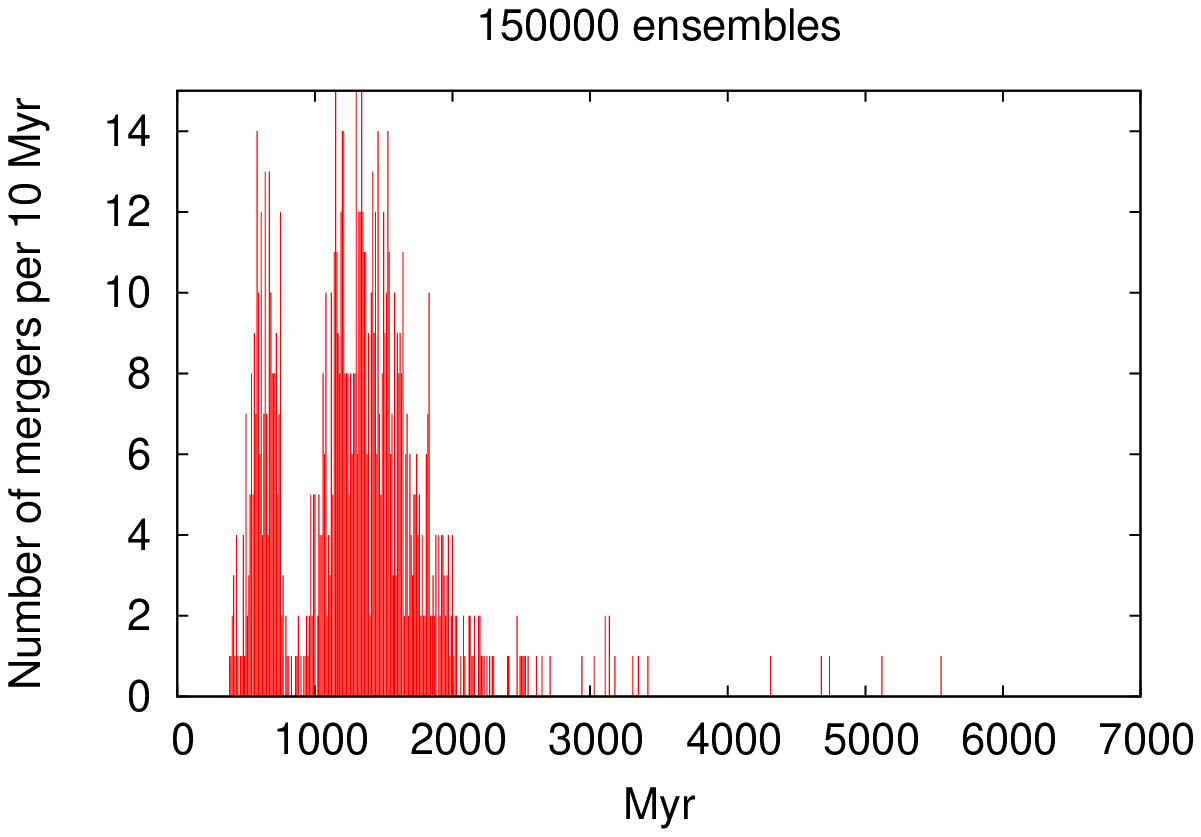}\\
\includegraphics[width=4cm]{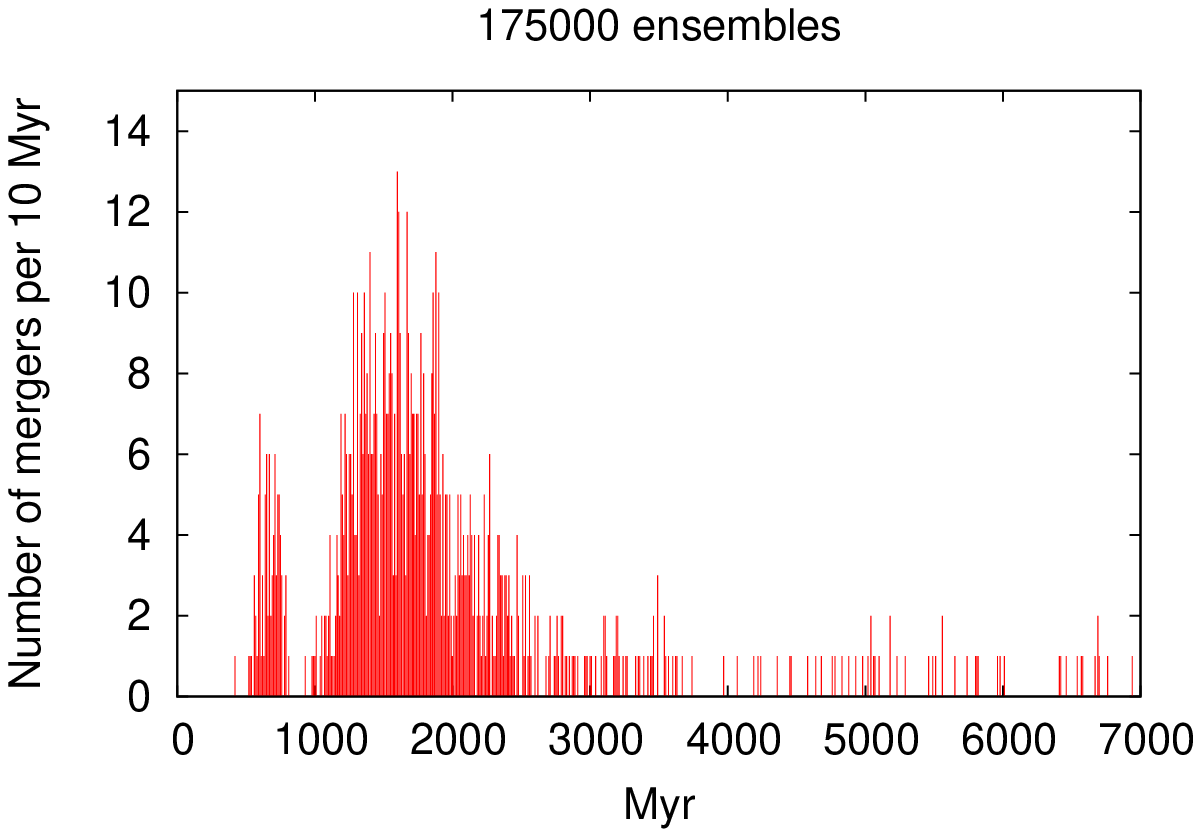} & \includegraphics[width=4cm]{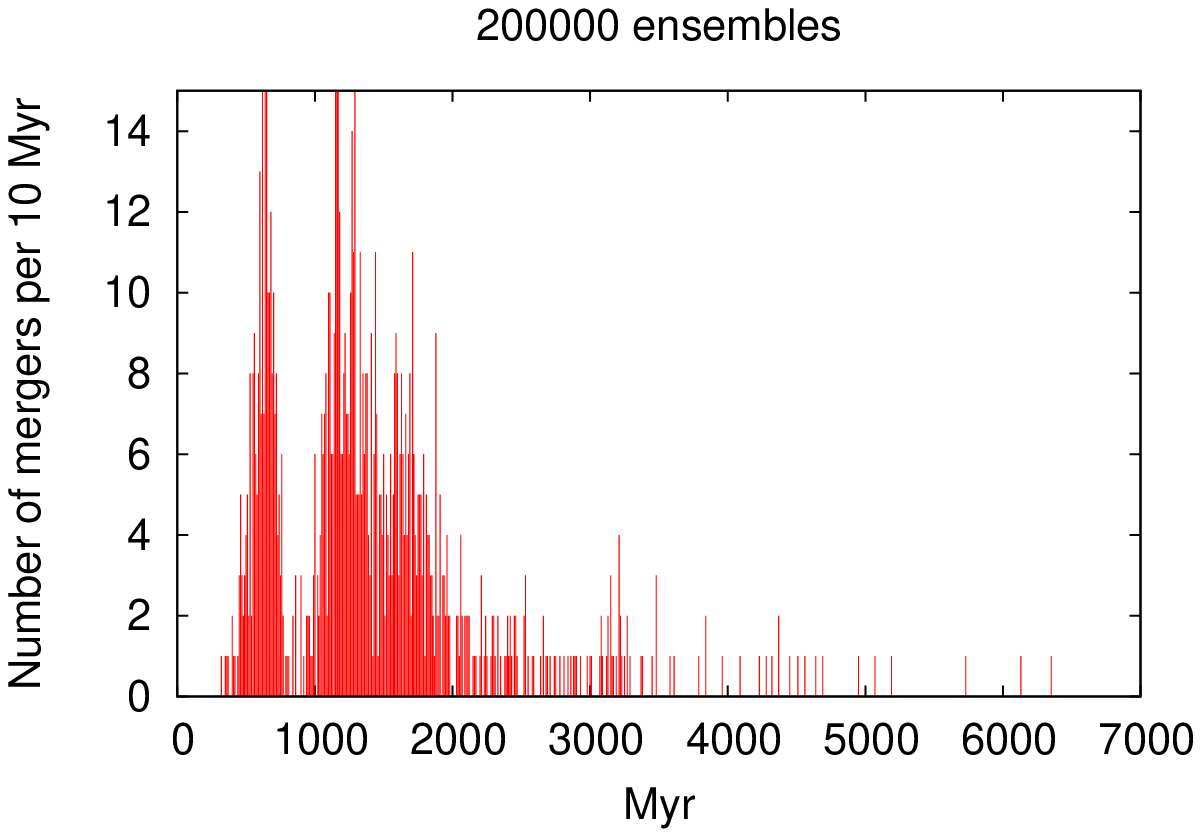}\\
\includegraphics[width=4cm]{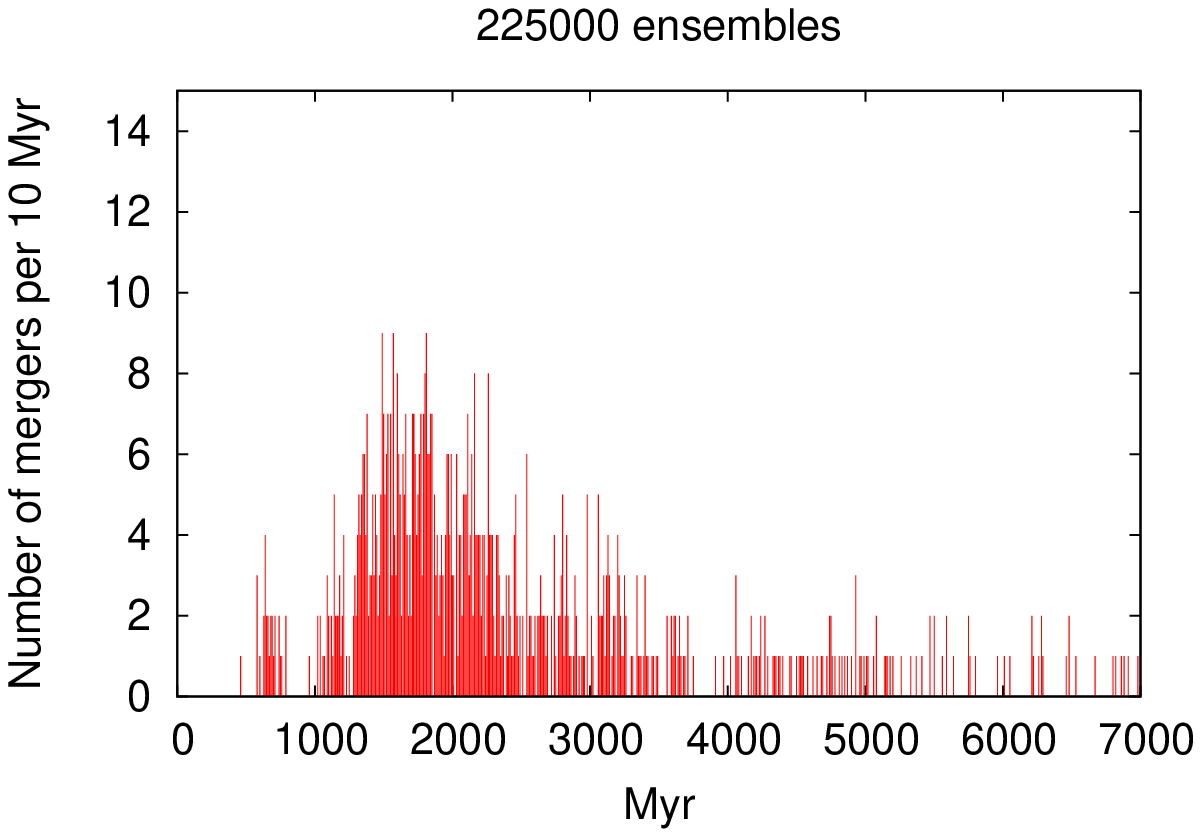} & \includegraphics[width=4cm]{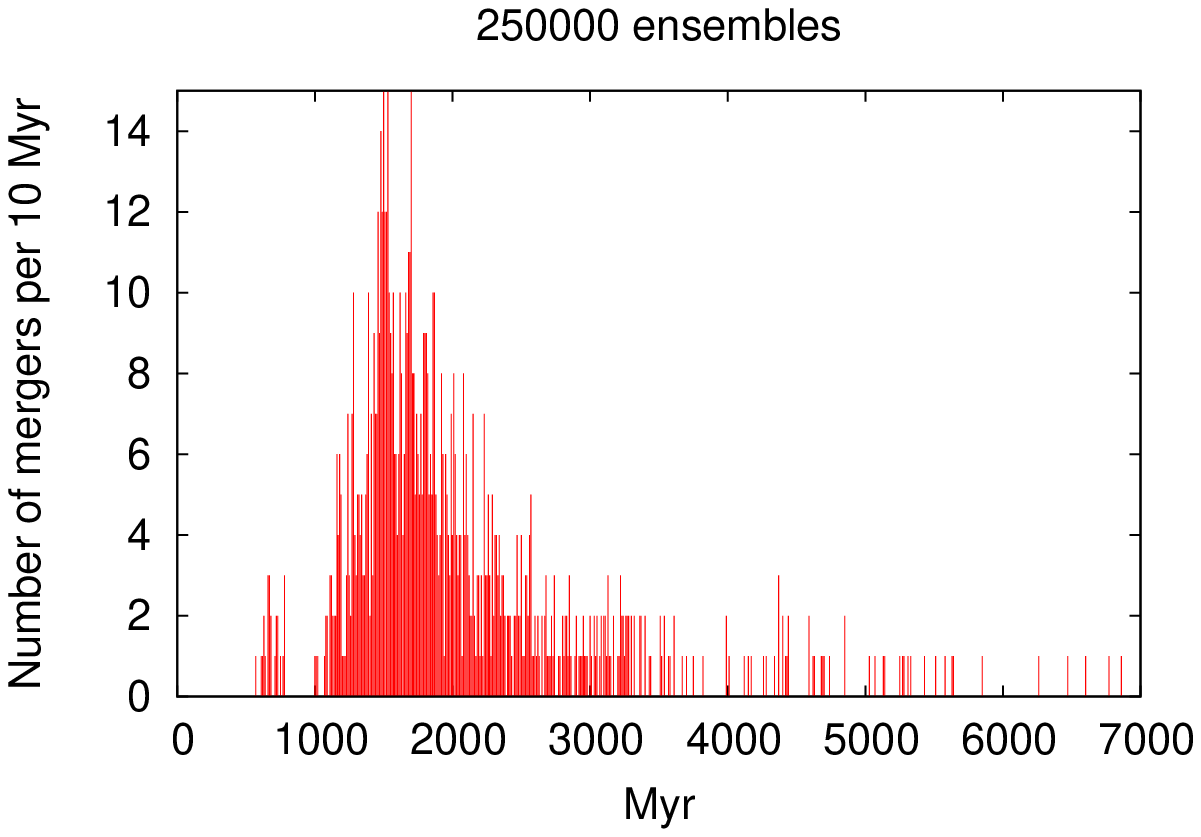}\\
\includegraphics[width=4cm]{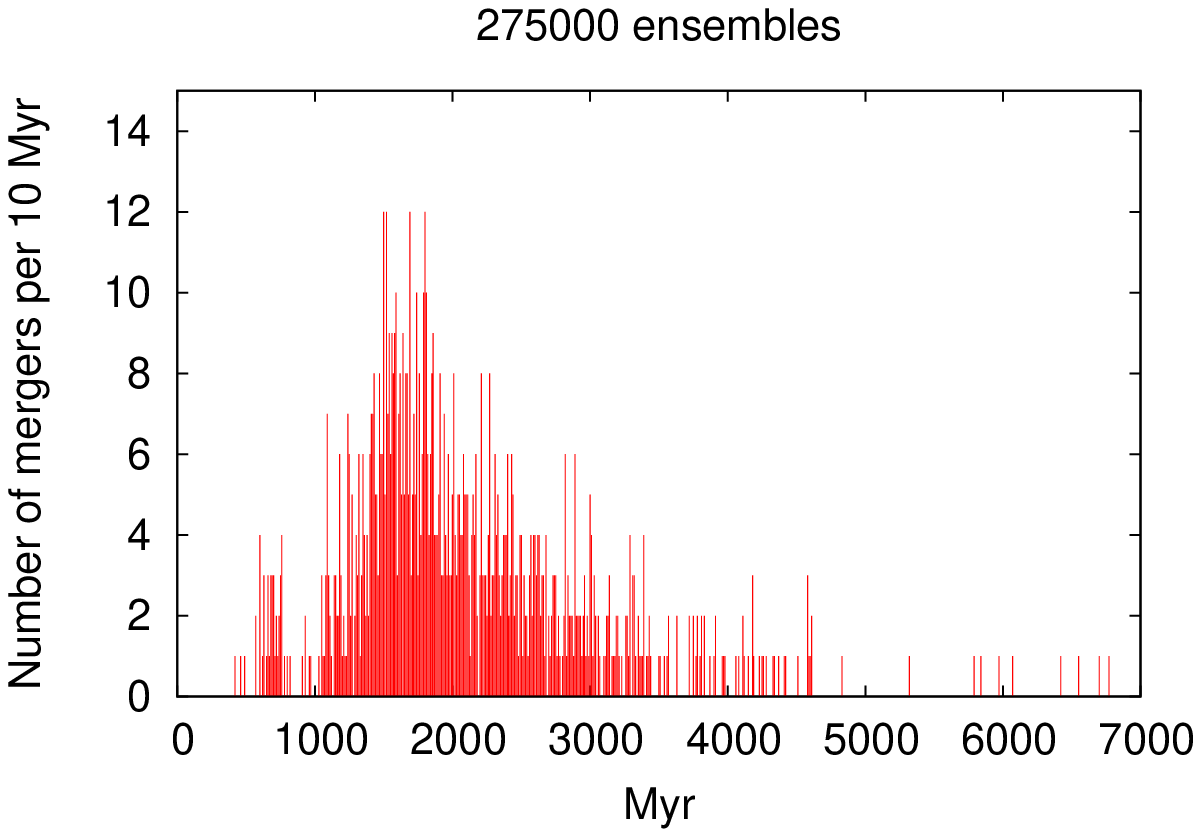} & \includegraphics[width=4cm]{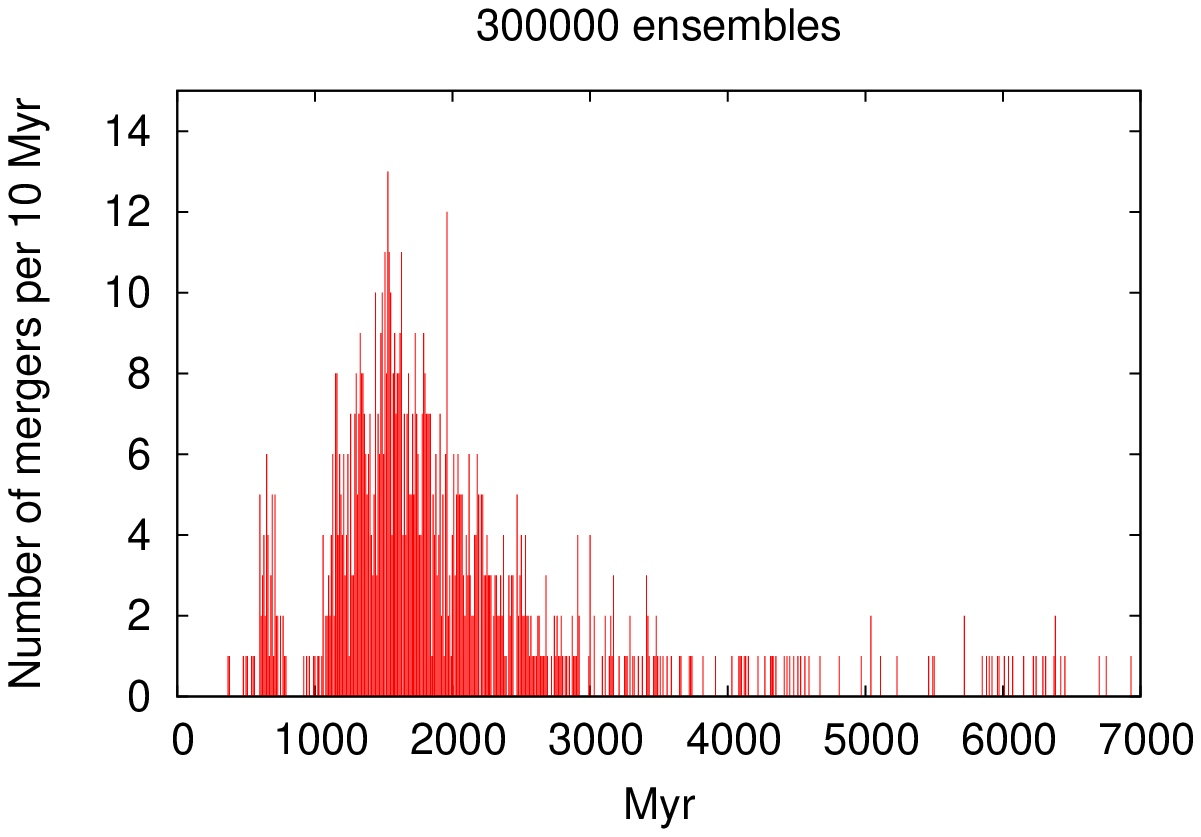}\\
\end{tabular}
\caption[]{MCMC: Merger rates based on the begin times as of the present 
displayed for the next 7~Gyr. (For further details refer to the text.)}  
\label{MC_merger}
\end{figure}

\begin{figure}
\centering
\begin{tabular}{cc}
\includegraphics[width=4cm]{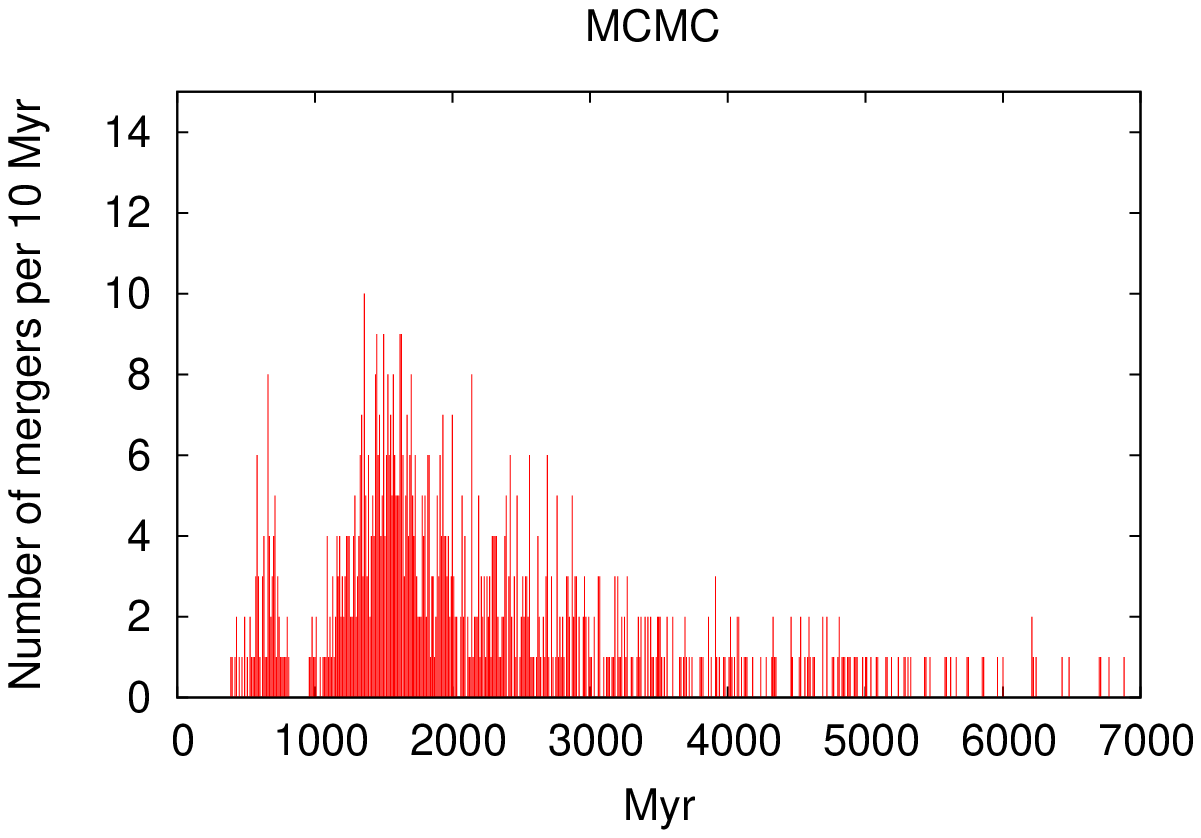} & \includegraphics[width=4cm]{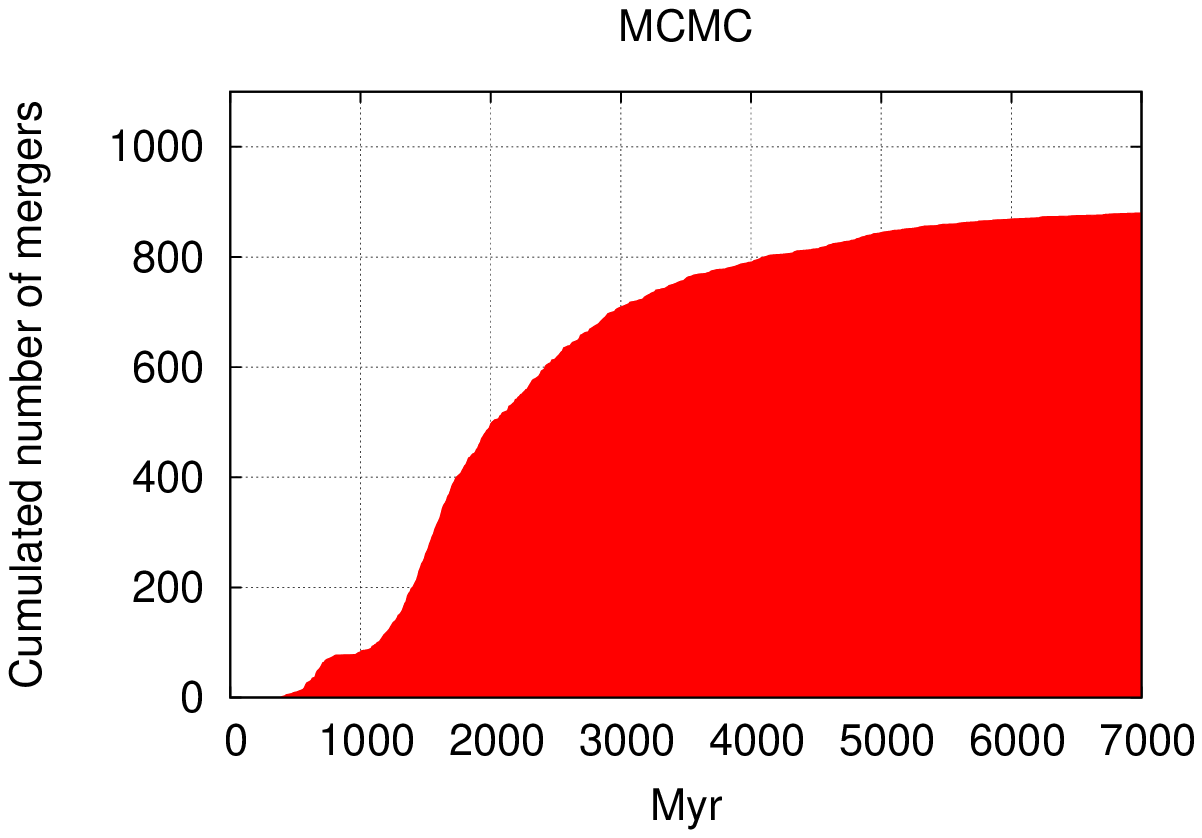}\\
\end{tabular}
\caption[]{MCMC: Merger rates and cumulated number of mergers 
displayed for the forthcoming 7~Gyr. In total 1000 walkers were considered 
by randomly extracting 100 walkers out of each of the ten follow-up sets.}  
\label{MC_cum}
\end{figure}

In accordance with the statements concerning the behaviour of the
autocorrelation functions we present the merger rates within the
forthcoming 7~Gyr in Figure~\ref{MC_merger}. Apparently no full
convergence is achieved, and the results appear "to walk around". As
an attempt to achieve meaningful results we calculated the averages of
the ten follow-up sets and show them in Figure~\ref{MC_cum} and
Table~\ref{tab:merger}.

Instead of applying clustering or alike methods in order to achieve
full convergence, we decided to employ another statistical method, the
genetic algorithm, which overcomes difficulties like the problem of
local maxima and the above mentioned one. As we will see, the results
obtained there are in good agreement with those obtained by the MCMC
method.

\section{Statistical Methods II: Genetic Algorithm}
\label{sec:GA}

General aspects of the genetic algorithm (GA) are explained in detail
in \citet{Charbonneau1995}. As pointed out there, the major advantage
of GA is perceived to be its capability of avoiding local maxima in
the process of searching the global maximum of a given function
(fitness function). A precise description of the algorithm, especially
instructions for the implementation, can also be found in
\citet{Theis2001}.

In our case each open parameter from Table~\ref{parameters} is mapped
to a 4-digit string ("gene") $\left[ abcd \right]_i$
($i\in\{1,...,6\}$):

\begin{equation}
\label{eq:gene}
P_i = (P_i)_{min} + \left[ abcd \right]_i  \cdot \left[  (P_i)_{max} - (P_i)_{min} \right] / 10\,000 \ . 
\end{equation}
Here the minimum and maximum values for the parameters $P_i$ are
choosen to be the same as in Section~\ref{sec:first}. All genes
together define the genotypes as 6$\times$4-digit strings

\begin{displaymath}
\ \ \ \ \ \ \ \ \ \ \ \ \ \ \ \ \ \ \ \ \ \ \ \ \fbox{$\left[ abcd \right]_1 ... \left[ abcd \right]_6$} \ ,
\end{displaymath}  
which, when creating a new generation, undergo the procedures of
cross-over, mutation and ranking within a population of $N_{pop}$
genotypes. The first generation is established by randomly creating
$N_{pop}$ 6$\times$4-digit strings.

\subsection{Definition of the Fitness Function}

In contrast to Section~\ref{sec:MCMC}, where the stretch moves may
take walkers outside the ranges specified in Table~\ref{coordinates},
we need not take care of those ranges explicitly because
Eq.~\ref{eq:gene} a~priori guarantees that the distances are confined
to their intervals. Otherwise we follow the goal to establish GA
evaluations compatible to what has been performed by means of
MCMC. Dropping Equation~\ref{eq:prio-z} we therefore arrive at the
following definition for the fitness function

\begin{equation}
\label{eq:fit-I}
\mathcal{F}(\vec{X} \mid D) = \prod_{i=2}^{3} \ p_{v}(\vec{X})_i \cdot P_{C}(\vec{X} \mid D)_i \cdot P_{u}(\vec{X} \mid D)_i \ ,
\end{equation} 
where the individual components are taken from
Equations~\ref{eq:prio-v},~\ref{eq:like-dist}
and~\ref{eq:like-v}. Like in Section~\ref{sec:MCMC}, $\vec{X}$ denotes
the parameter vector while further entities incorporated are indicated
by \emph{D}.

\subsection{Results}
\label{sec:ga-results}

We generated a set of 1000 solutions fulfilling condition COND. Upon
detection of a first solution the GA procedure is being repeated by
randomly creating a new population as a starting point, as explained
above, until 1000 appropriate three-body orbits are collected.

\begin{figure}
\centering
\begin{tabular}{cc}
\includegraphics[width=4cm]{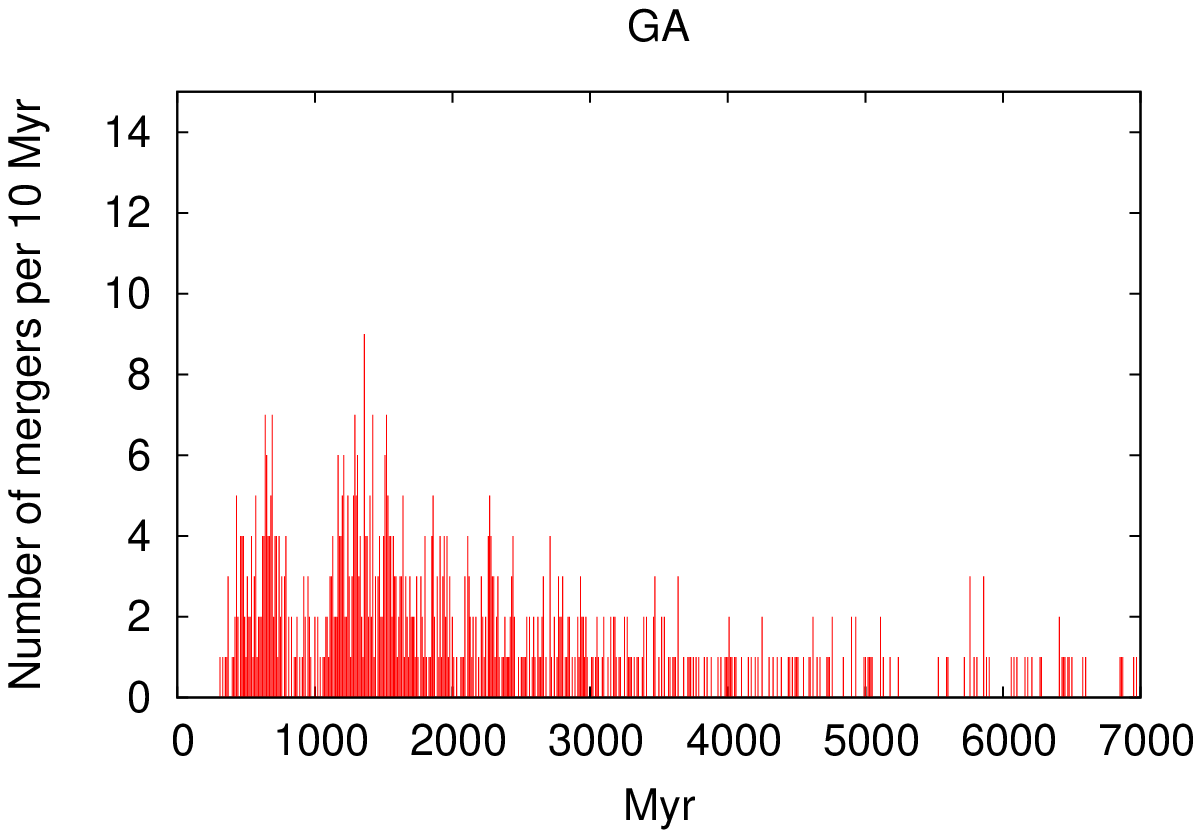} & \includegraphics[width=4cm]{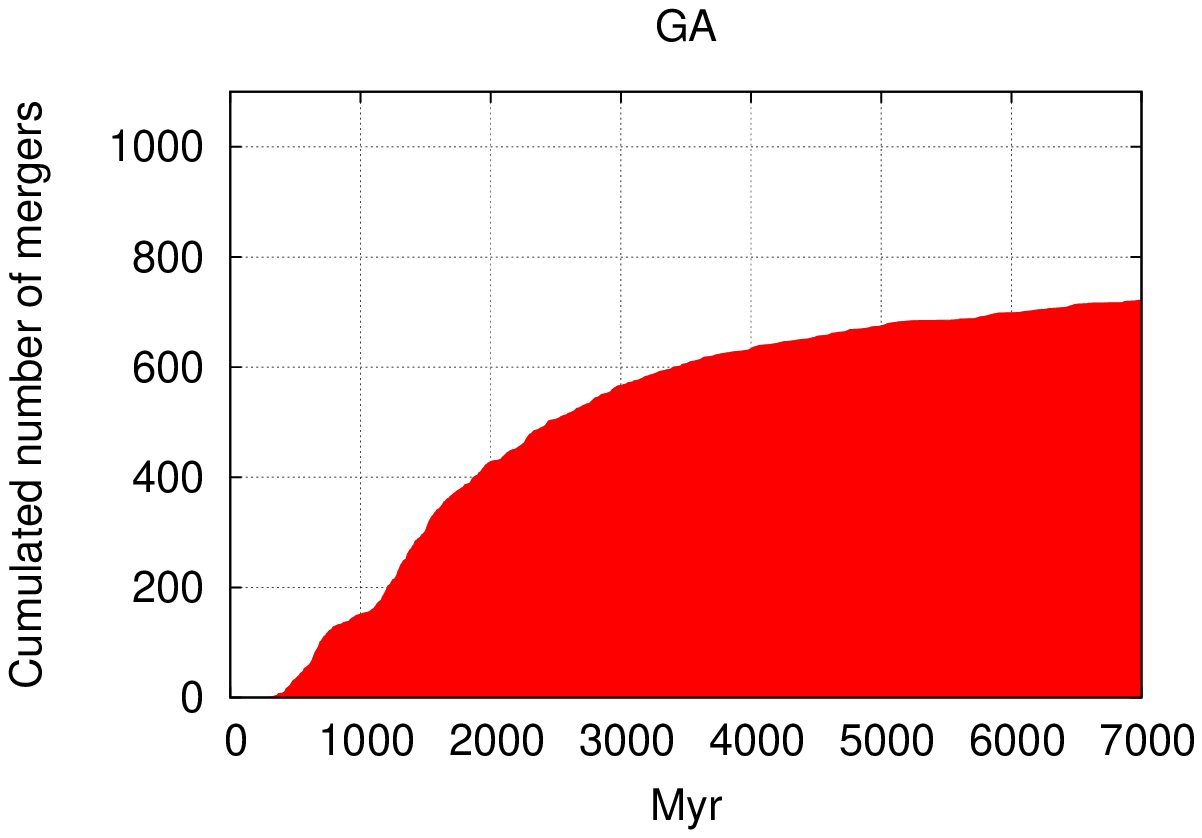}\\
\end{tabular}
\caption[]{GA: Merger rates and cumulated number of mergers 
displayed for the forthcoming 7~Gyr, based on a population of 1000 solutions.}  
\label{GA_I}
\end{figure}

The fitness function corresponds to the posterior
probability density of Section~\ref{sec:MCMC}. In Figure~\ref{GA_I}
the merger events occurring within the next 7~Gyr are presented for
that model. The comparison of Figures~\ref{MC_cum} and~\ref{GA_I}
confirms the statement that both statistical methodes applied, namely
MCMC and GA, yield similar results with respect to the merger
rates, and key numbers are presented in Table~\ref{tab:merger}.

\begin{table*}           
\centering                        
\begin{tabular}{c c c}        
\hline\hline                  
                                                                     & MCMC           & GA                   \\
\hline 
solutions not merging within next 7 Gyr               & 118               &  278                 \\ 
\hline               
solutions not merging  within next 7 Gyr and:       &                      &                         \\                     
neither M82 nor N3077 bound to M81 7 Gyr ago         & 117               &  276                 \\ 
\hline               
solutions not merging  within next 7 Gyr and:       &                      &                         \\                     
one companion bound to M81 7 Gyr ago              & 1                   &  2                    \\ 
\hline           
solutions for:                                               &                       &                          \\                          
M82 and N3077 bound to M81 7 Gyr ago      &  66                 &   70                  \\ 
\hline                                   
longest lifetime from today for:                   &                       &                          \\ 
M82 and N3077 bound to M81  7 Gyr ago                    &  $2.7$ Gyr     &   $2.8$ Gyr      \\ 
\hline                                   
average lifetime from today for:                 &                       &                          \\ 
M82 and N3077 bound to M81  7 Gyr ago                  &  $1.7$ Gyr      &   $1.3$ Gyr       \\ 
\hline
\end{tabular}
\caption{Key numbers for both statistical methods MCMC and GA, based
  on populations of 1000 solutions in either case. Actually, the three
  solutions not merging within the next 7~Gyr where one companion is
  bound to M81 merge after 7.3~Gyr (MCMC), and 7.8 and 8.2~Gyr (GA).}
\label{tab:merger}      
\end{table*}                

\begin{figure*}
\centering
\begin{tabular}{cc}
\includegraphics[width=8.0cm]{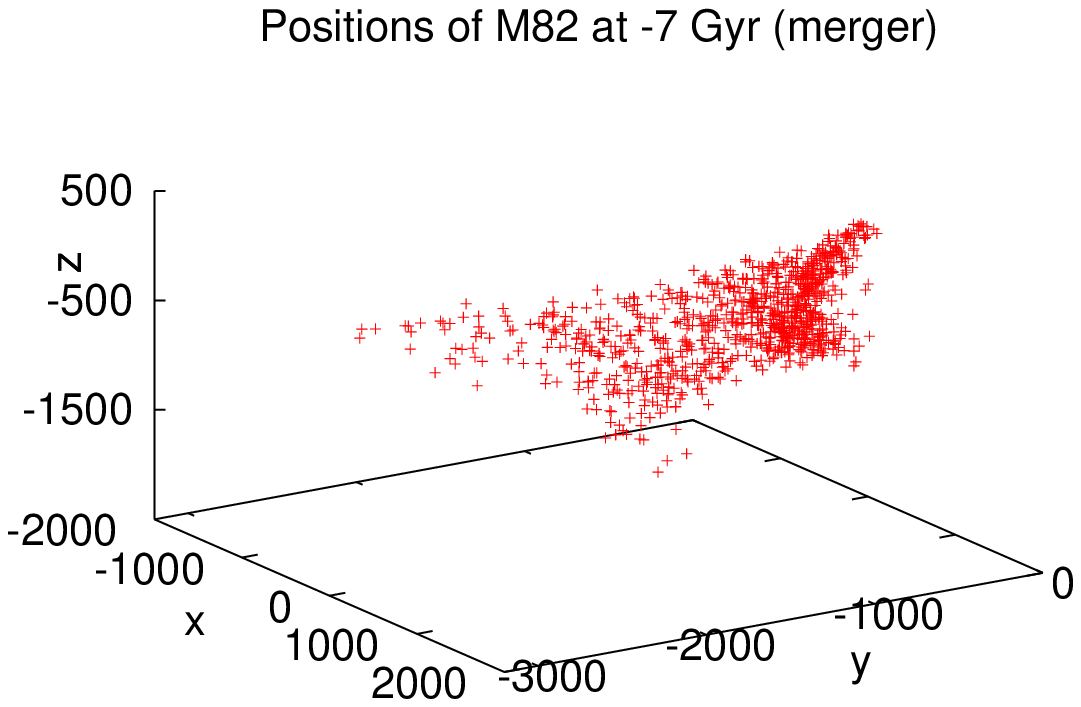} & \includegraphics[width=8.0cm]{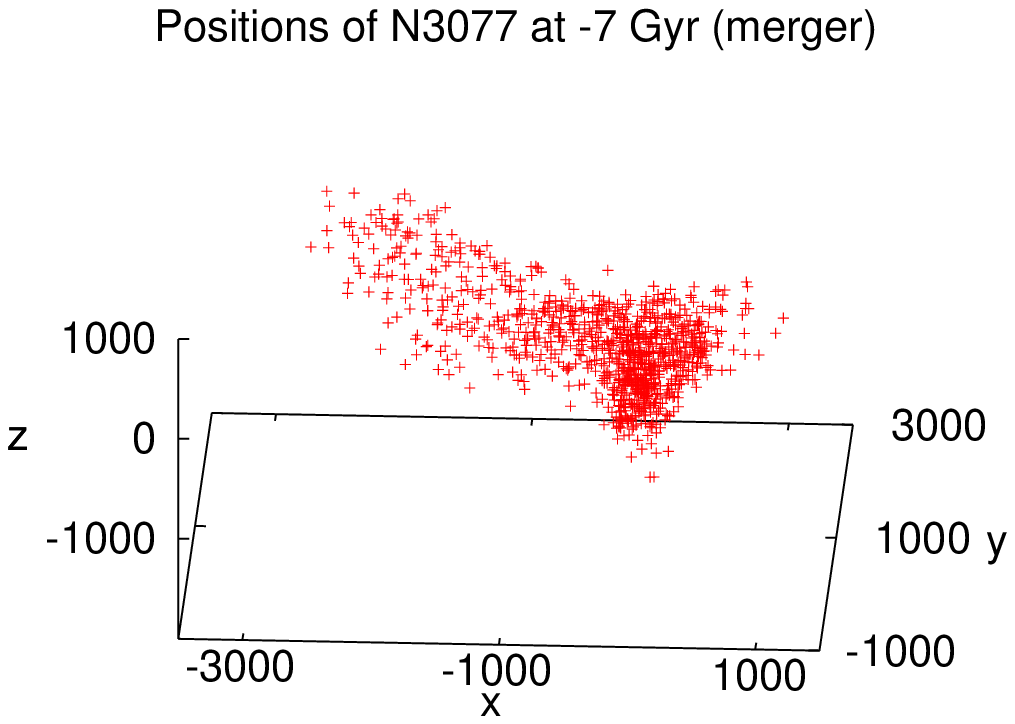} \\
\includegraphics[width=8.0cm]{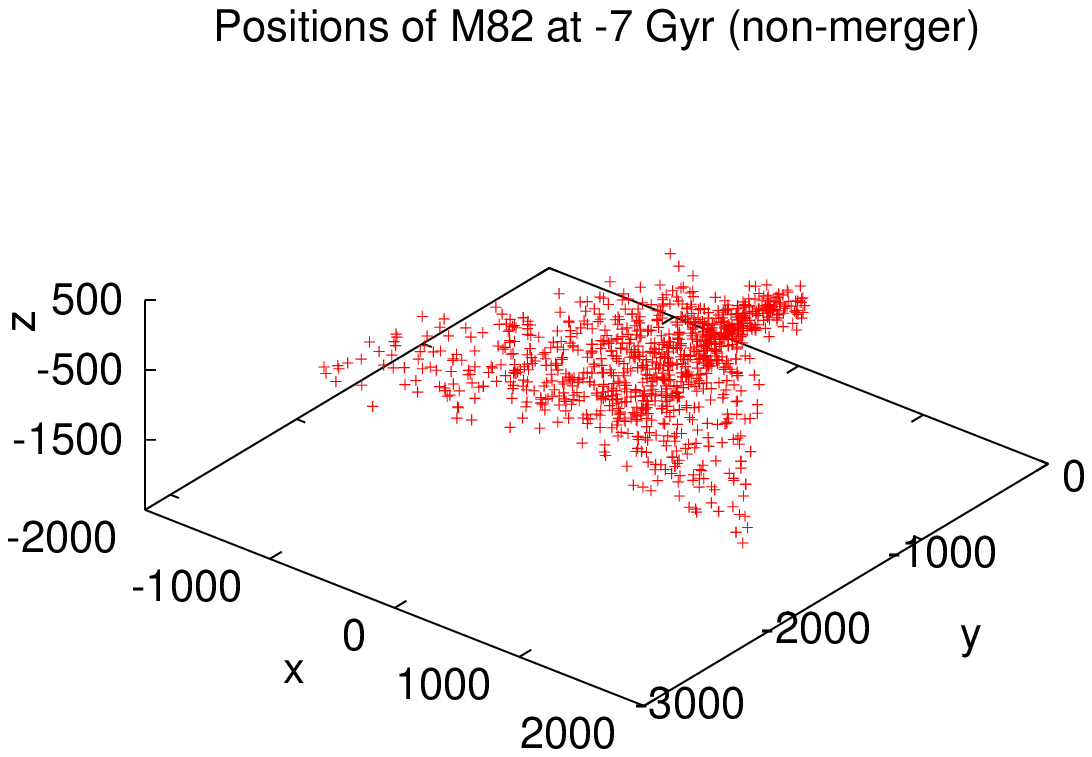} & \includegraphics[width=8.0cm]{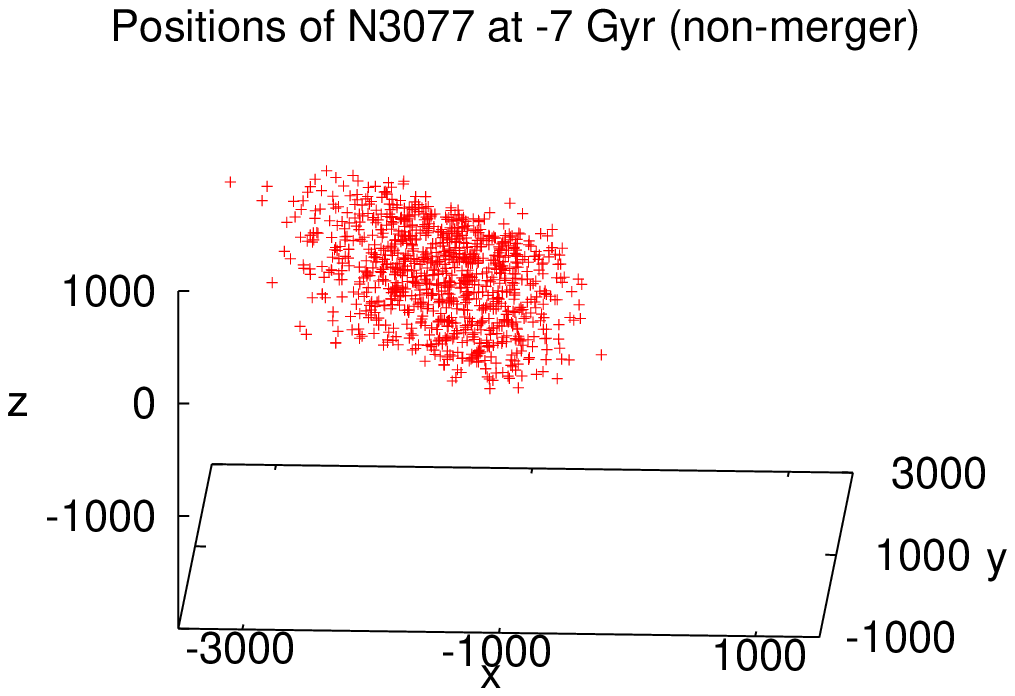} \\
\includegraphics[width=8.0cm]{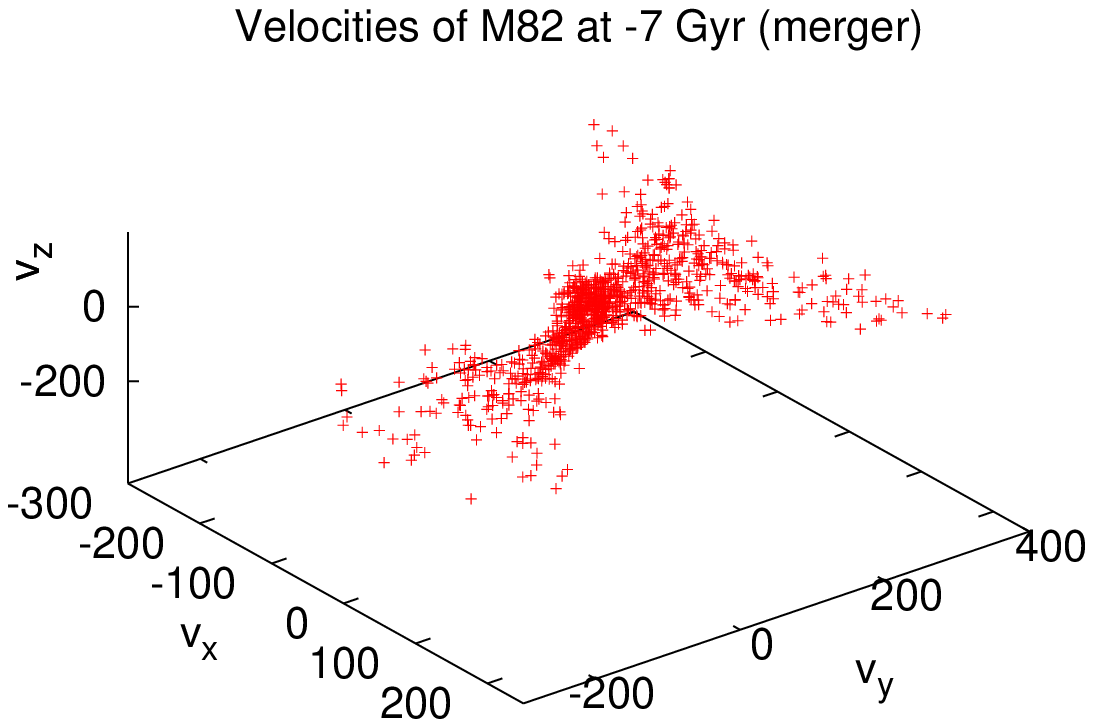} & \includegraphics[width=8.0cm]{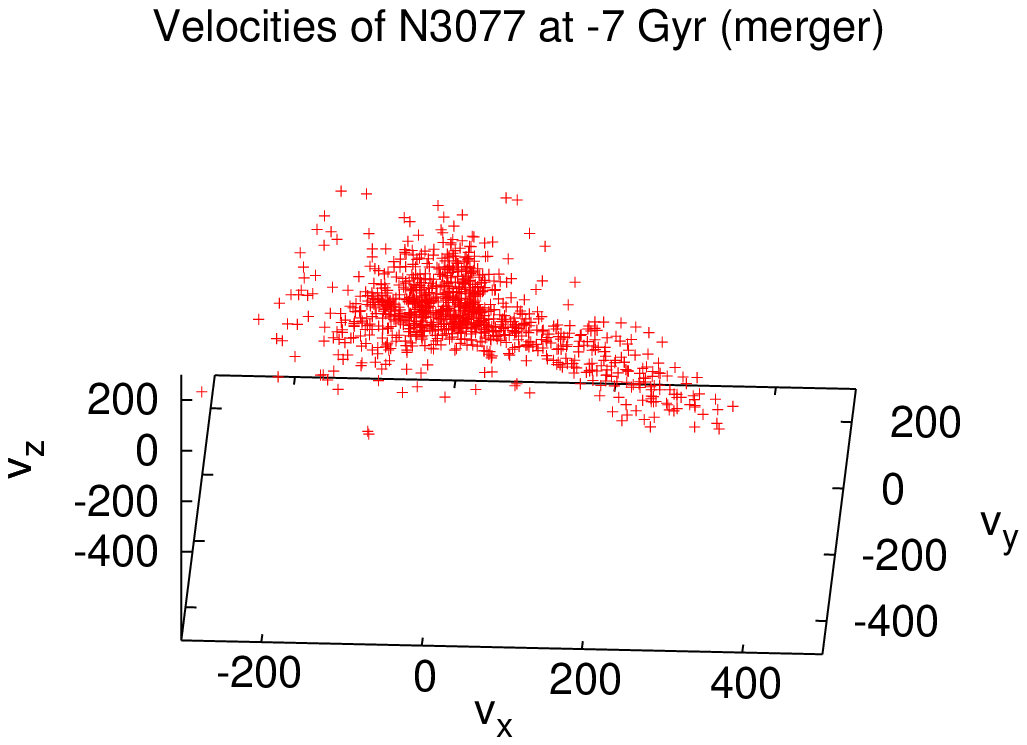} \\
\includegraphics[width=8.0cm]{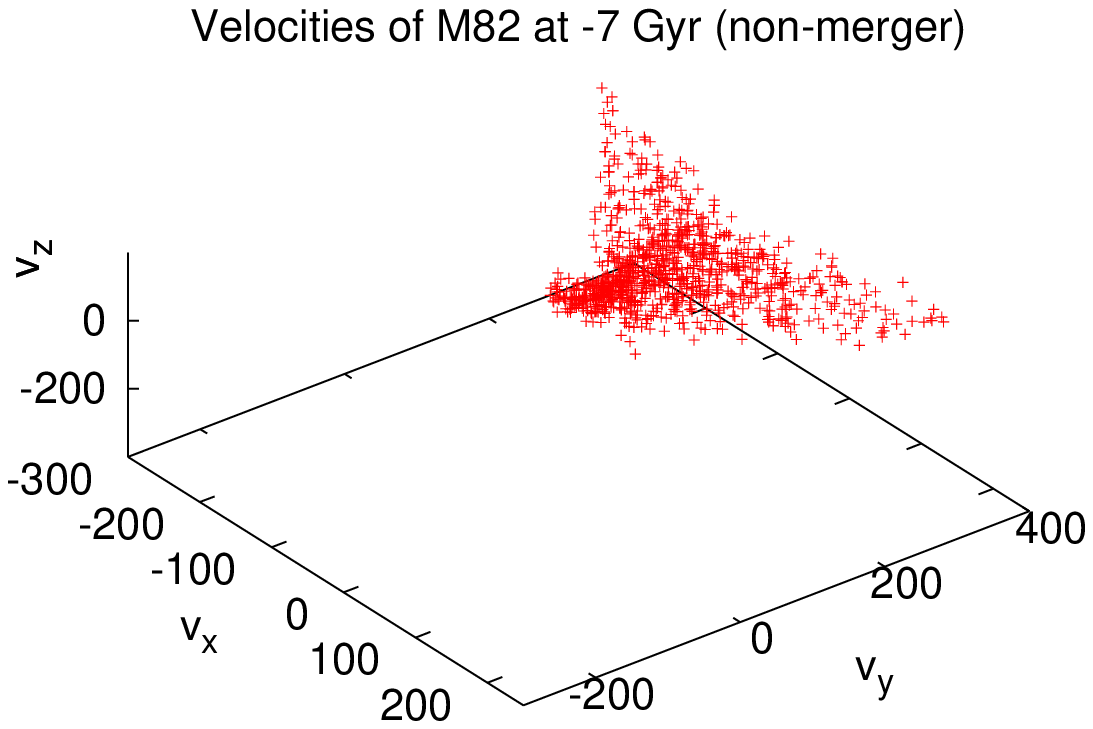} & \includegraphics[width=8.0cm]{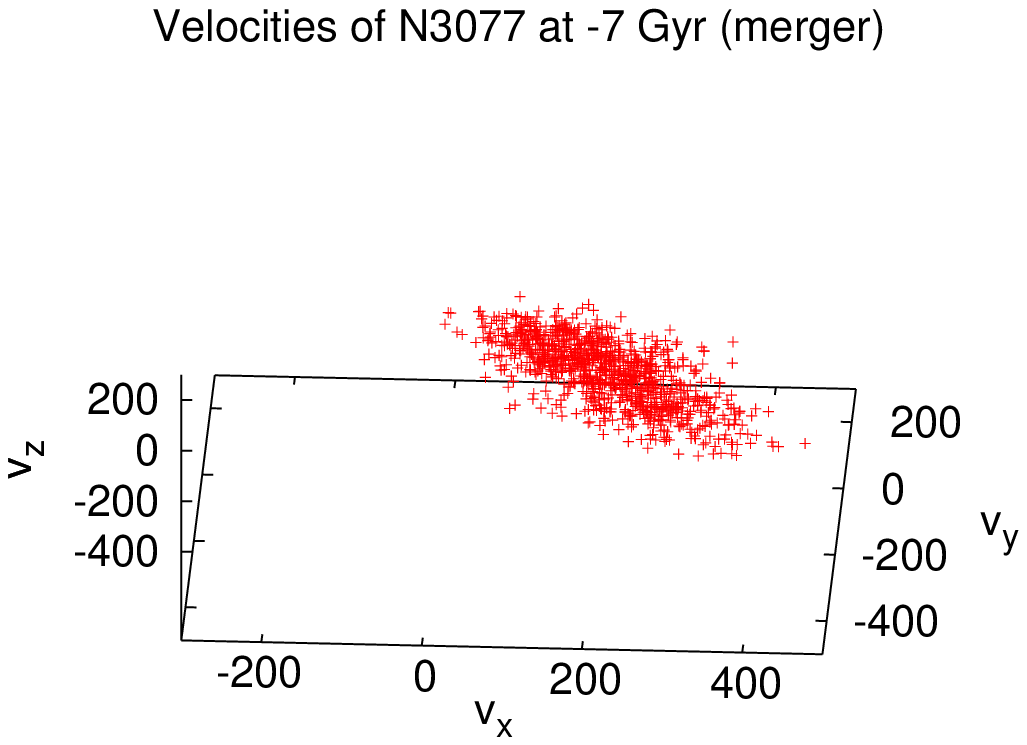} \\
\end{tabular}
\caption{GA: Distribution of the pre-infall phase space at -7~Gyr for
  M82 and NGC~3077 in today's M81 reference frame. A population of
  1000 solutions merging within the forthcoming 7~Gyr 
  (i.e. within 14~Gyr since -7~Gyr) is displayed in
  the first and the third row for the coordinates (kpc) and the
  velocities (pc/Myr), respectively. Accordingly the second and fourth
  row refer to a population of 1000 solutions not merging.}
\label{3D}
\end{figure*}

\begin{figure*}
\centering
\begin{tabular}{ccc}
\includegraphics[width=5.5cm]{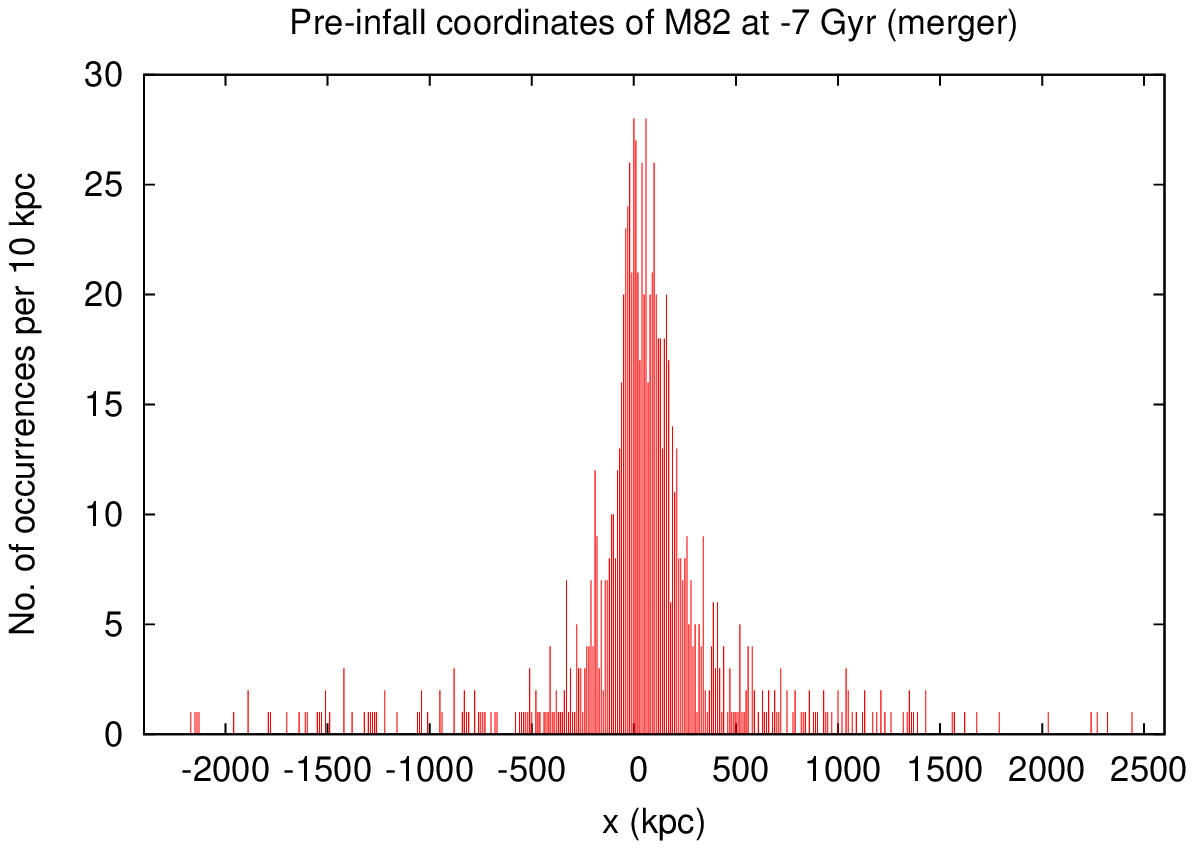} & \includegraphics[width=5.5cm]{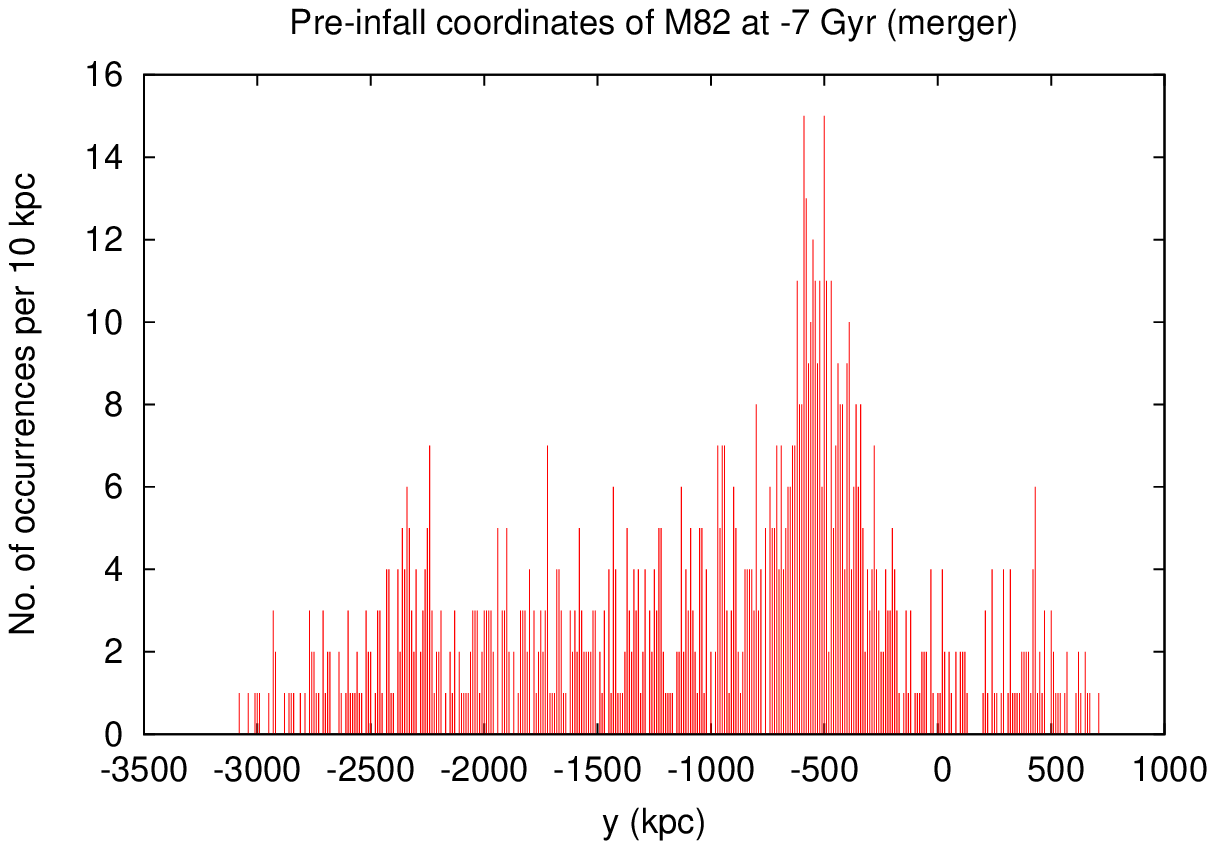} 
                                                                       & \includegraphics[width=5.5cm]{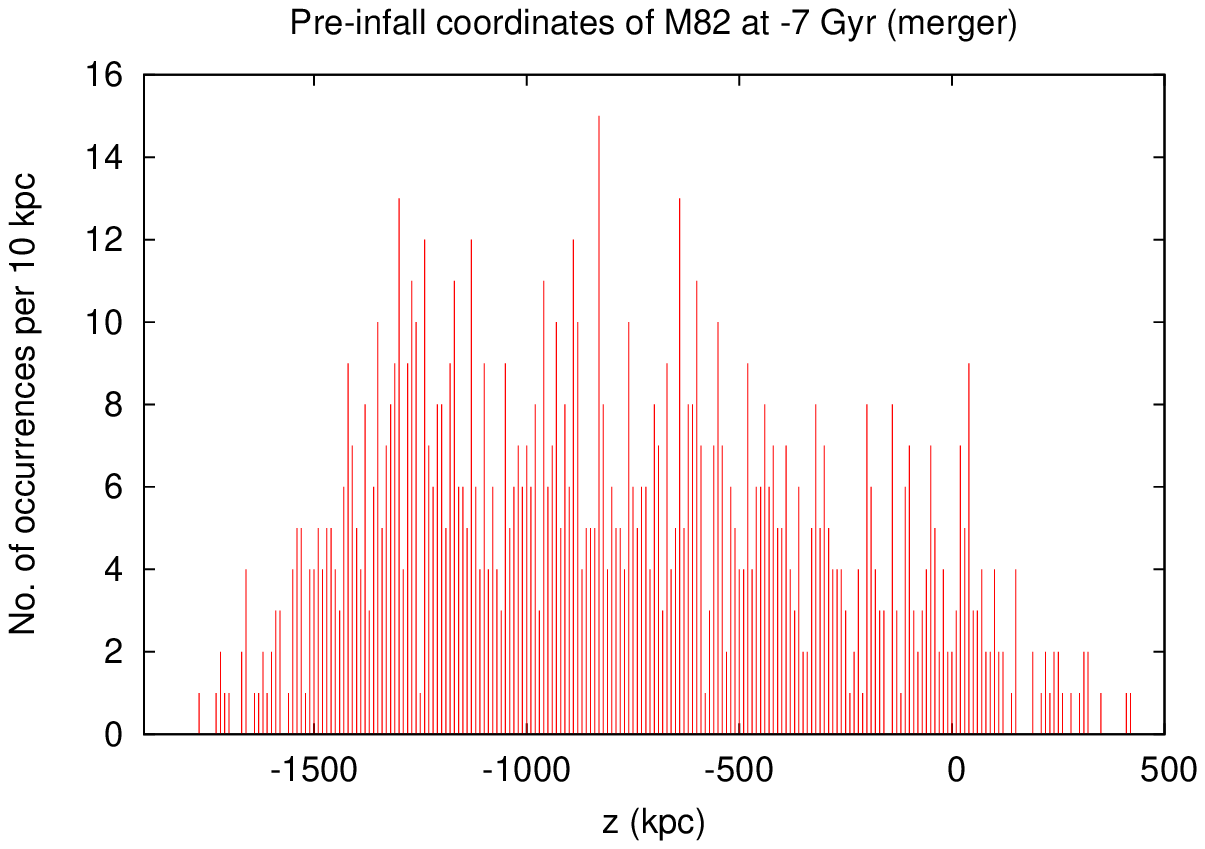}\\
\includegraphics[width=5.5cm]{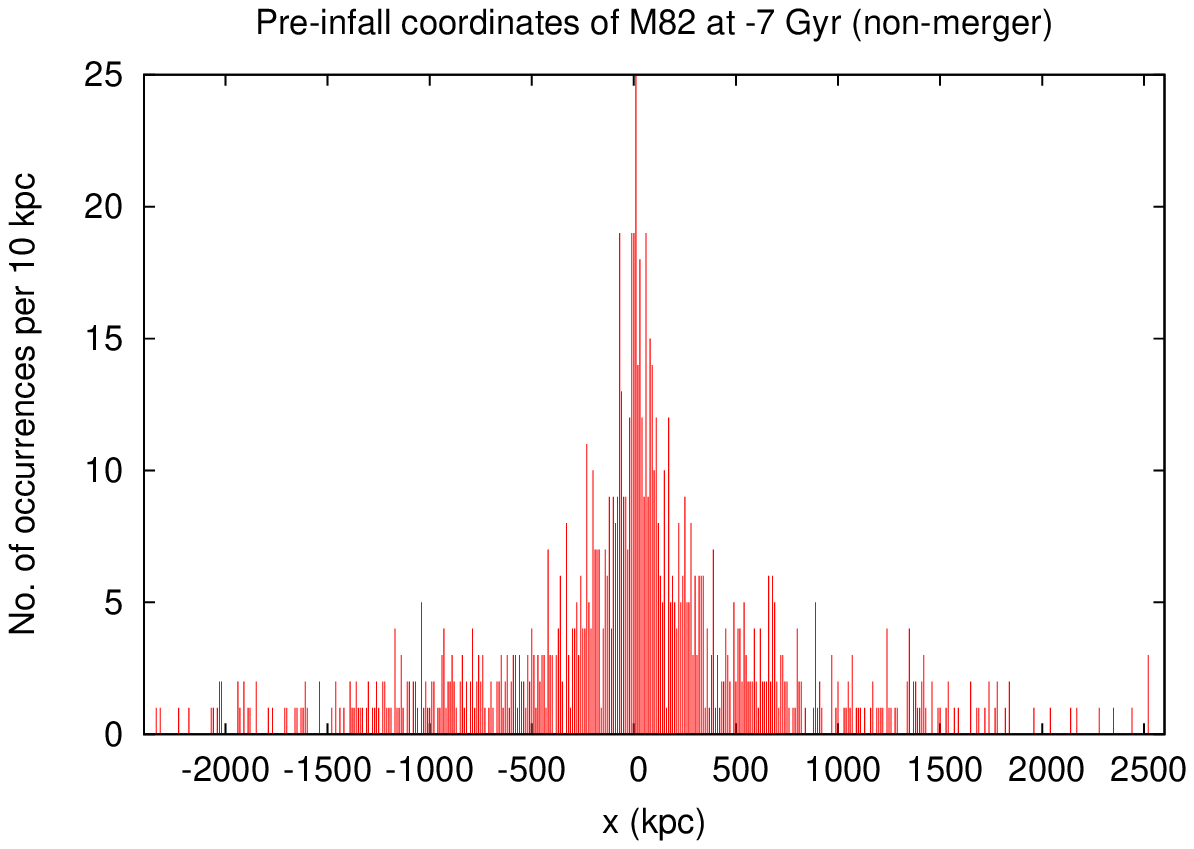} & \includegraphics[width=5.5cm]{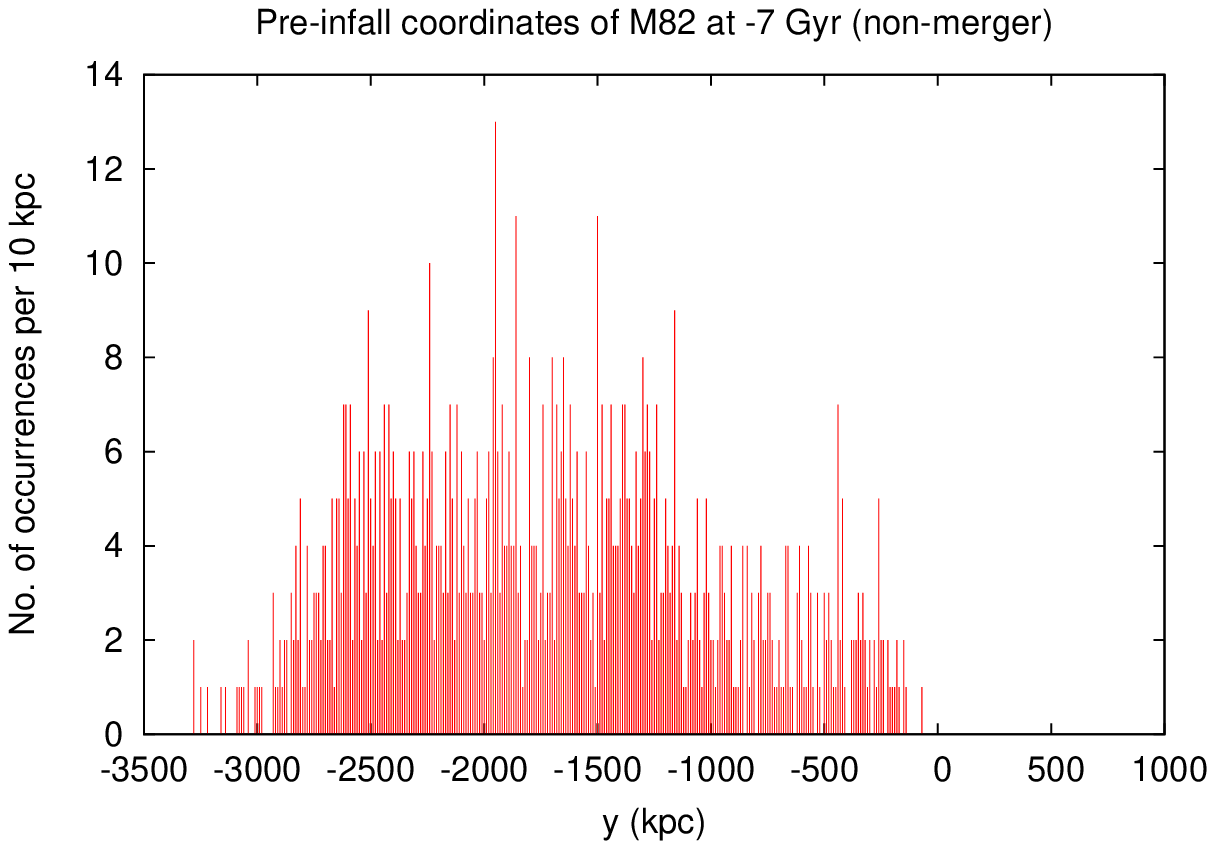} 
                                                                       & \includegraphics[width=5.5cm]{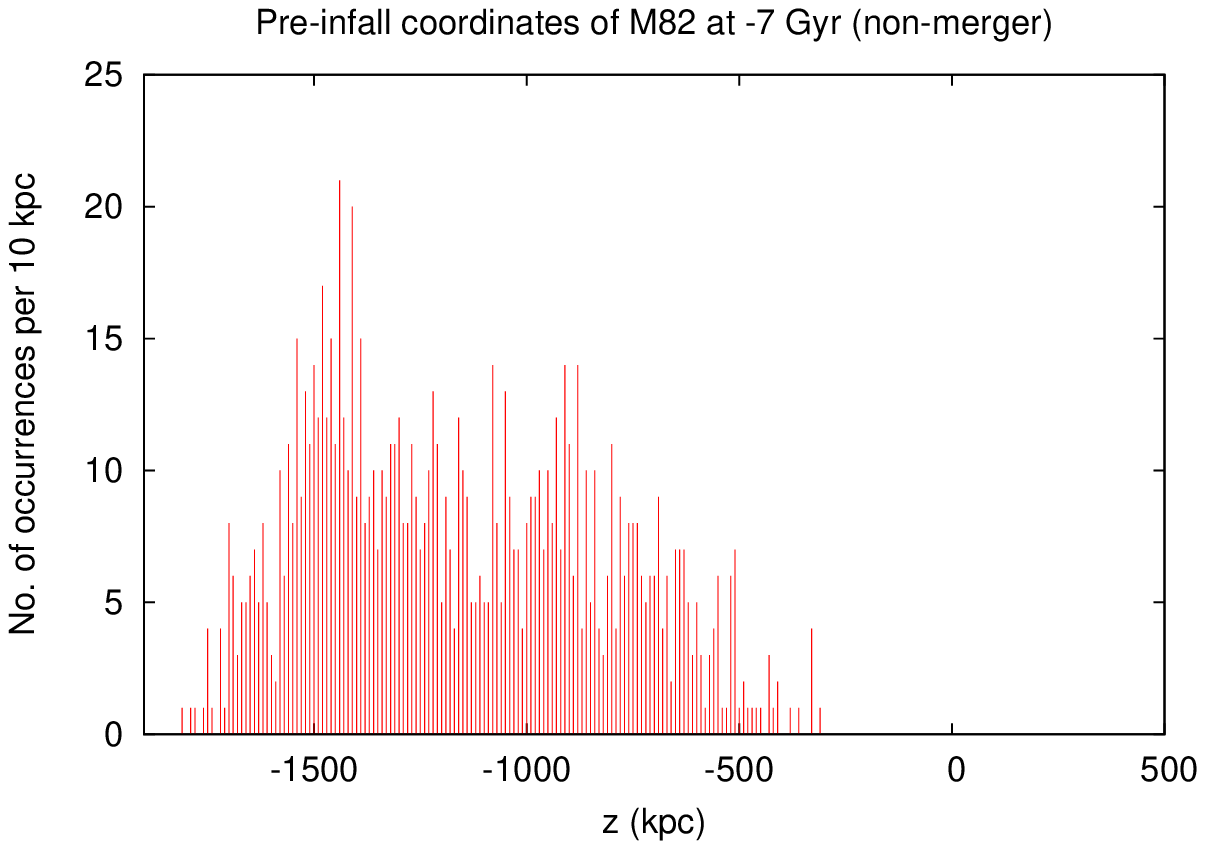}\\
\includegraphics[width=5.5cm]{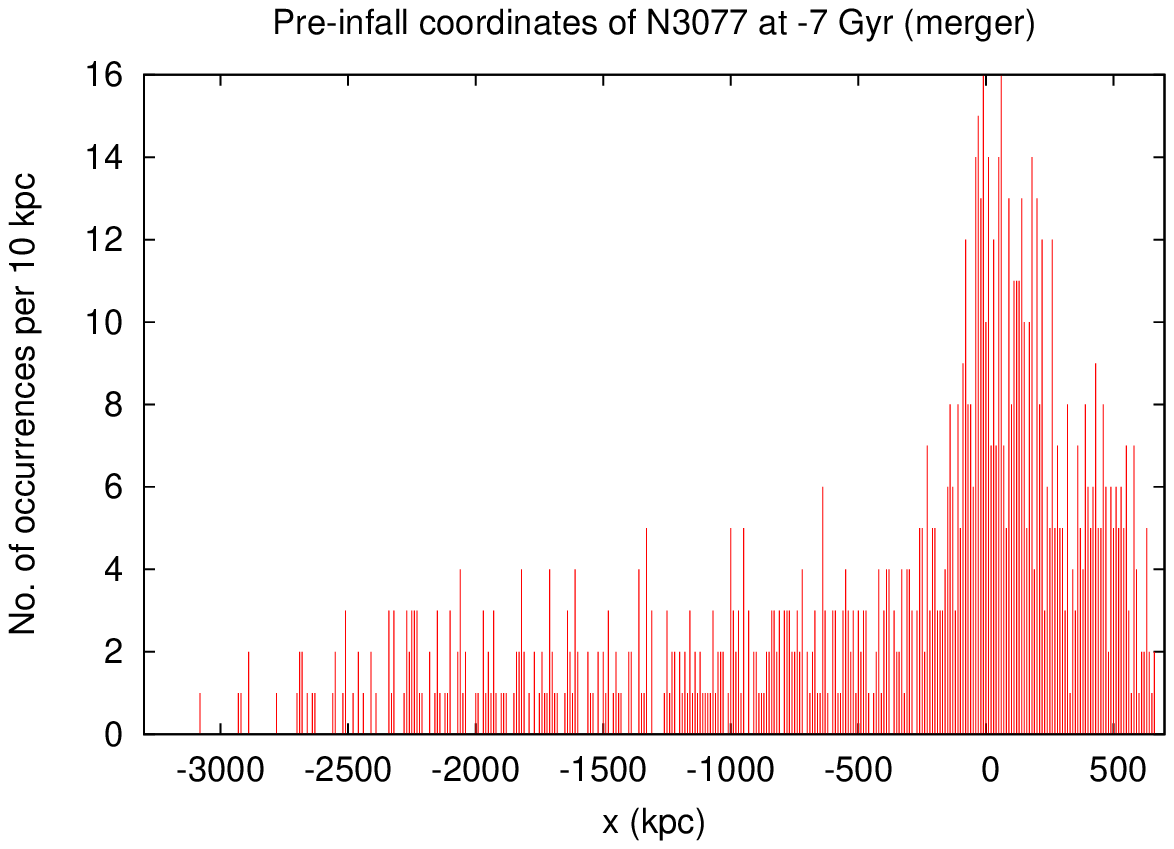} & \includegraphics[width=5.5cm]{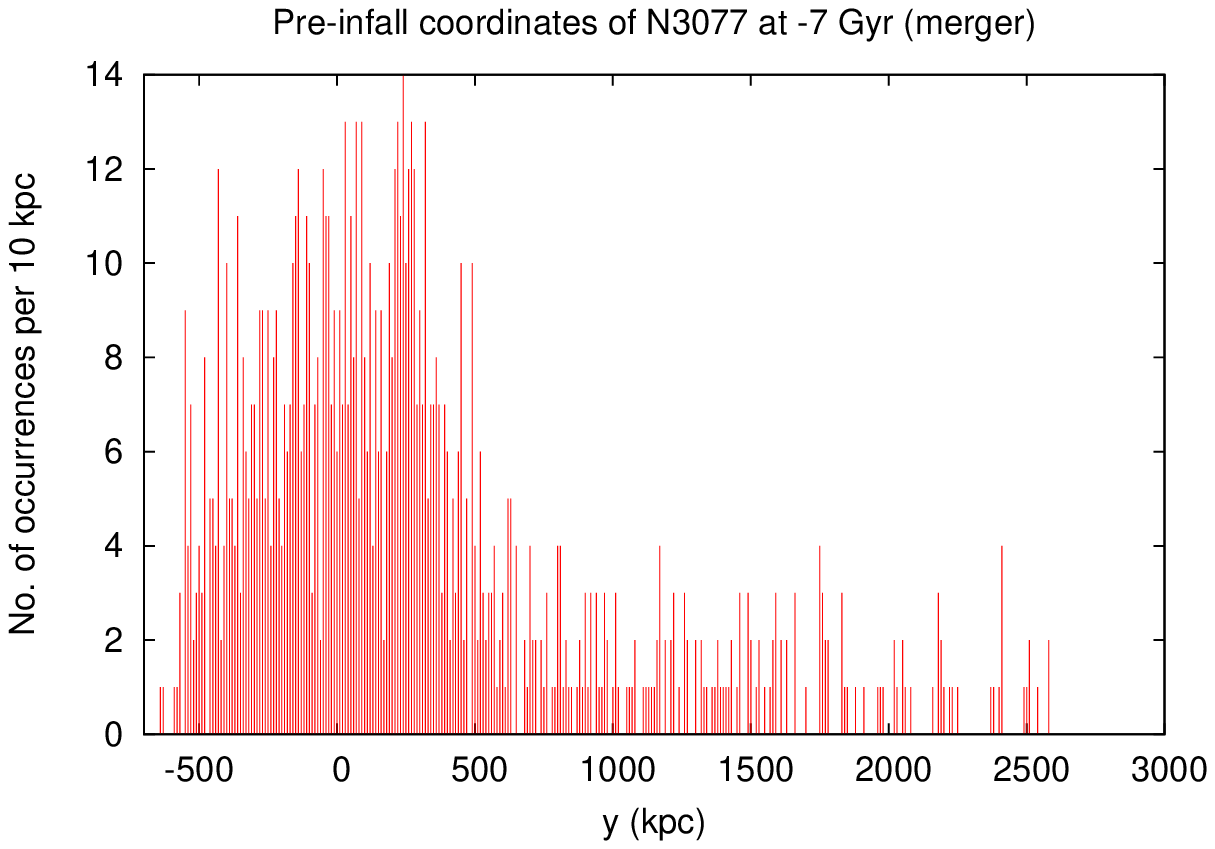} 
                                                                       & \includegraphics[width=5.5cm]{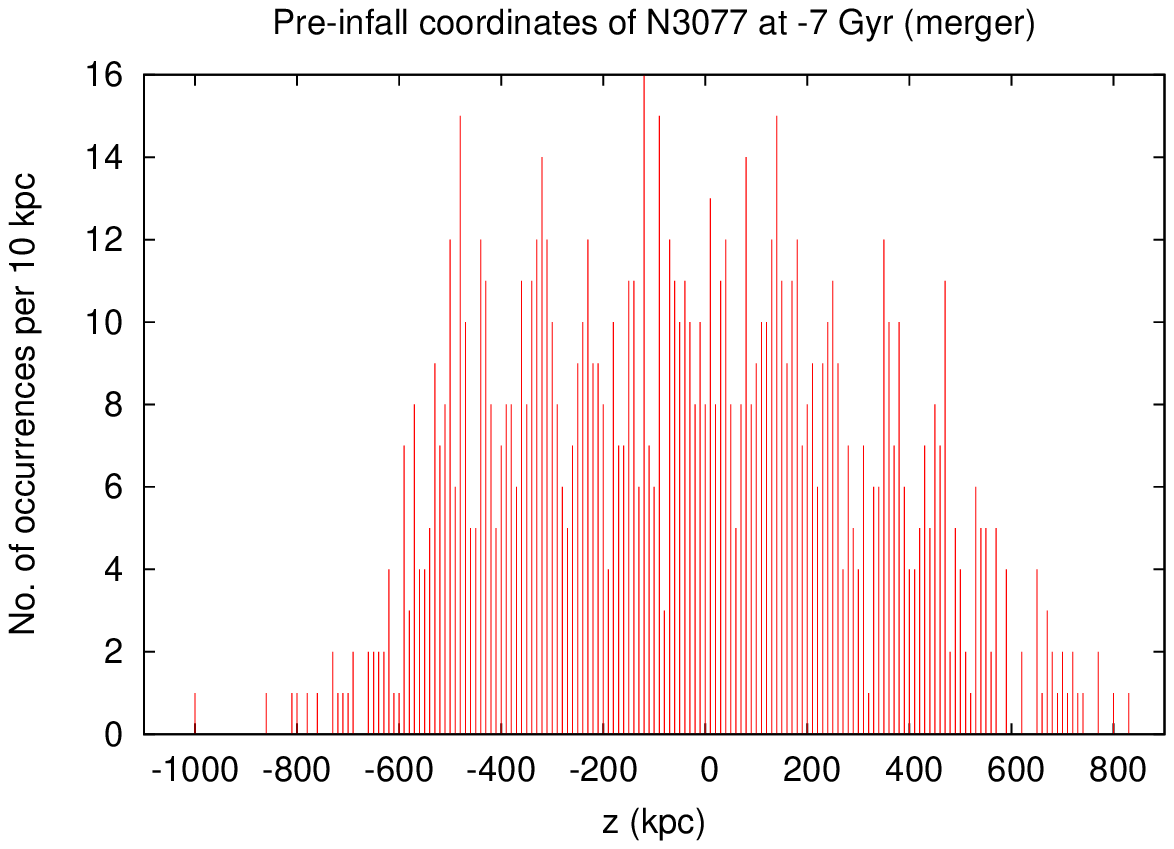}\\
\includegraphics[width=5.5cm]{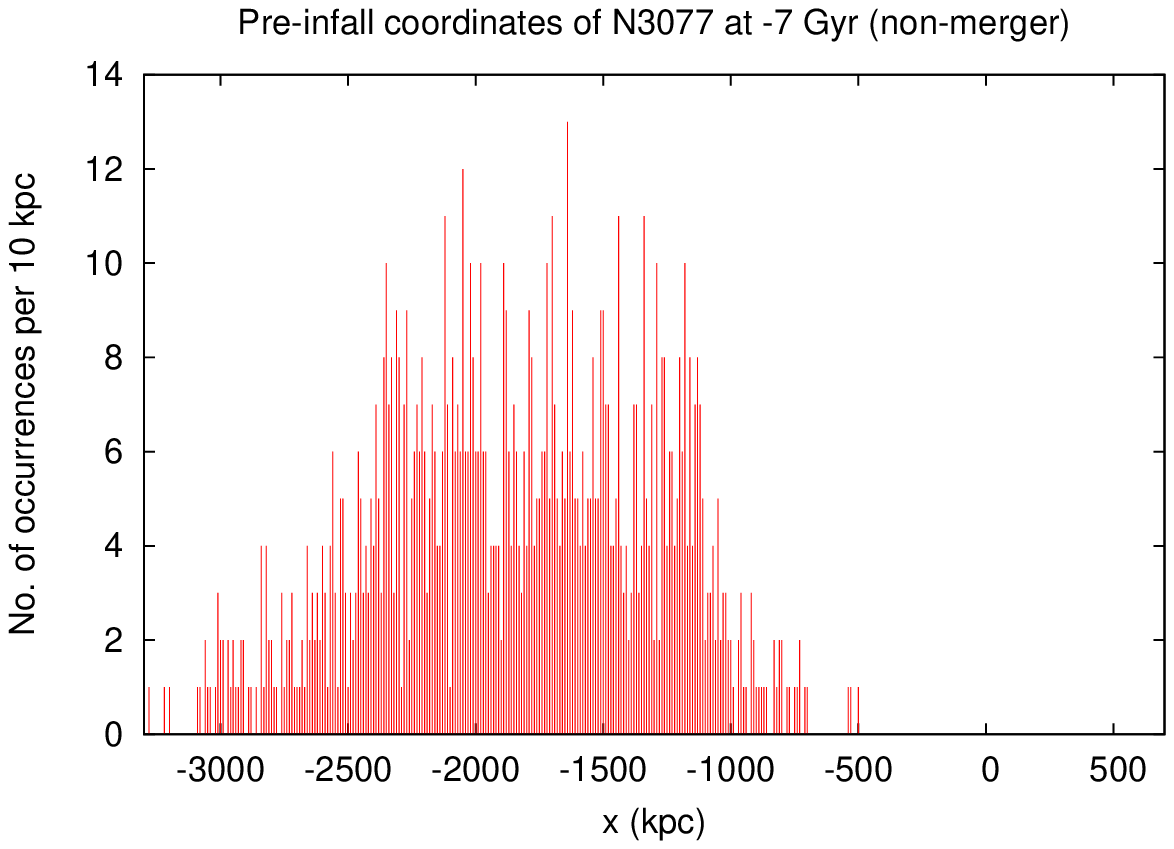} & \includegraphics[width=5.5cm]{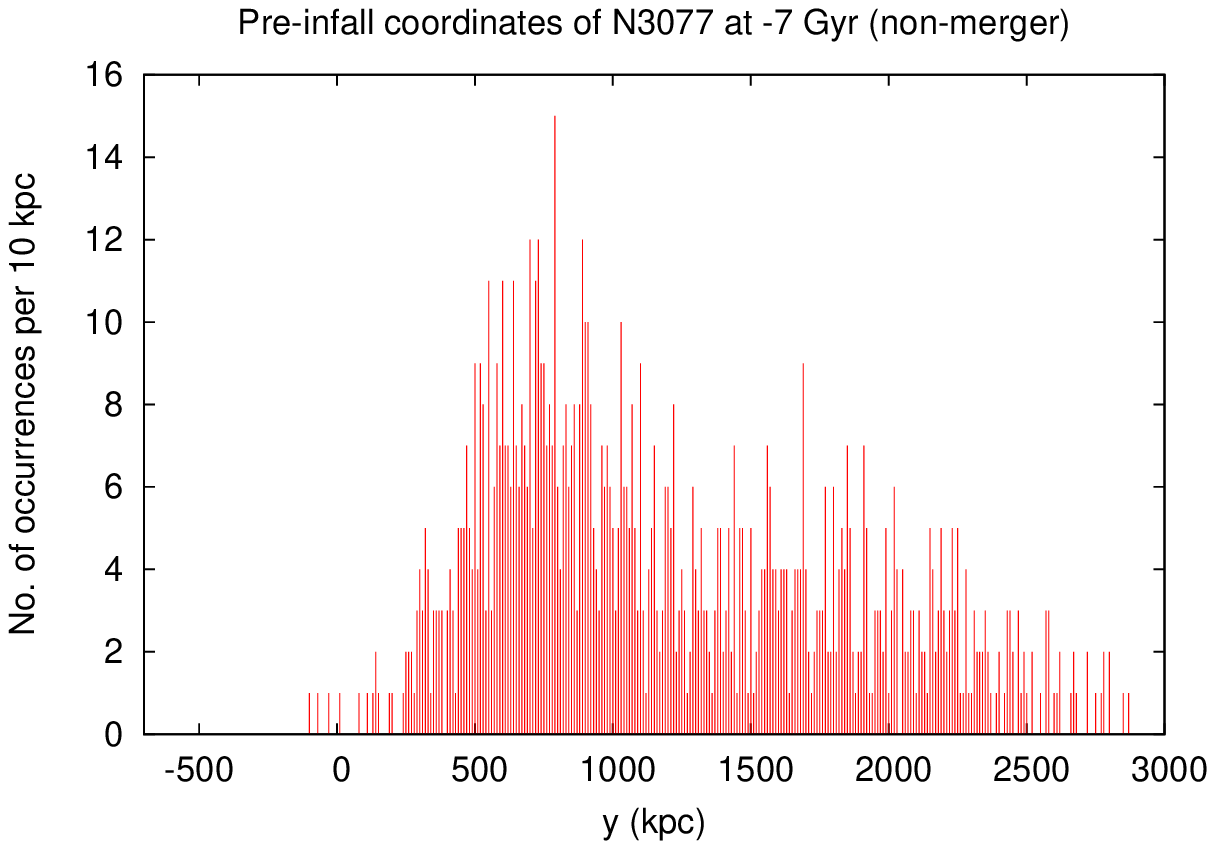} 
                                                                       & \includegraphics[width=5.5cm]{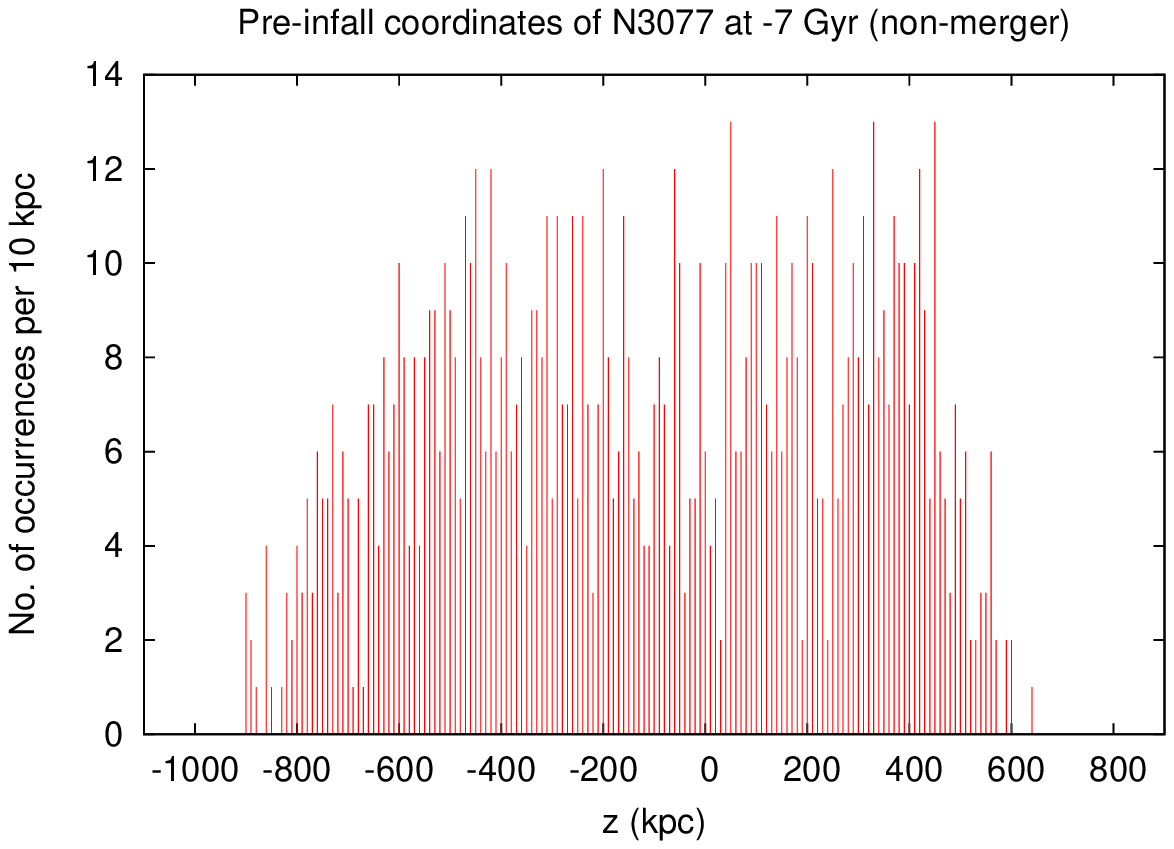}\\
\end{tabular}
\caption{GA: Distribution of the pre-infall coordinates at -7~Gyr for
  M82 and NGC~3077 in today's M81 reference frame. A population of
  1000 solutions merging within the forthcoming 7~Gyr 
  (i.e. within 14~Gyr since -7~Gyr) is displayed in
  the first and the third row for M82 and NGC~3077,
  respectively. Accordingly the second and fourth row refer to a
  population of 1000 solutions not merging.}
\label{ps-x}
\end{figure*}

\begin{figure*}
\centering
\begin{tabular}{ccc}
\includegraphics[width=5.5cm]{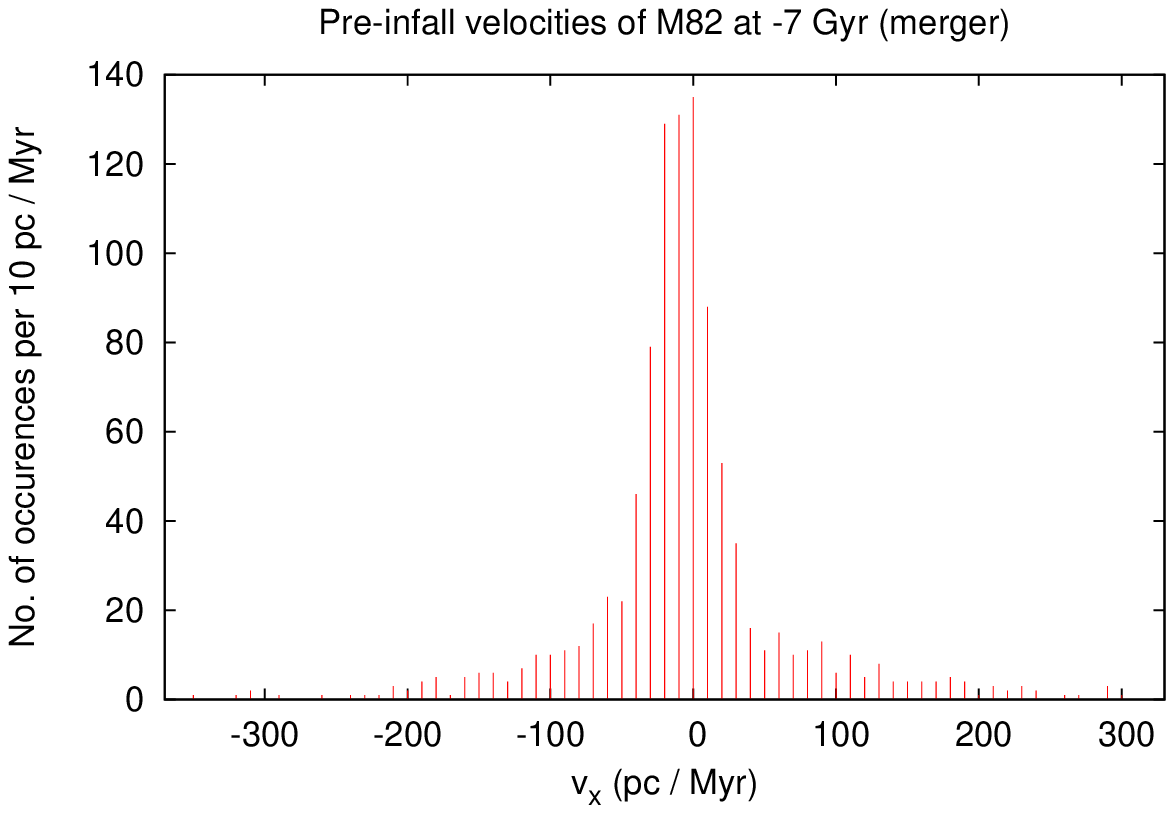} & \includegraphics[width=5.5cm]{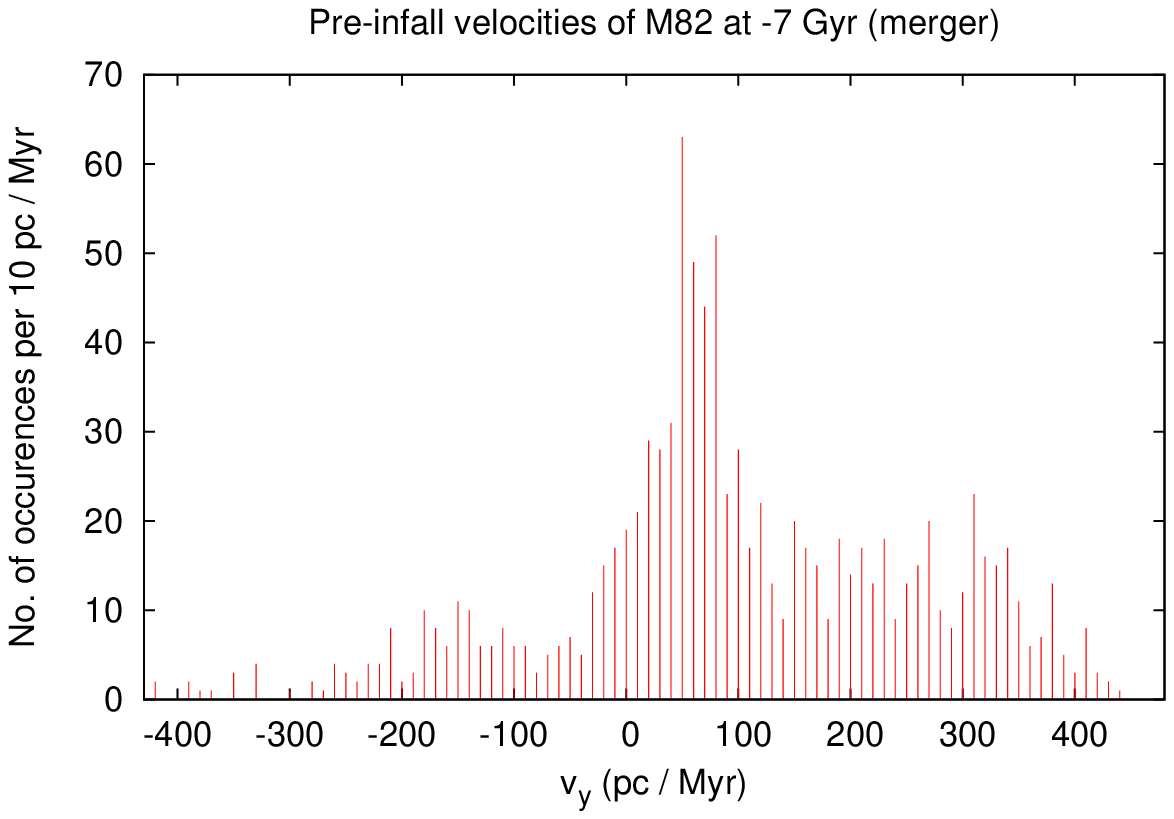} 
                                                                         & \includegraphics[width=5.5cm]{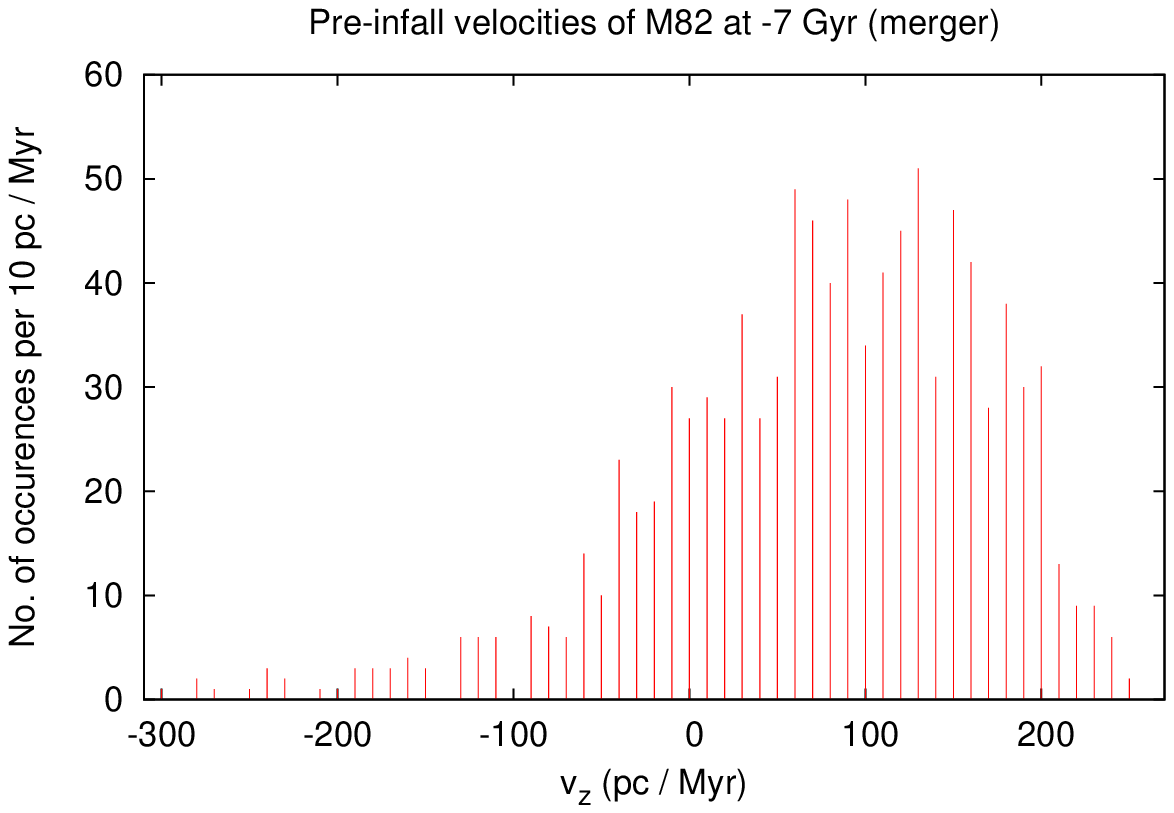}\\
\includegraphics[width=5.5cm]{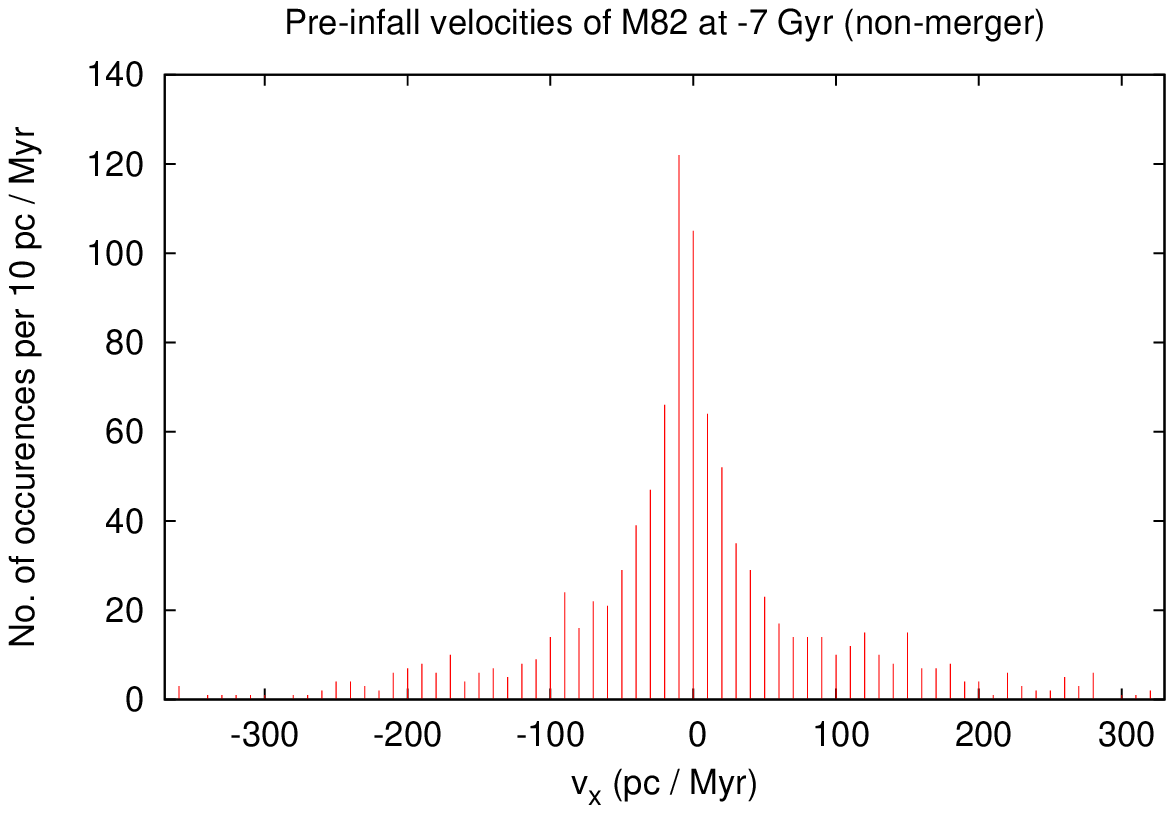} & \includegraphics[width=5.5cm]{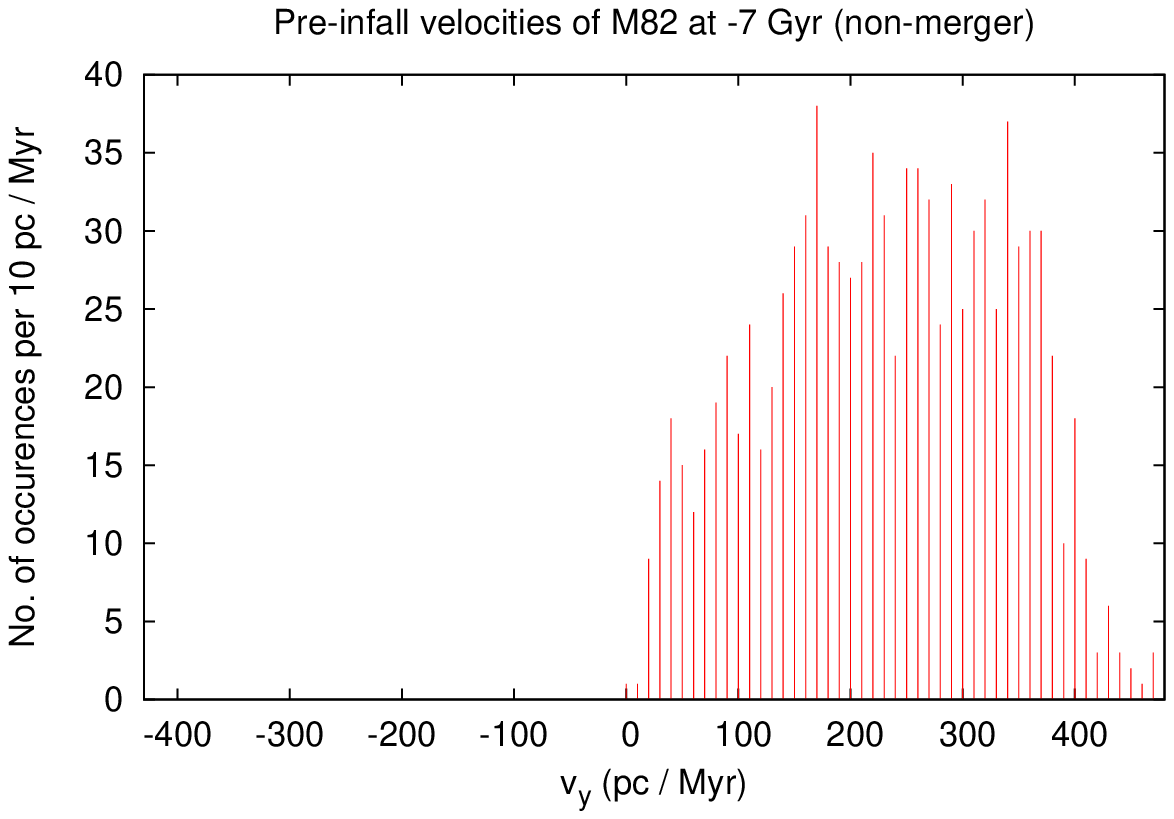} 
                                                                          & \includegraphics[width=5.5cm]{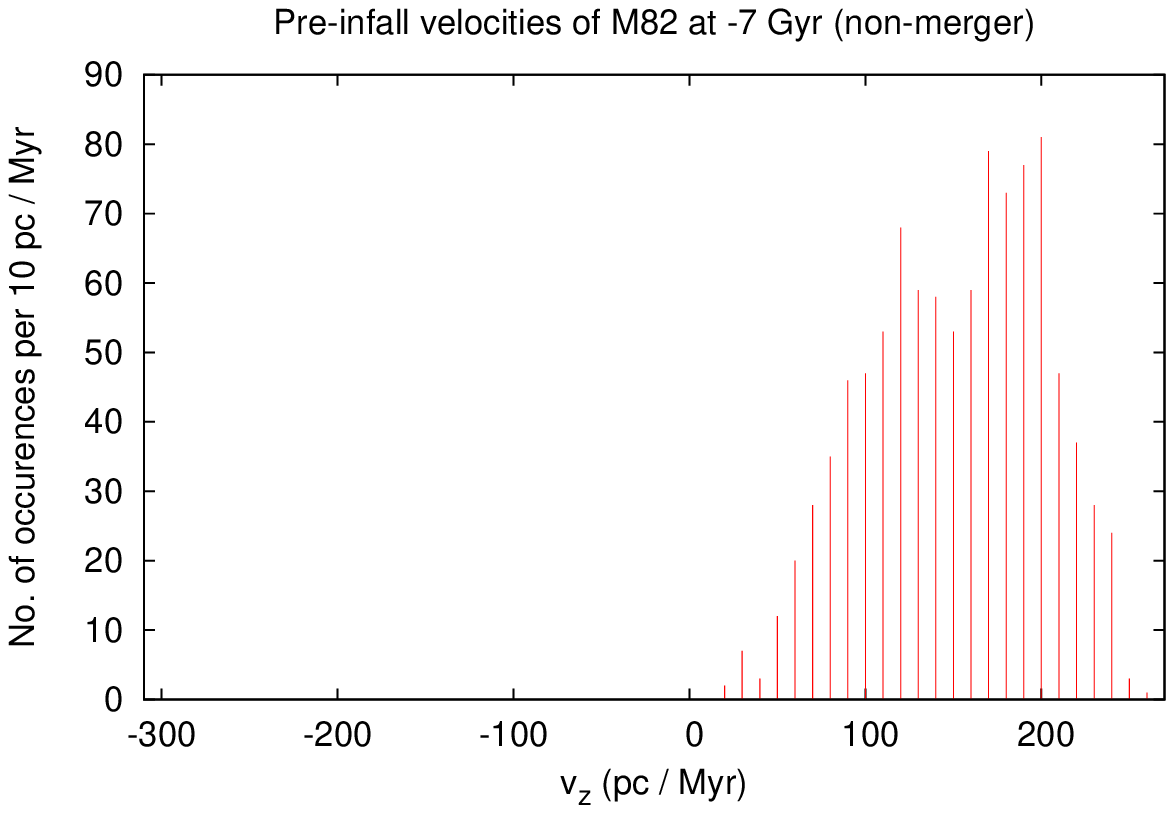}\\
\includegraphics[width=5.5cm]{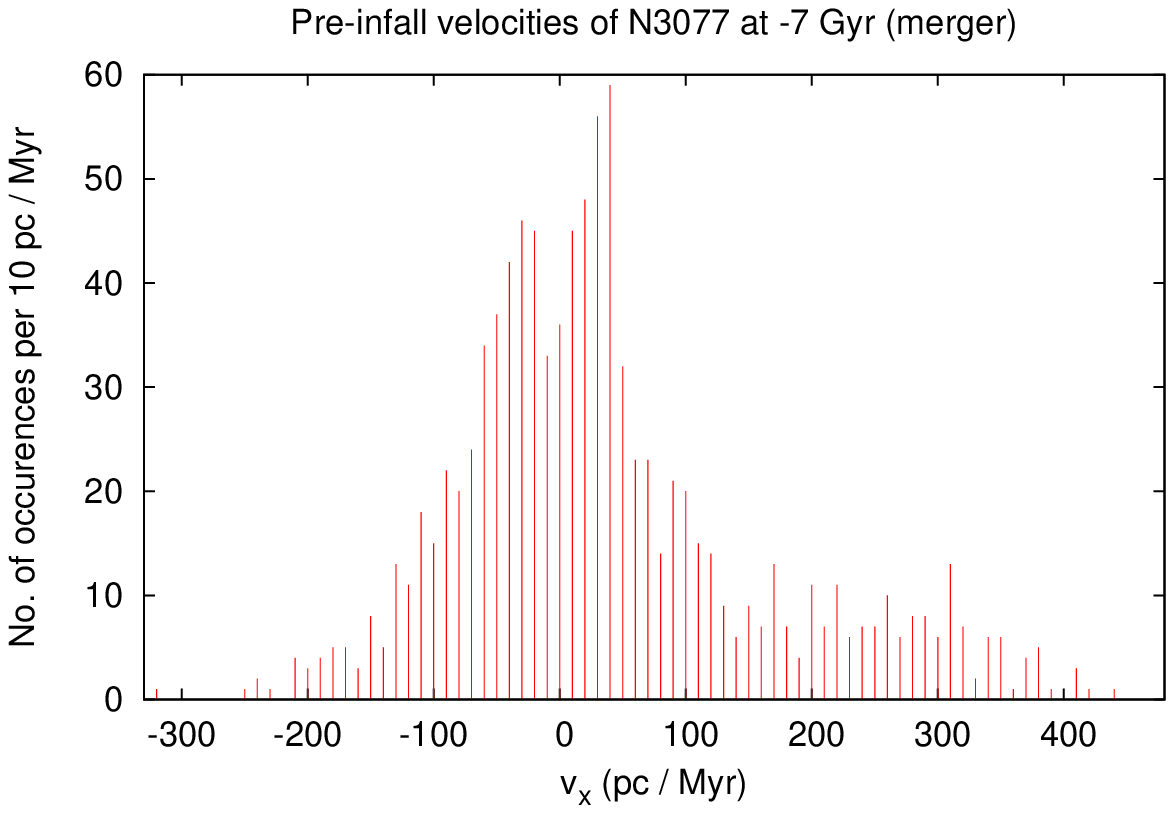} & \includegraphics[width=5.5cm]{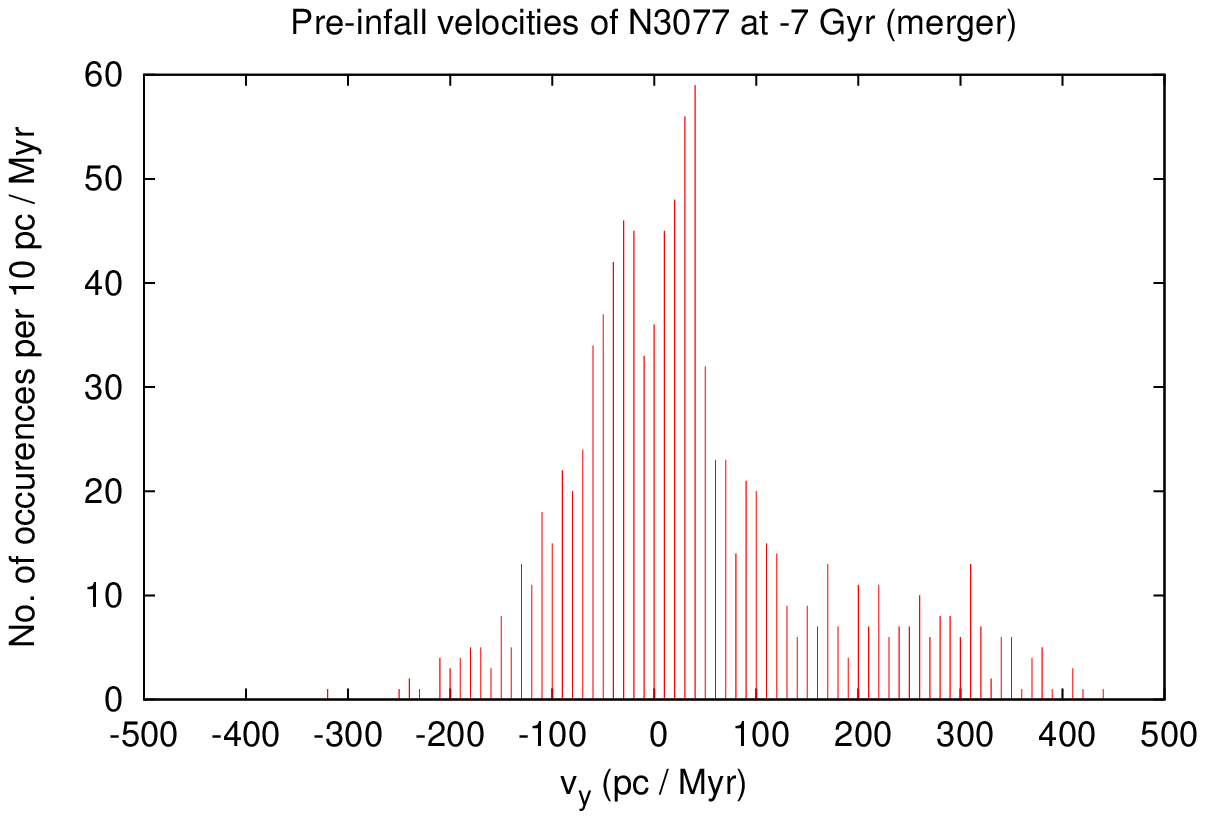} 
                                                                         & \includegraphics[width=5.5cm]{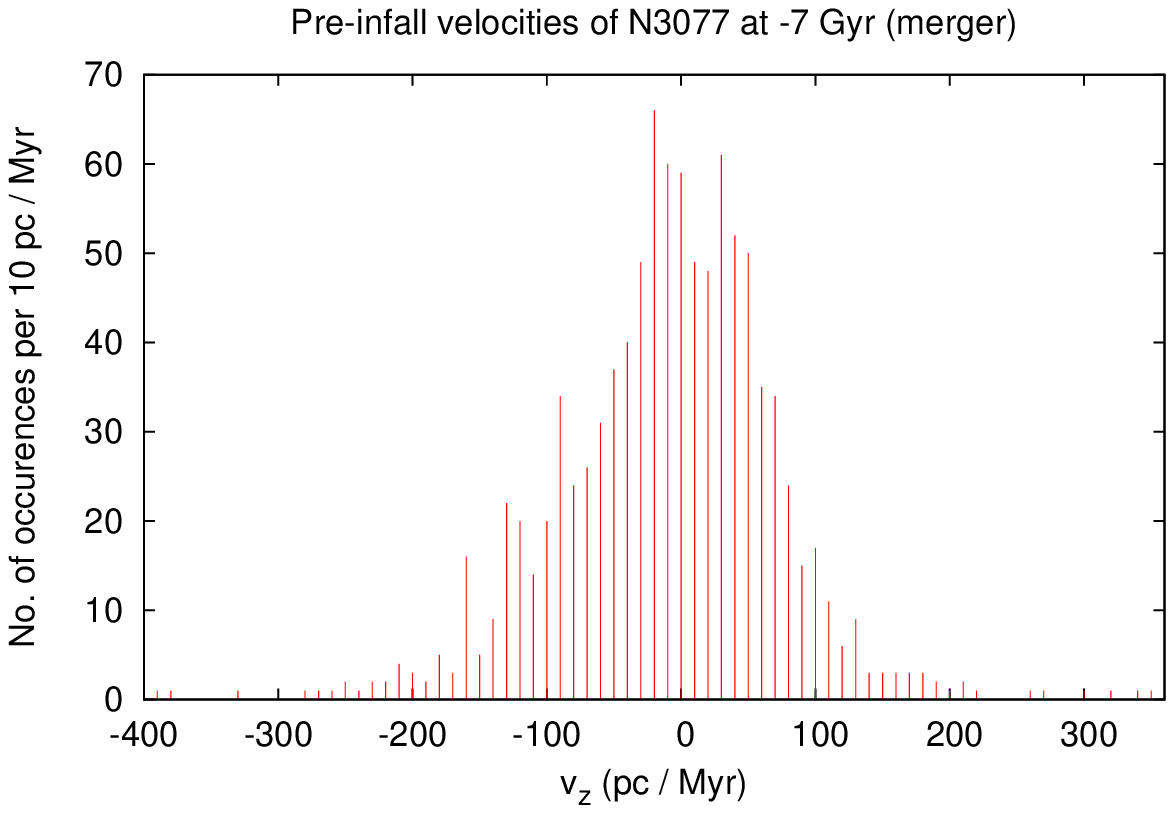}\\
\includegraphics[width=5.5cm]{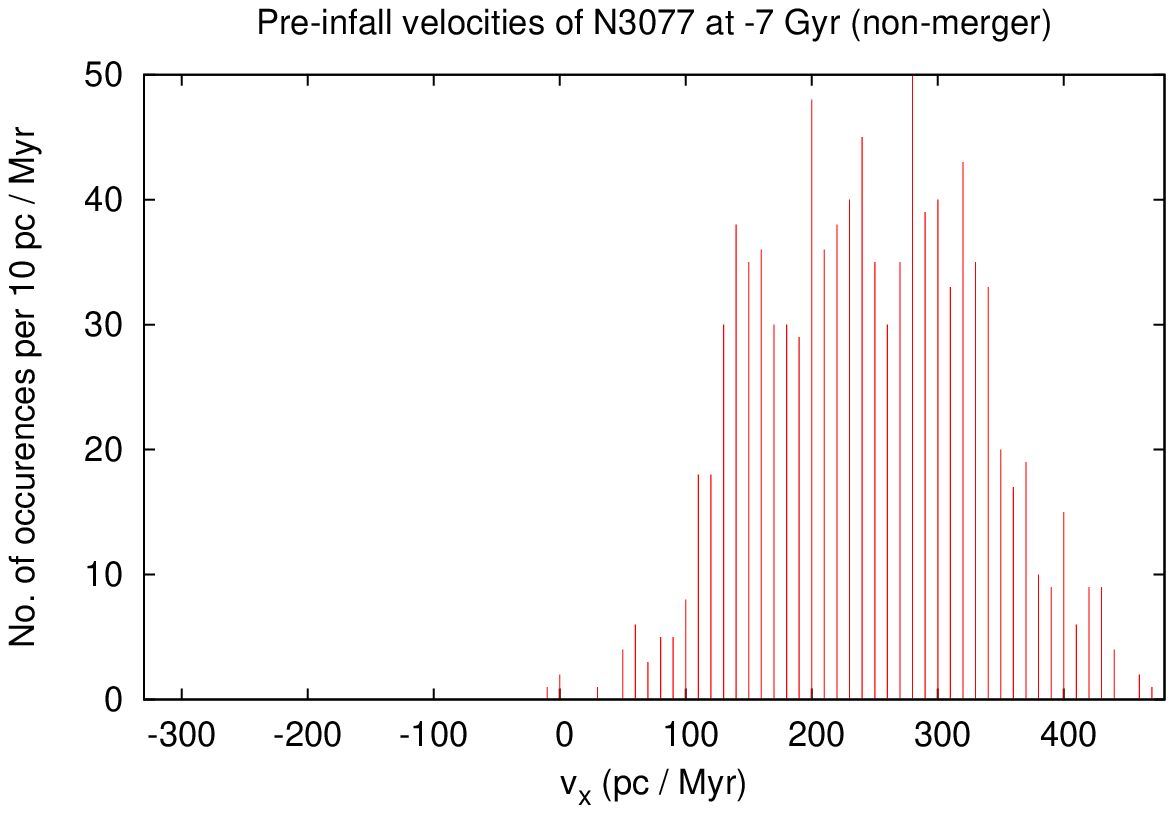} & \includegraphics[width=5.5cm]{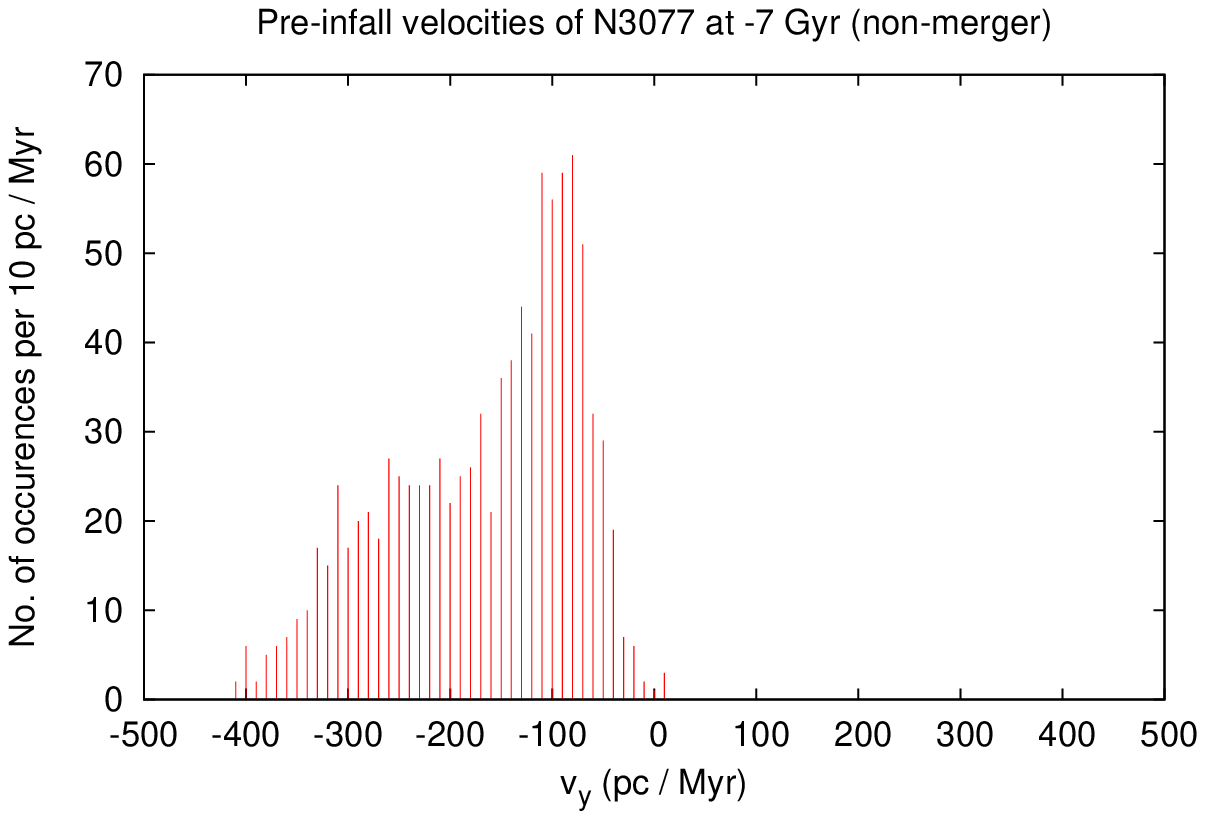} 
                                                                          & \includegraphics[width=5.5cm]{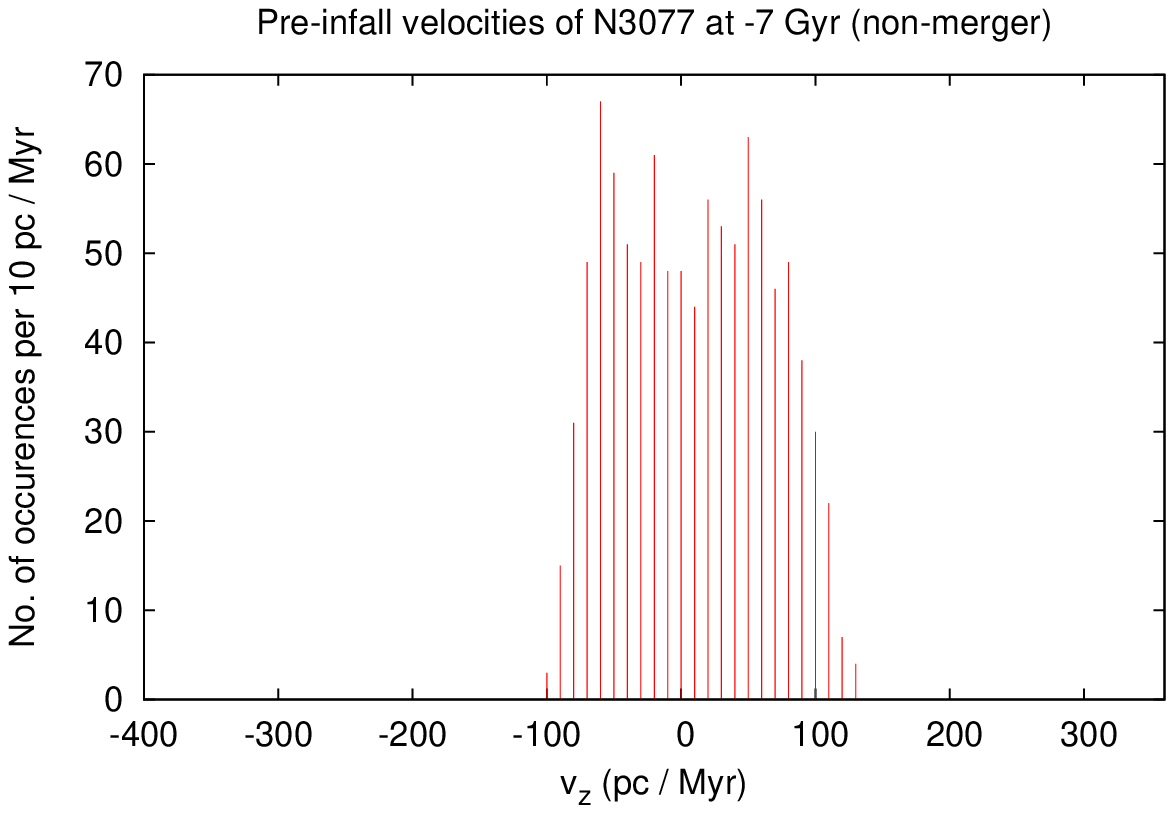}\\
\end{tabular}
\caption{GA: Same as Figure~\ref{ps-x}, but for the velocity components.}
\label{ps-v}
\end{figure*}

It is worthwhile to investigate the pre-infall phase space at
-7~Gyr. For this purpose we continued the GA-runs until 1000~merging
solutions and 1000~solutions not merging were
collected. Figure~\ref{3D} shows the corresponding 3D-plots which were
rotated in a way that structures are efficiently visualized. Roughly
evaluating Figure~\ref{ps-x} for the spatial coordinates and
Figure~\ref{ps-v} for the velocity components of M82 and NGC~3077 one
might draw the conclusion that the phase space volume for the
population of the merging solutions fairly exceeds the phase space
volume for the population of the non-merging solutions. 
However, for the population of the merging solutions a higher correlation
in phase space is evident from Figure~\ref{3D} 
(especially for NGC~3077).  
Therefore a solid statement regarding the ratio
of the phase space volume cannot  be readily extracted.

\subsection{Extended Fitness Function}
\label{sec:eff}

\begin{figure}
\centering
\begin{tabular}{cc}
\includegraphics[width=4cm]{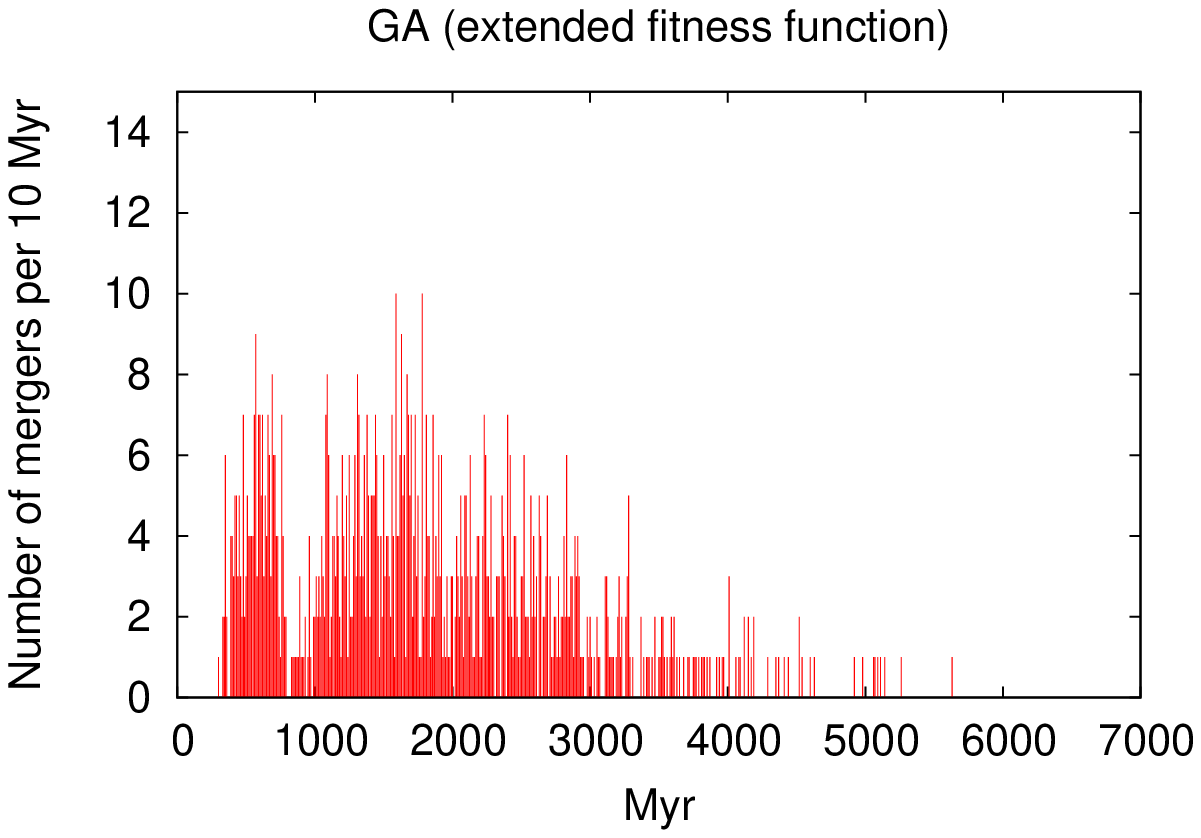} & \includegraphics[width=4cm]{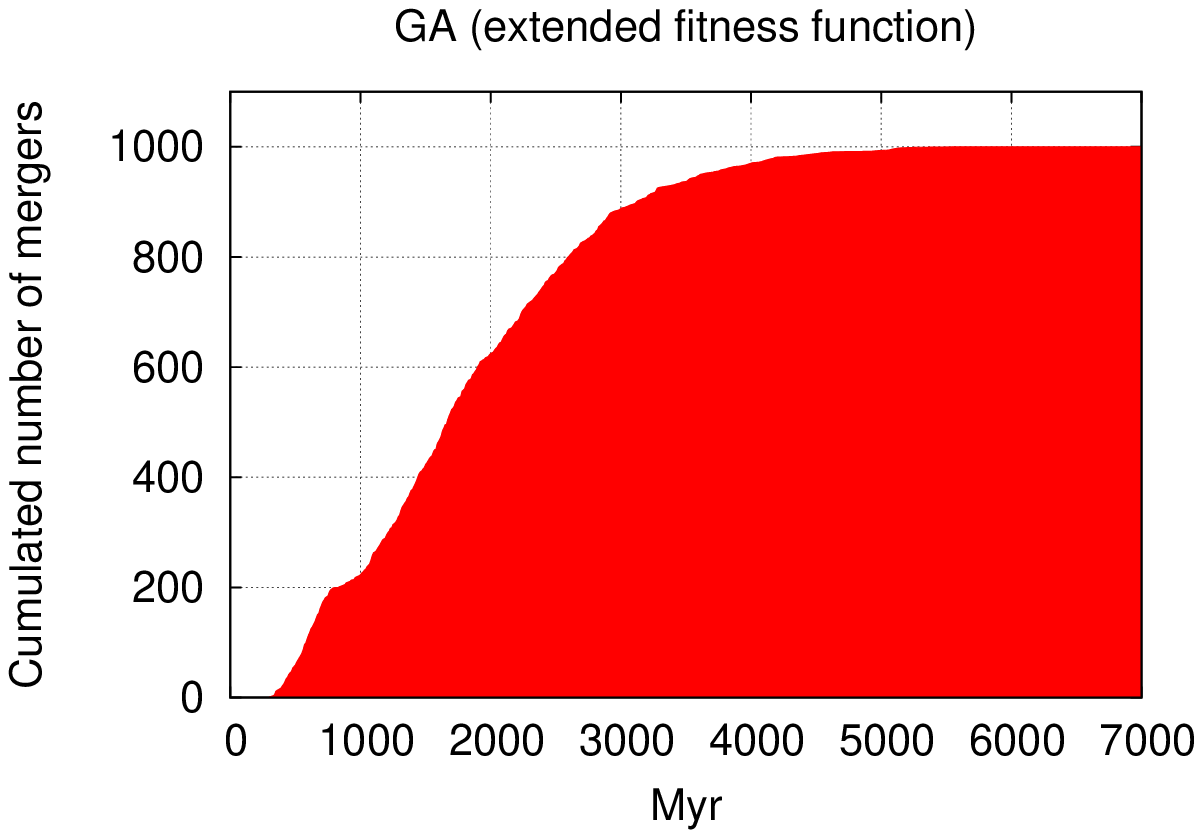}\\
\end{tabular}
\caption[]{Same as Figure~\ref{GA_I}, but fitness function $\mathcal{F}_E$ (Eq.~\ref{eq:fit-II}).}  
\label{GA_II}
\end{figure}

In order to facilitate that the GA evaluations preferably generate
solutions where at least one of the companions M82 and NGC~3077 is
bound to M81, we enhance the fitness function by a function depending
on the three-body energy $E$ in the centre-of-mass system before the
encounter

\begin{equation}
\label{eq:fit-E}              
f_E = \left\{
\begin{array}{ll}
1, &  E \le E_{min}/ 2\ , \\
\exp\left({-{\displaystyle\frac{\left( {E}-{E_{min}/ 2} \right)^2}{2\cdot {(E_{min}/ 2)}^2}}}\right), &   E > E_{min}/ 2 \ .
\end{array} \right. 
\end{equation}
Here $E_{min}$ is the minimum three-body energy possible, depending on
the particular distances of M82 and NGC~3077 resulting from the
genotypes. All together the extended fitness function reads

\begin{equation}
\label{eq:fit-II}
\mathcal{F}_E(\vec{X} \mid D) = f_E \cdot \mathcal{F}(\vec{X} \mid D) \ .
\end{equation} 
Thus avoiding the unlikely constellations that both companions are
arriving from far distances fulfilling condition COND 
by collecting solutions with a three-body
energy $E$~$\le$~$0$ in the centre-of-mass system at $-$~7~Gyr, the
resulting merger rates for this fitness function are presented in
Figure~\ref{GA_II}. After $12.6$~Gyr, i.e. $5.6$~Gyr as from today, all solutions will have merged.

\subsection{Proper motions}

\begin{figure*}
\begin{center}
\begin{tabular}{cc}
\includegraphics[width=8cm]{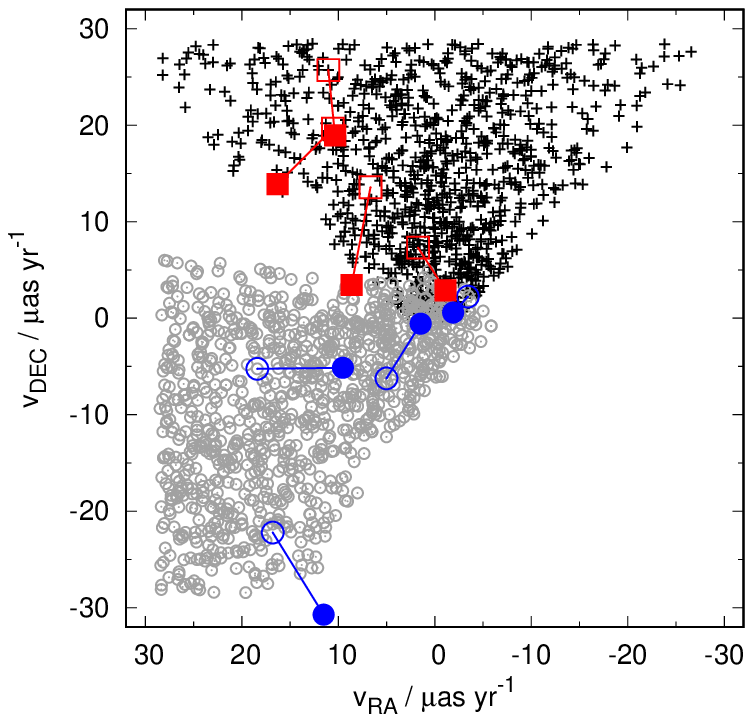} & \includegraphics[width=8cm]{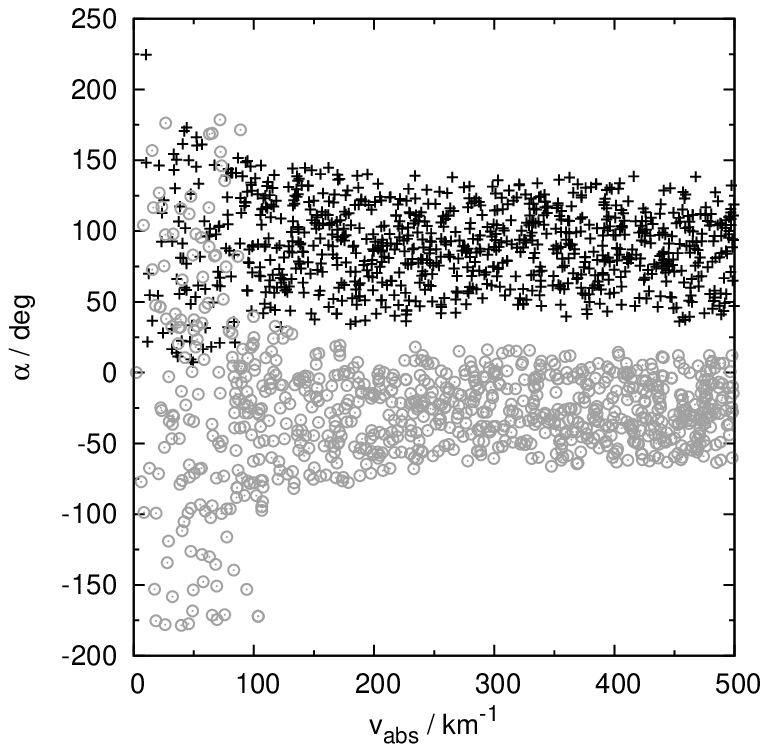}\\
\end{tabular}
\caption{\label{pmotion}Proper motions of M82 (crosses) and NGC~3077
  (open circles) of the GA solutions, all with respect to
  M81. {\bf Left panel:} Proper motions in RA and DEC in
  $\mu\textrm{as\,yr}^{-1}$. Note that the maximum velocity of
    the solutions in each x,y component is $\pm500$~$\textrm{km\,s}^{-1}$,
  corresponding to about $\pm30$~$\mu\textrm{as\,yr}^{-1}$. The solutions computed with RAMSES
  are shown as filled squares (M82) and filled circles (NGC~3077) while the corresponding
  GA solutions are depicted with open symbols, mutually connected by lines.
  {\bf Right panel:} The distribution of the absolute velocities,
  $v_\textrm{abs}$ in $\textrm{km\,s}^{-1}$, and the position angles,
  $\alpha$, in degrees, measured counterclockwise from the North
    Celestial Pole.}
\end{center}
\end{figure*}

The proper motions of the galaxies are extracted from the available
solutions.  Since the center-of-mass motion of the M81 group is a free
parameter in these calculations the proper motions of M82 and NGC~3077
in the solutions for GA are calculated with respect to M81 at $t=0$
(i.e. present day) in $\mu\textrm{as\,yr}^{-1}$
($1\,\mu\textrm{as\,yr}^{-1}$ corresponds to 17.2
$\textrm{km\,s}^{-1}$ based on the distance of M81 being 3.62~Mpc).
The results are shown in Fig. \ref{pmotion}.  The velocities are
clearly limited to certain directions, at least for higher
velocities. In the low-velocity end the direction angles are rather
randomly distributed. The distribution of solutions in
  Fig.~\ref{pmotion} allows to place additional constraints on the
  models with dark matter, once the proper motion of~M82 and/or
  of~NGC3077 have been measured.  In addition, the solutions computed
  with RAMSES (Sec.~\ref{sec:N-body}) are included as filled squares
  (M82) and filled circles (NGC~3077) while the corresponding GA
  solutions are depicted with open symbols, mutually connected by
  lines. The dynamically self-consistent RAMSES solutions are in good
  agreement with the GA solutions.

\section{Statistical Methods III: Discussion of MCMC vs. GA}
\label{sec:comp}

As presented in Section~\ref{sec:MCMC-results} apparently no full convergence 
is achieved in our approach of applying the MCMC method to the inner M81 group. 
The autocorrelation functions keep varying, though at low values (Figure~\ref{ACF}), 
and the merger rates appear "to walk around" (Figure~\ref{MC_merger}).

We expect that this phenomenon is caused by our choice of the various
contributions to the posterior probability density. To be precise, all
of them consist partly of a region with a constant value~1, simply
because we do not really know, and any procedure defining preferred
values there would mean a non-justified interference. As a result, for
each open parameter the individual walkers have no direction so long
they find themselves in regions of constant value, and unless they are taken
out from there again by a stretch move. The missing directions "where
to move to" in regions of constant value probably prevents, in our regard, 
the MCMC formalism from full convergence.

As already indicated at the end of Section~\ref{sec:MCMC-results} we sacrificed
on-top methodoligies to MCMC in order to overcome this problem. The transfer
to the genetic algorithm, implemented as an additional independent statistical method,
turned out to be extremely valuable. For, this formalism apparently delivers stable 
statistical results under circumstances mentioned above. To establish this assesment we 
repeated the GA-runs for a second time and present the results in Table~\ref{repeat}. 
Only slight deviations of the merger rates in comparison to the first evaluation are 
produced by the second one.

However as a matter of fact, at the beginning of our implementation we started with the genetic algorithm. 
Since, by means of this algorithm,  it appeared to be diffcult to generate solutions at all while establishing our concepts 
we decided to implement the Markov chain Monte Carlo method, too. And in fact we were
able to generate solutions by employment of the MCMC method quickly. In our experience, 
MCMC is the stronger search engine in comparison to GA and therefore helped signifficantly to "get feet 
on ground" during the early phase of our project.

To summarize our assessment: Both statistical methods turned out to be very valuable in 
our regard. MCMC helped to establish the concepts, and at the end GA delivered more 
stable results.

\begin{table*}           
\caption{GA: Percentages of mergers for selected periods of time
from the present until maximally 7~Gyr. 
The original GA-evaluations are denoted by \#1, and the repitition runs by \#2.}
\label{repeat}      
\centering                        
\begin{tabular}{c c c c c c c c c}        
\hline\hline                  
Fitness function    & GA-run   & 0-1 Gyr   & 0-2 Gyr    & 0-3 Gyr    & 0-4 Gyr   & 0-5 Gyr   & 0-6 Gyr    & 0-7 Gyr\\ 
\hline            
Eq.~\ref{eq:fit-I} &  \#1       & 15.1\%   & 42.8\%  & 56.6\%    & 63.3\%   & 67.5\%   & 69.8\%     & 72.2\%\\
                             &  \#2       & 13.9\%   & 40.4\%  & 54.0\%     & 61.1\%   & 67.2\%   & 70.9\%    & 73.2\%\\
\hline            
Eq.~\ref{eq:fit-II} &  \#1       & 22.3\%   & 62.5\%  & 88.8\%   & 96.8\%   & 99.3\%   & 100\%     &  100\%\\
                              &  \#2       & 23.2\%   & 60.9\%  & 86.1\%    & 96.8\%   & 99.2\%   & 99.9\%    &  100\%\\
\hline
\end{tabular}
\end{table*}                

\section{Numerical simulations with RAMSES}
\label{sec:N-body}

\begin{table*}
\begin{center}
\begin{tabular}{ccccccc}\hline
            &$x$&$y$&$z$&$v_x$&$v_y$&$v_z$\\\hline
\multicolumn{7}{c}{Model 1728-2}\\ 
M81         &153.33&-116.46&499.62&-19.04&16.67  &-62.78 \\
M82         &268.33&-385.66&-1143.&-38.02&46.11  &146.66 \\
NGC3077     &-1350.&1439.94&200.22&178.35&-185.38&-32.09 \\\hline
\multicolumn{7}{c}{Model 1728-6}\\ 
M81         &13.04  &409.78 &475.99 &-3.82 &-49.76&-61.98\\
M82         &-166.98&-1100.0&-1009.9&16.39 &139.63&132.02\\
NGC3077     &317.92 &534.88 &10.61  &-18.97&-78.76&-2.57 \\\hline
\multicolumn{7}{c}{Model 1728-16}\\ 
M81         &265.72 &597.69 &256.07 &-27.57&-81.71 &-31.57\\
M82         &-245.68&-1731.1&-567.83&29.68 &230.47 &72.63 \\
NGC3077     &-719.28&1068.89&61.64  &65.10 &-132.00&-13.59\\\hline
\multicolumn{7}{c}{Model 1729-633}\\ 
M81         &-18.84&205.98 &283.40 &-2.07 &-12.41&-20.41\\ 
M82         &68.10 &-467.72&-530.50&-13.74&31.52 &38.45 \\
NGC3077     &-64.56&74.58  &-155.10&41.29 &-12.12&10.59 \\\hline
\end{tabular}
\end{center}
\caption{Initial positions, and velocities of the halos hosting the
  galaxies M81 (virial mass $=1.17\cdot10^{12}\,M_\odot$), M82
  ($5.54\cdot10^{11}\,M_\odot$), and NGC3077
  ($2.43\cdot10^{11}\,M_\odot$). The positions are in kpc, the
  velocities in km/s. Each halo is modelled by
  particles of $1\cdot10^{6}\,M_\odot$, i.e., the M81 halo contains
  1.17 million particles, M82 554,000 particles and NGC3077 243,000
  particles.}
\label{halodata}
\end{table*}

\begin{figure*}
\begin{center}
\includegraphics[width=17.5cm]{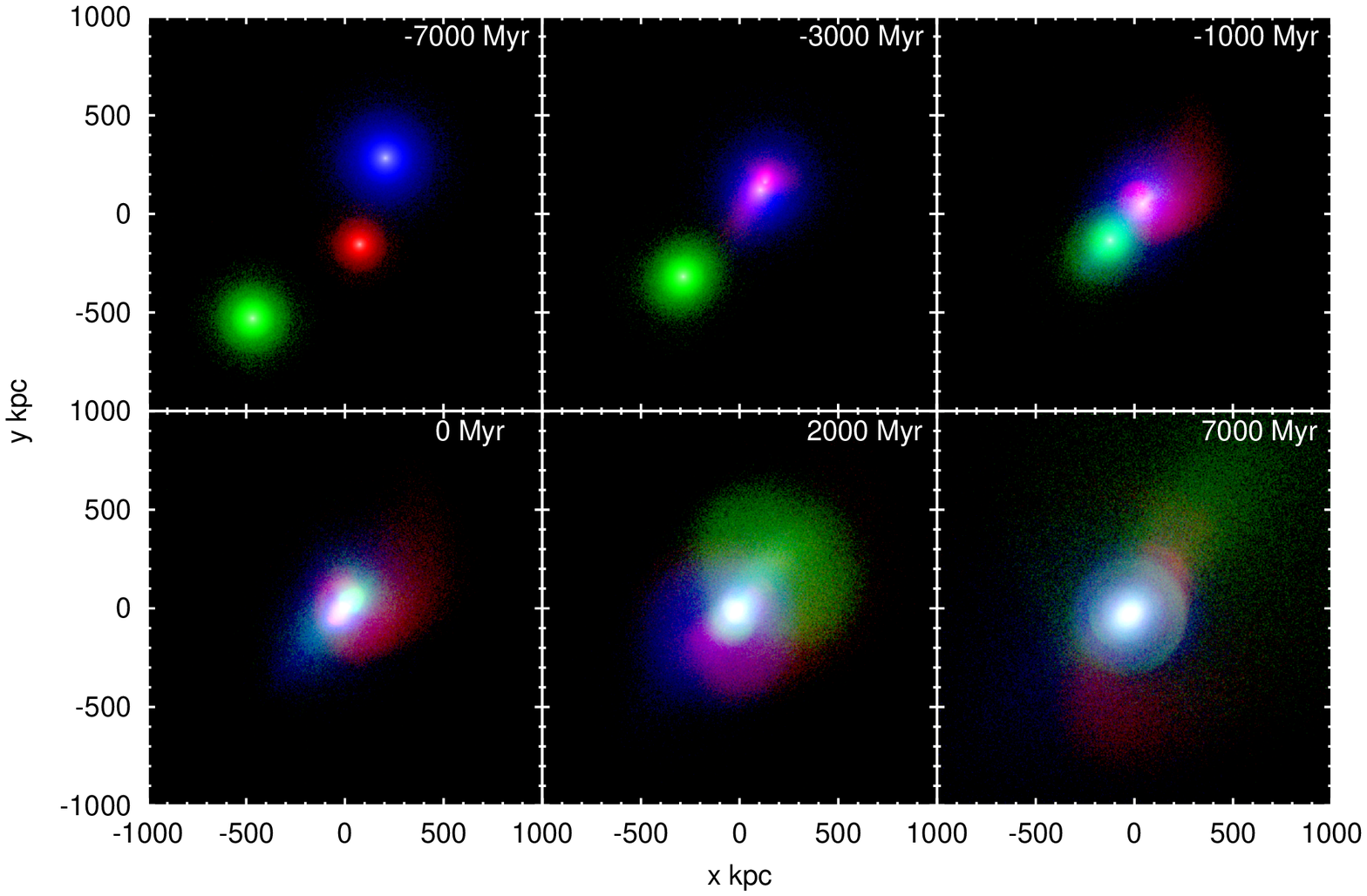}
\caption{\label{snapshots}Snapshots of the merging halos of NGC~3077
  (red) and M82 (green) with M81 (blue). The initial conditions
  correspond to model 1729-633.  It can be seen that the density
  centres merge within several Gyr after their first encounter with
  M81. Both halos lose substantial amounts of mass, partially ejected
  as tidal streams (NGC~3077), partially as a wide-angle spray cloud
  (M82). Also shell-like DM structures form (lower right panel).}
\end{center}
\end{figure*}

To validate the semi-analytic approach described in the previous
sections, we computed numerical models with the well-tested adaptive
mesh refinement (AMR) and particle-mesh code RAMSES
\citep{Teyssier2002}.  The halos are set up as NFW profiles
\citep*{NFW:1996} as in the semi-analytic model described in Section
\ref{sec:model}.  The virial masses are the same as in the
semi-analytic model (see Tables \ref{coordinates} and \ref{NFW}). The
total masses are taken as the DM halo masses, i.e. neglecting the
baryonic content.  The initial positions and velocities are listed in
in Table~\ref{halodata}.  In total 1.967 million particles are used
for the models shown here, with each particle representing
$10^6\,M_\odot$.  Each simulation takes between about 10 and 30 hours
with 32 parallel threads used.  Additional runs with 196,700 and
19,670 particles have been performed to confirm the models to be
essentially independent of the particle number, which is indeed the
case. The halos are allowed to settle for several Gyr before being
included in the initial conditions for the simulations. The position
of each halo is represented by the density centre calculated via the
method described by \citet {Casertano+Hut:1985}.  The density centre
method rather than the centre of mass has been chosen to determine the
positions and mutual distances of the halos since the in-falling halos
lose large amounts of mass upon encounter (see
Fig. \ref{snapshots}). The resulting steady shift of the centre of
mass would render its position useless for the purpose of this
study. In addition, the density centres are the best representations
of the positions of the embedded galaxies which can be observed today.

\begin{figure*}
\centering
\begin{tabular}{cc}
\includegraphics[width=8.0cm]{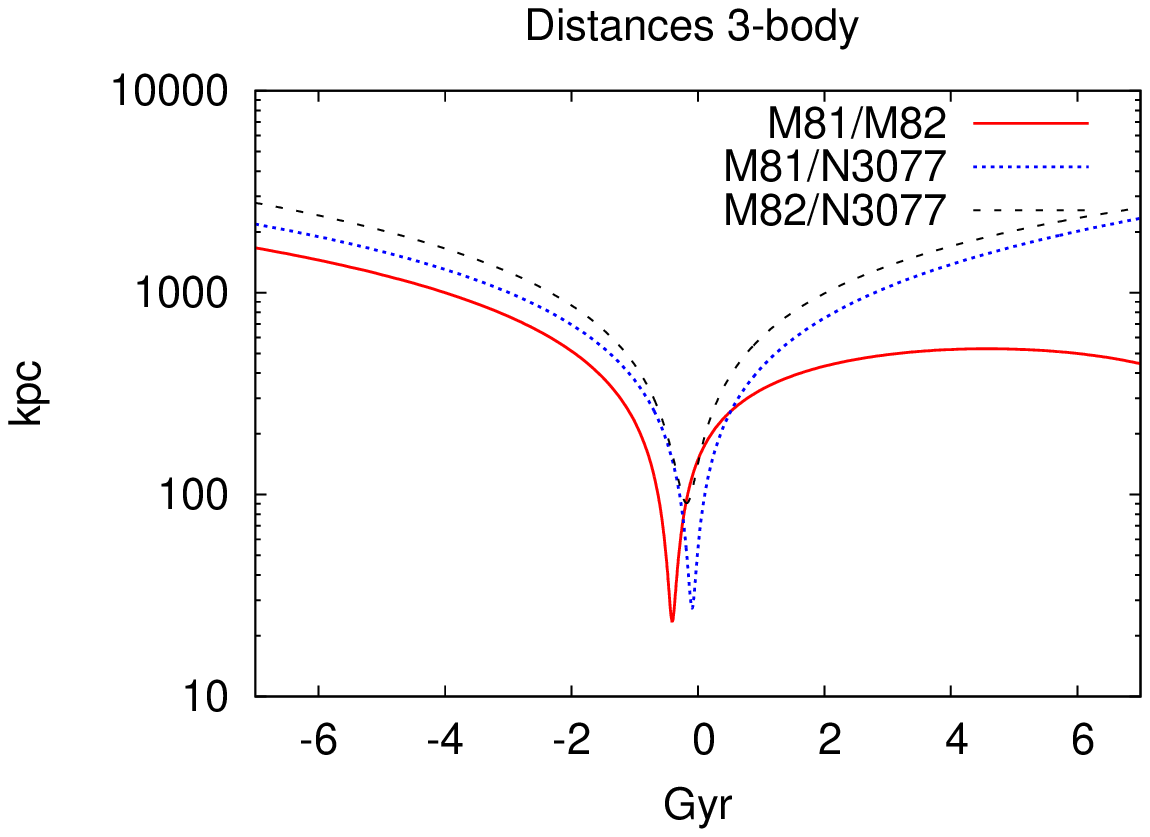} & \includegraphics[width=8.0cm]{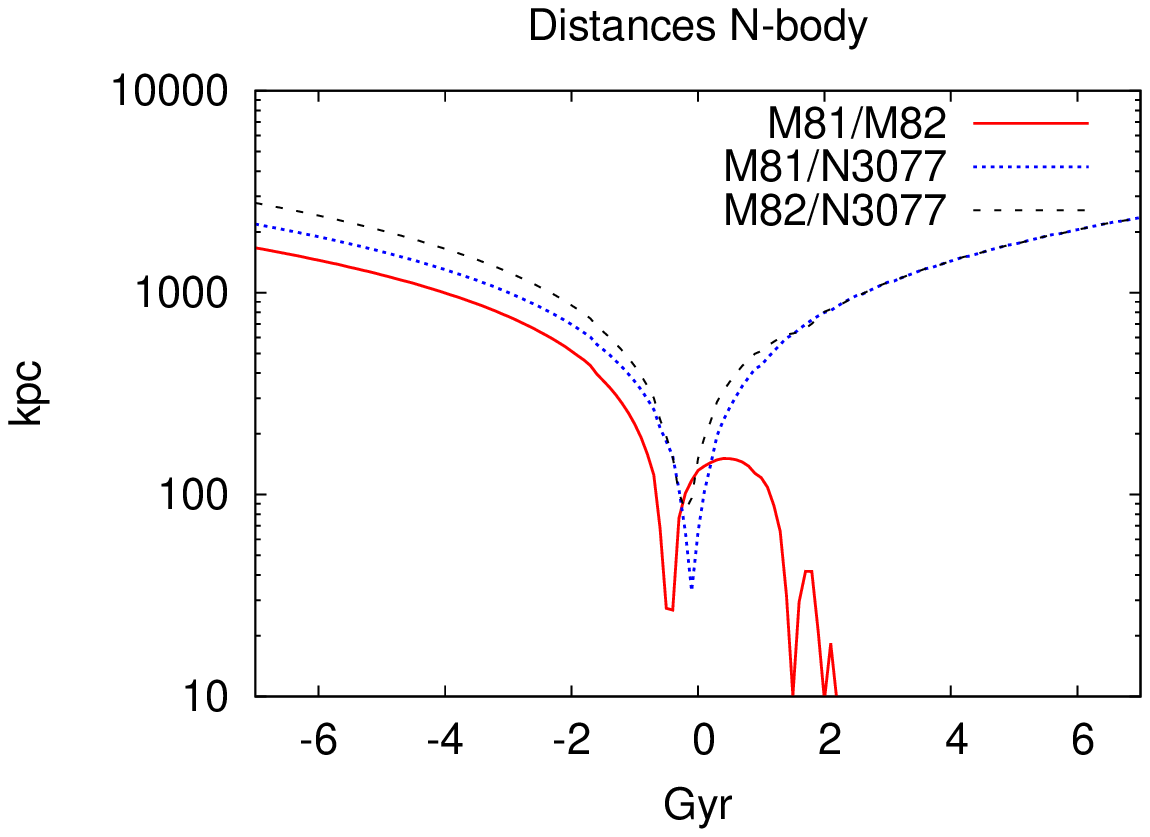} \\
\includegraphics[width=8.0cm]{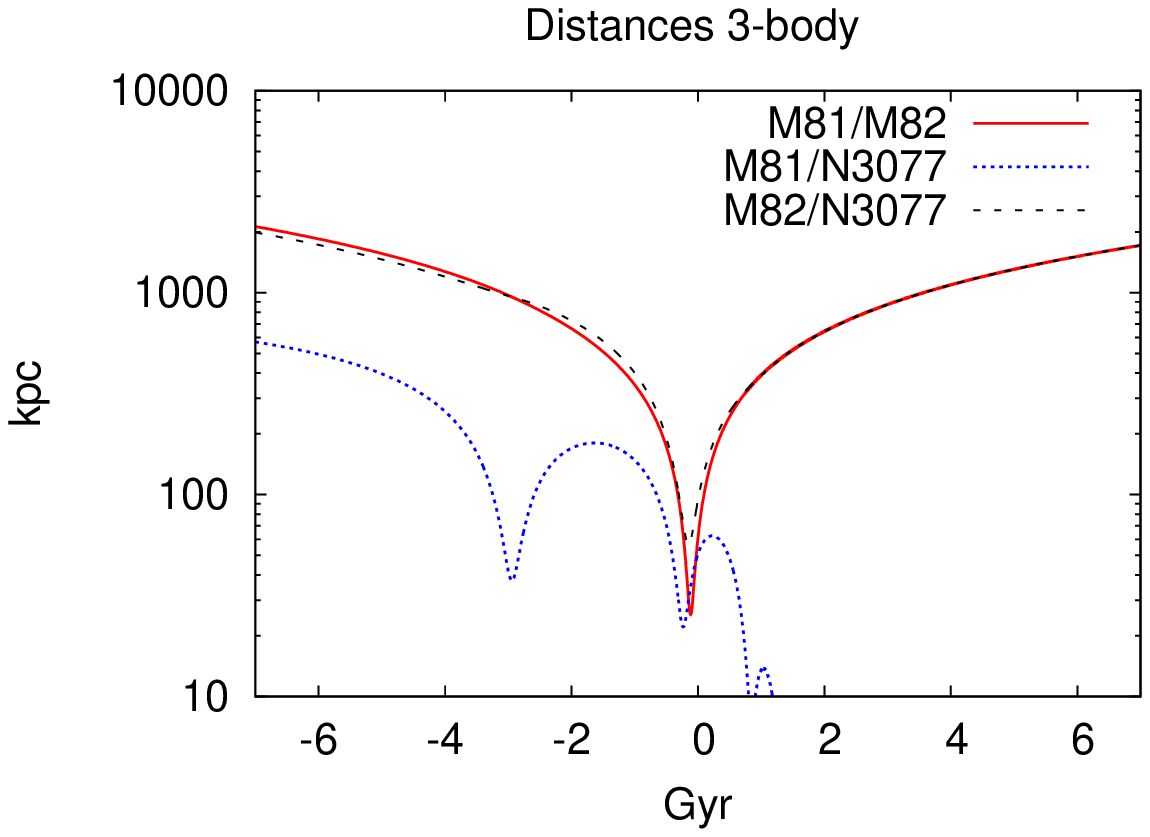} & \includegraphics[width=8.0cm]{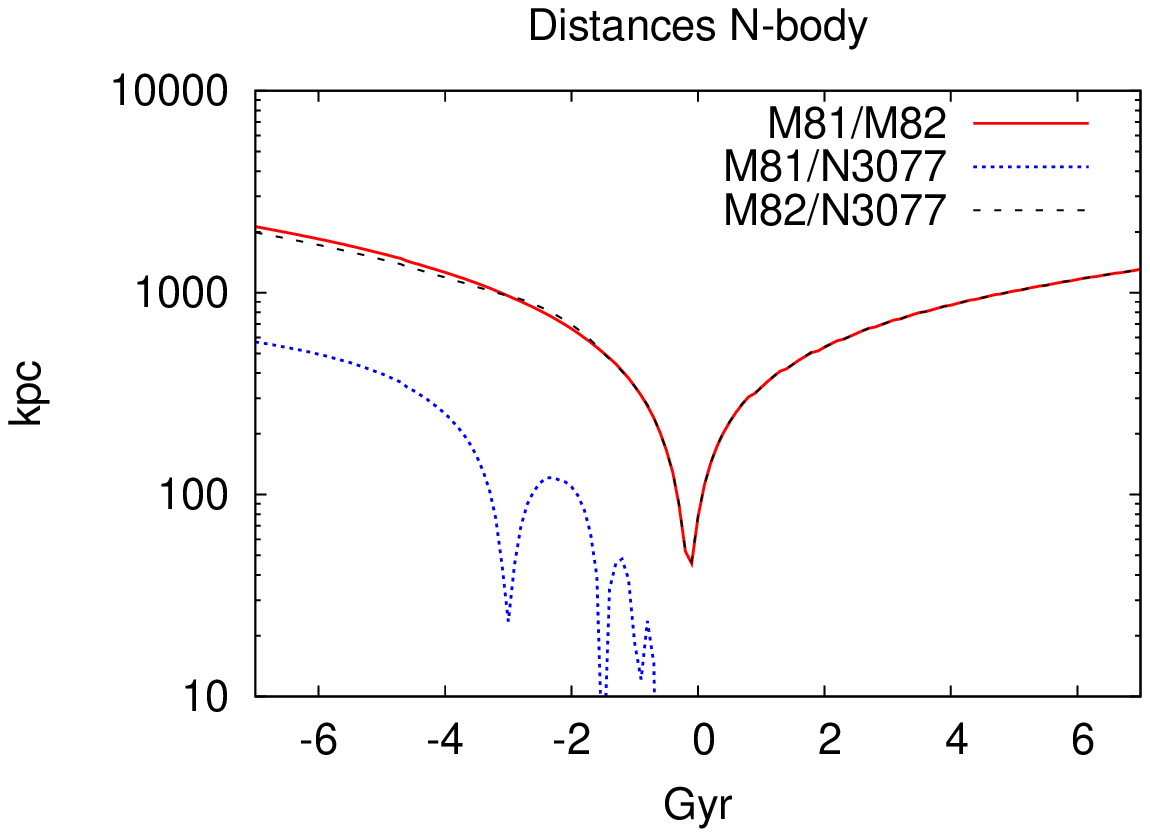} \\
\includegraphics[width=8.0cm]{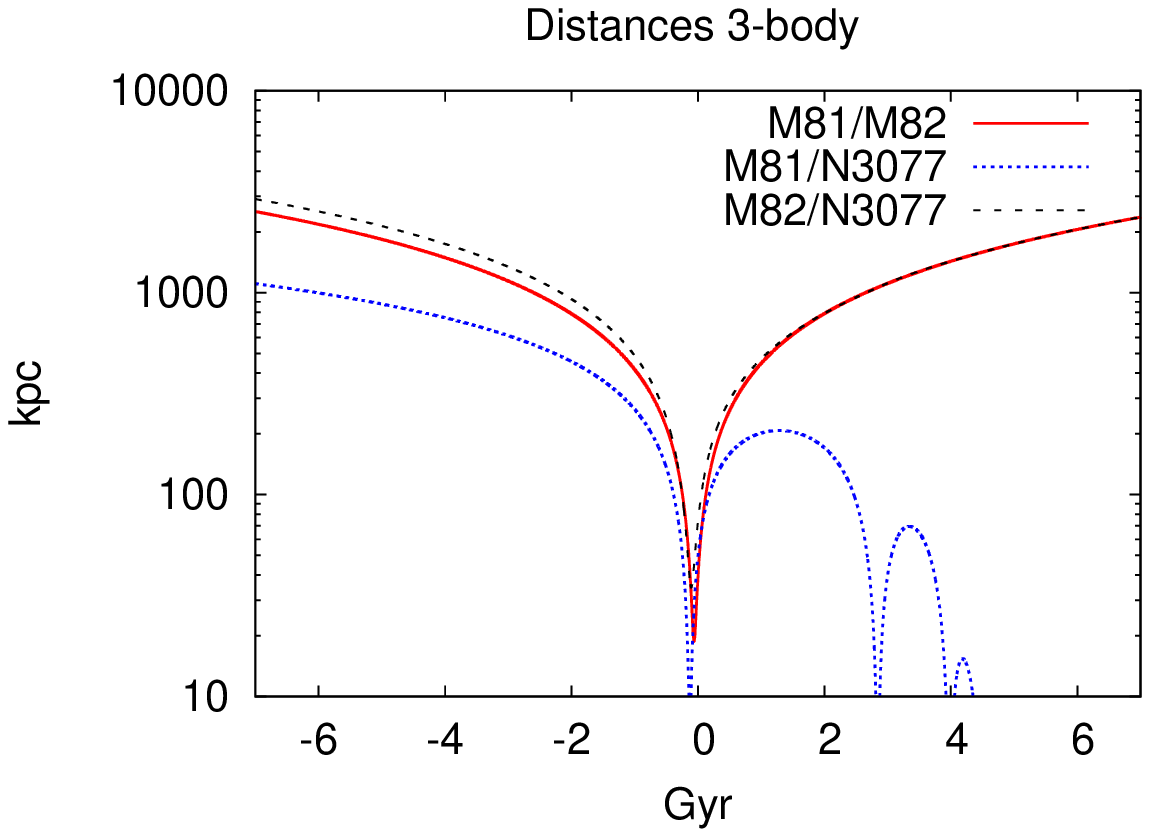} & \includegraphics[width=8.0cm]{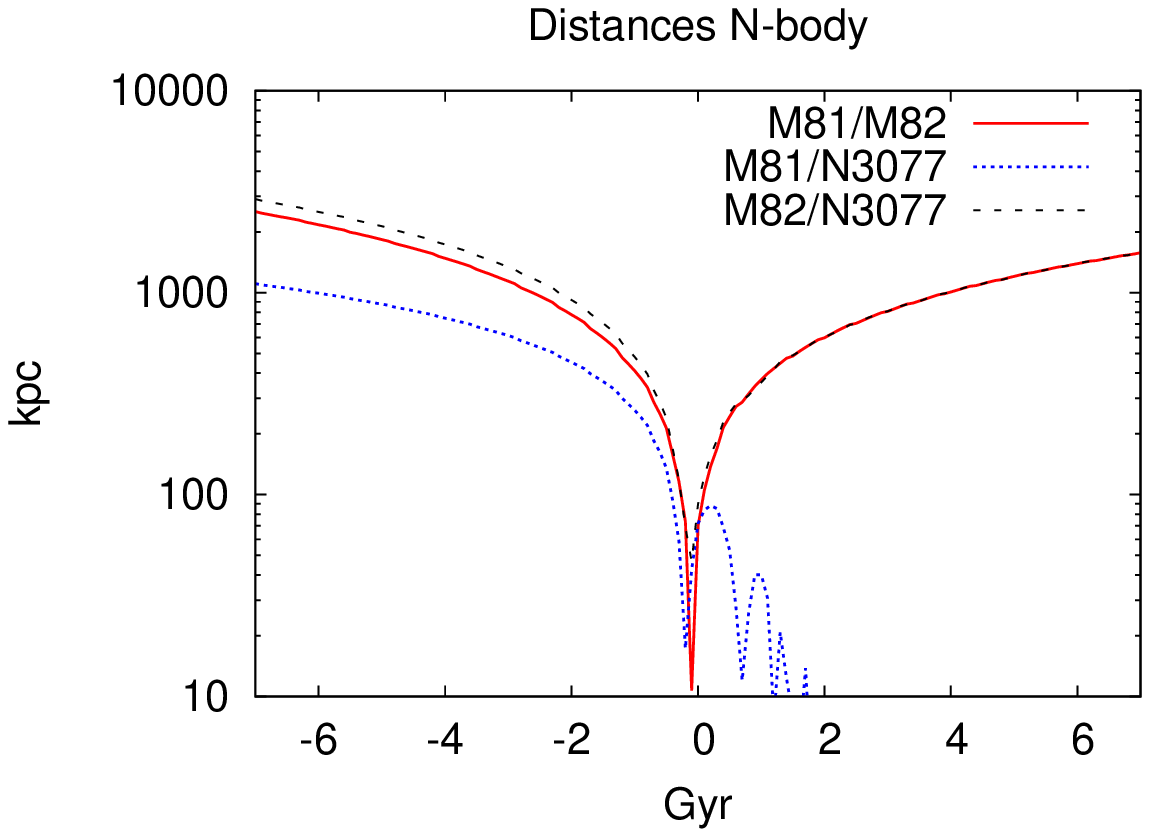} \\
\includegraphics[width=8.0cm]{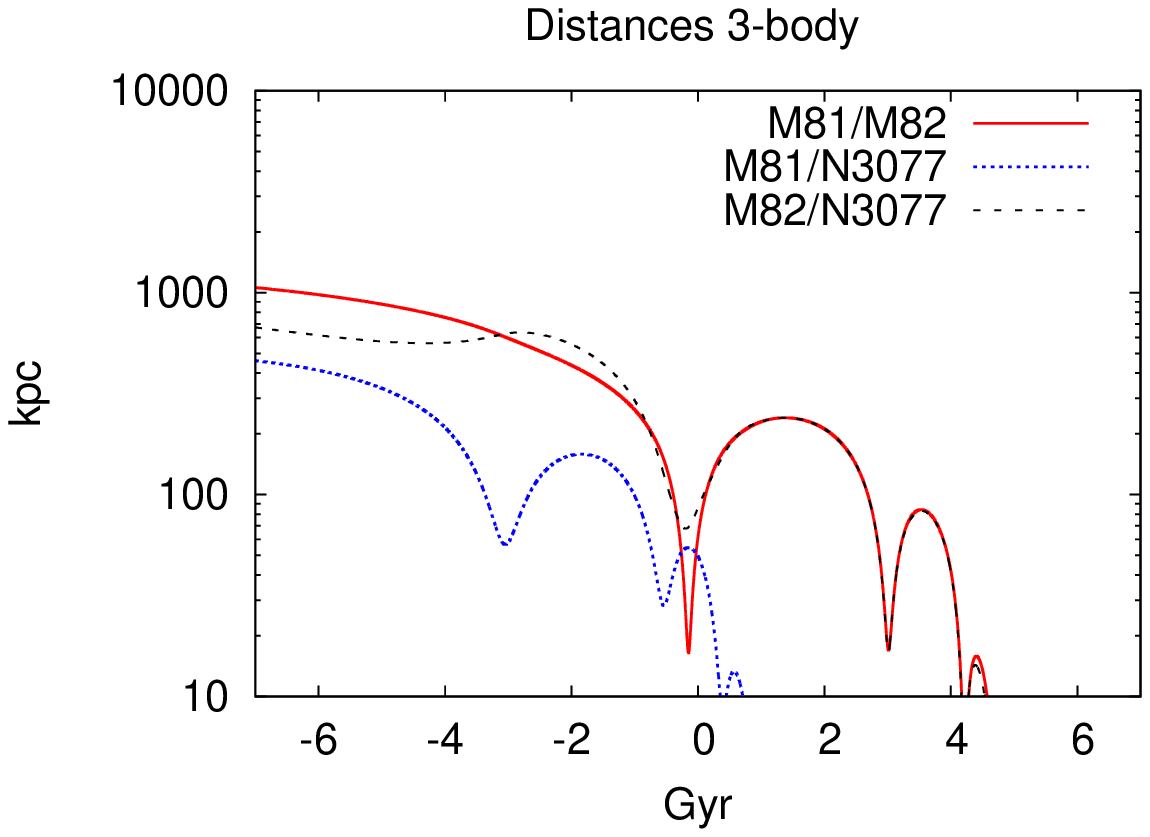} & \includegraphics[width=8.0cm]{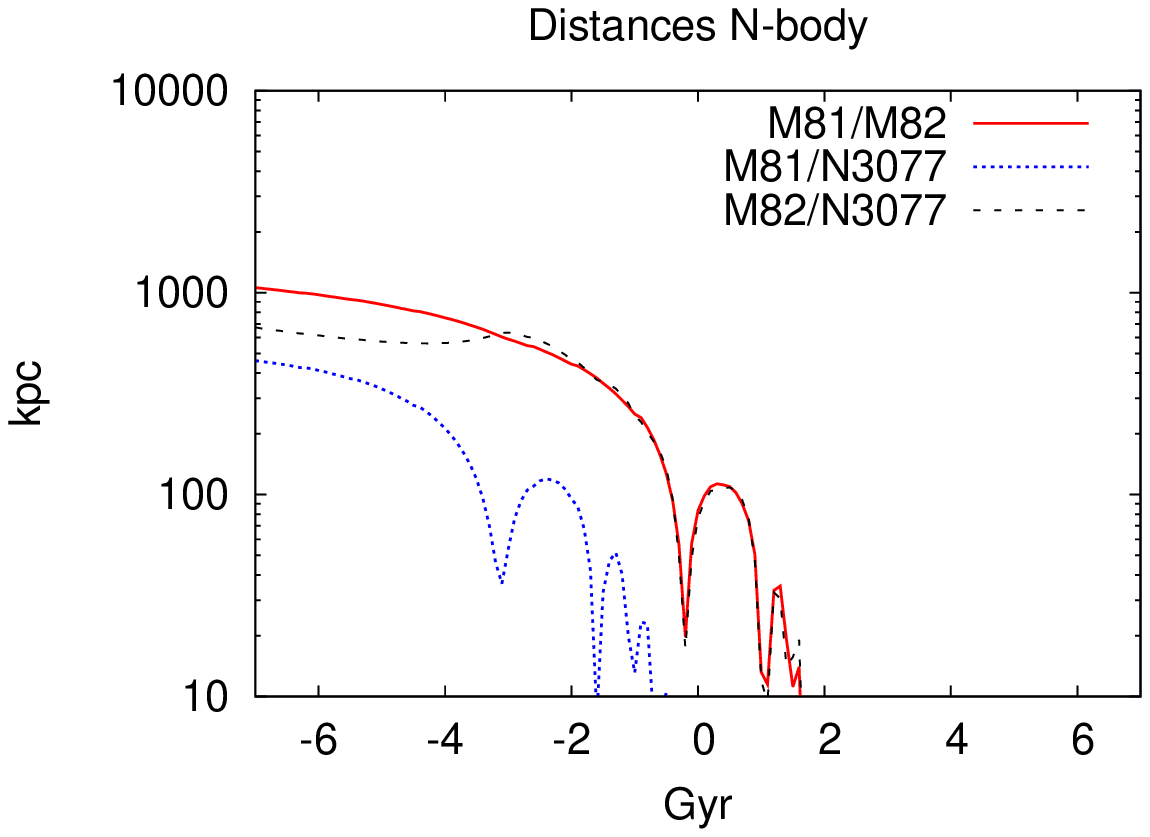} \\
\end{tabular}
\caption{Comparison between the results obtained by the 3-body model
  (left) and the N-body calculations (right). Four solutions were
  selected randomly out of the population created by the GA
  method, fitness function Eq.~\ref{eq:fit-I}. 
  First row (solution 1728-2): no merger within the forthcoming 7~Gyr. 
  Second row (1728-6): rapid merger ($0.8$~Gyr). 
  Third row (1728-16): merger after $4.2$~Gyr. 
  Fourth row (1729-633): both companions are merging with M81.}
\label{solutions}
\end{figure*}

\begin{figure*}
\centering
\begin{tabular}{cc}
\includegraphics[width=8.0cm]{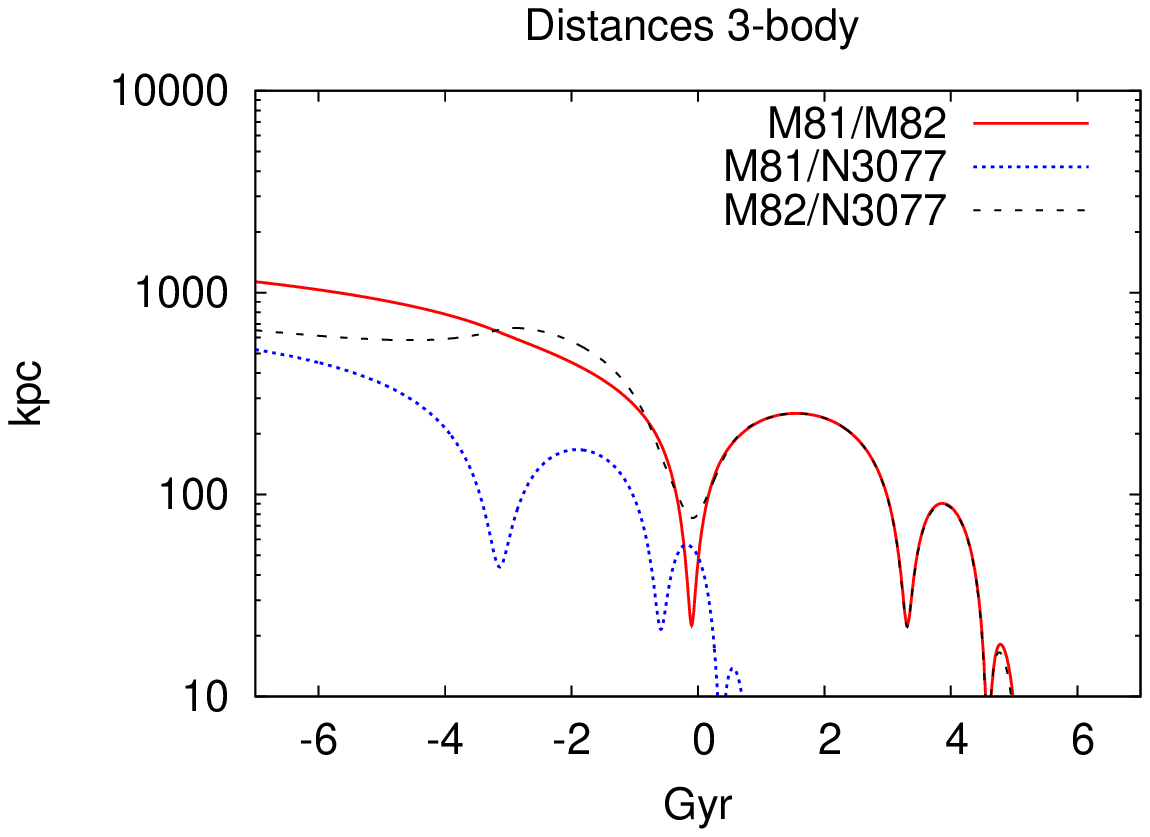} & \includegraphics[width=8.0cm]{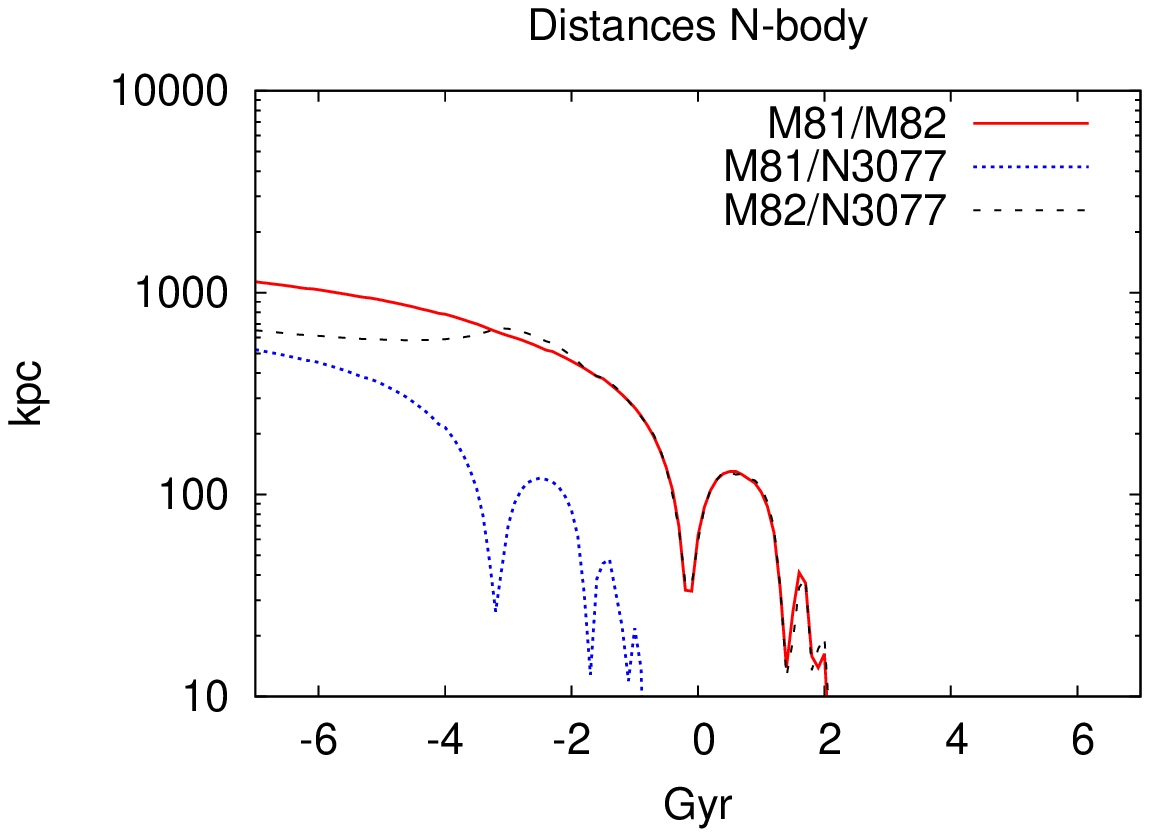} \\
\includegraphics[width=8.0cm]{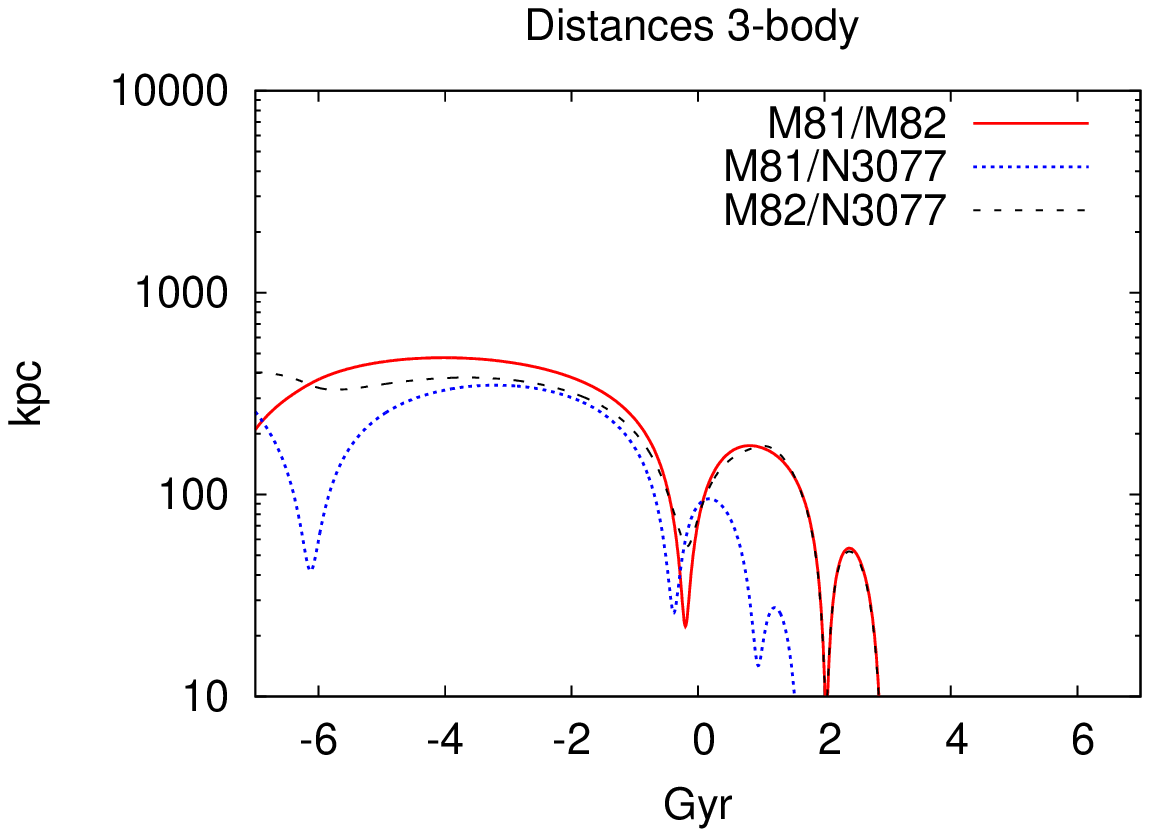} & \includegraphics[width=8.0cm]{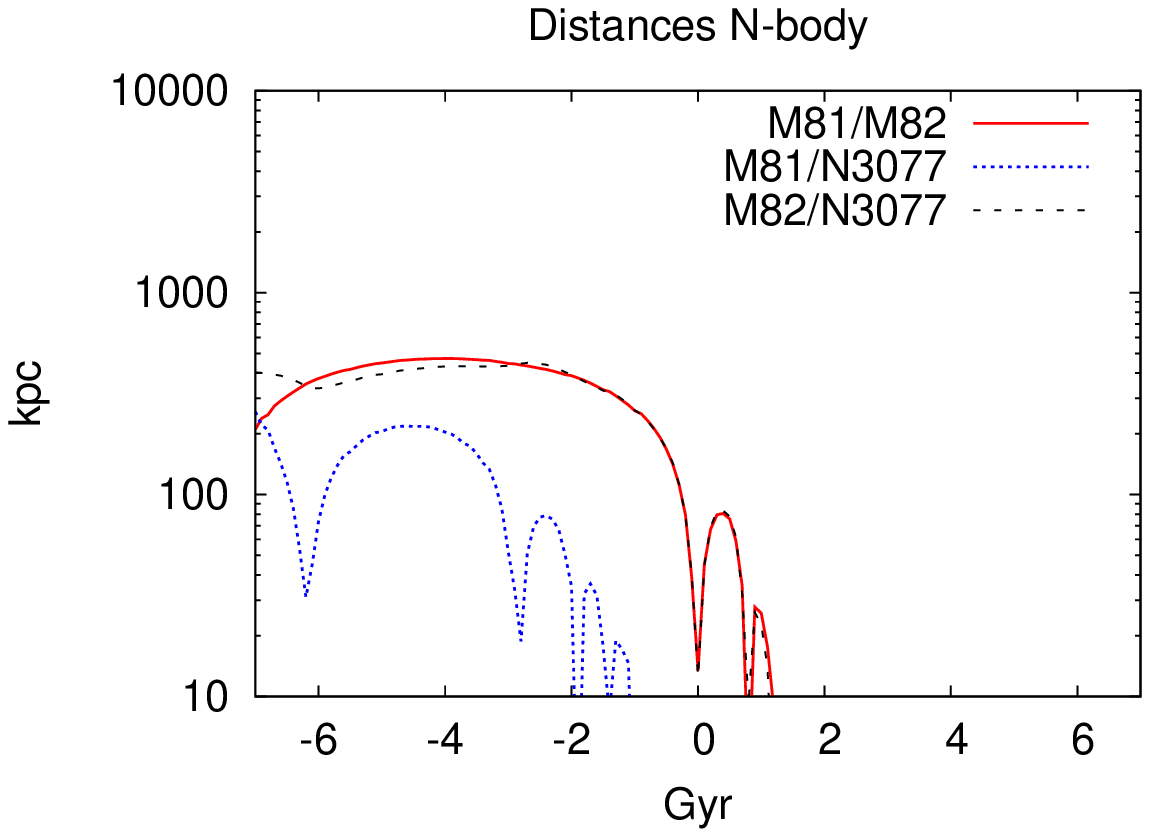} \\
\includegraphics[width=8.0cm]{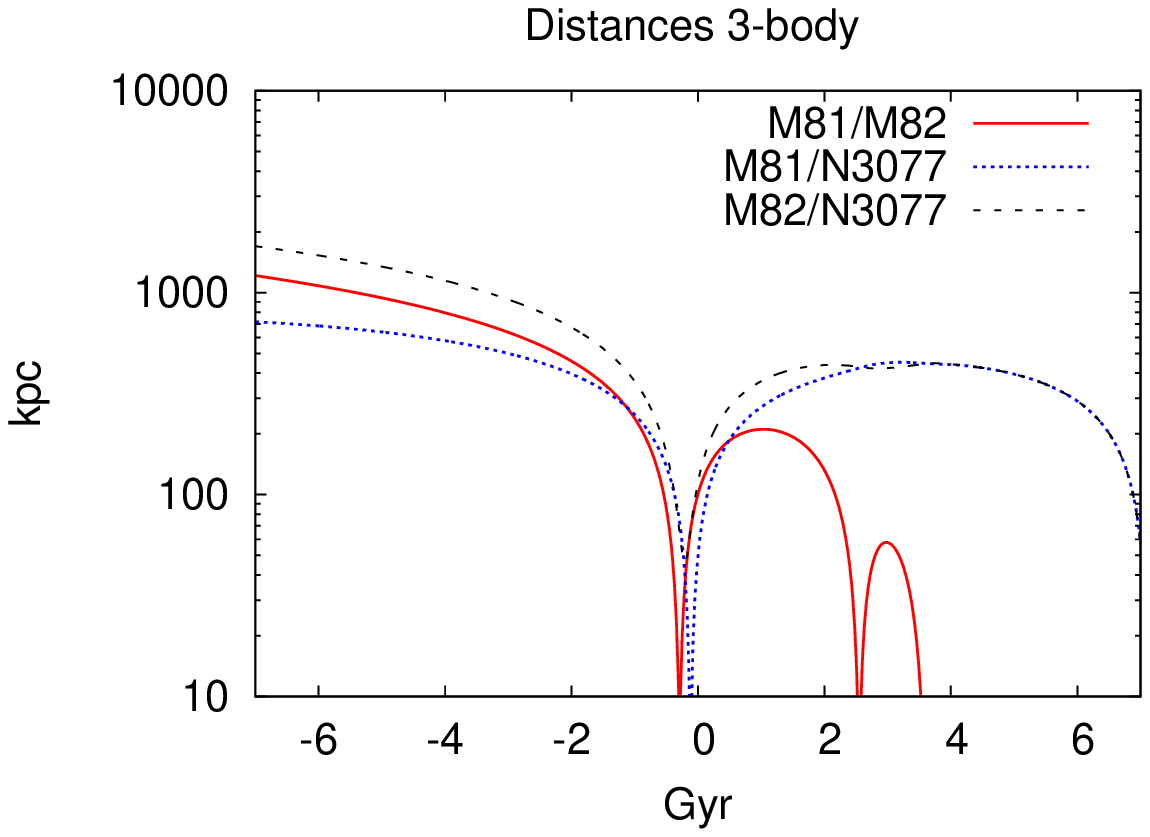} & \includegraphics[width=8.0cm]{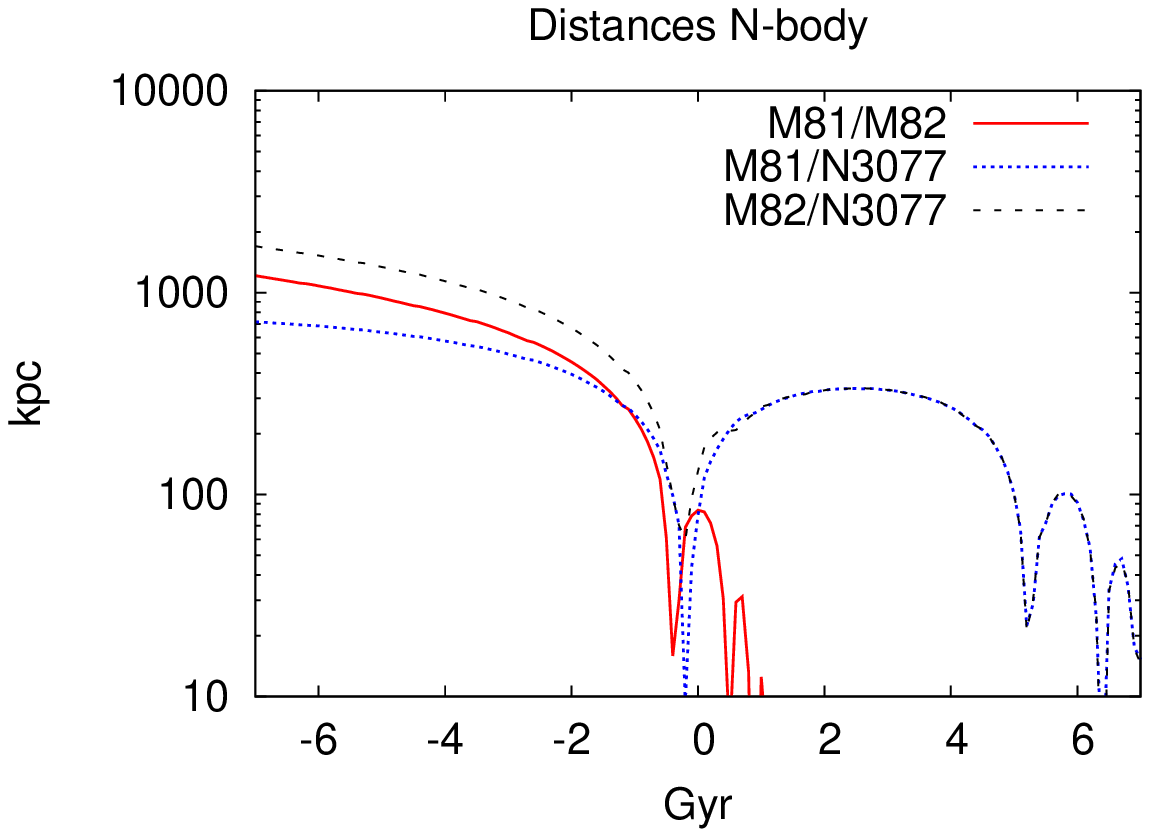} \\
\includegraphics[width=8.0cm]{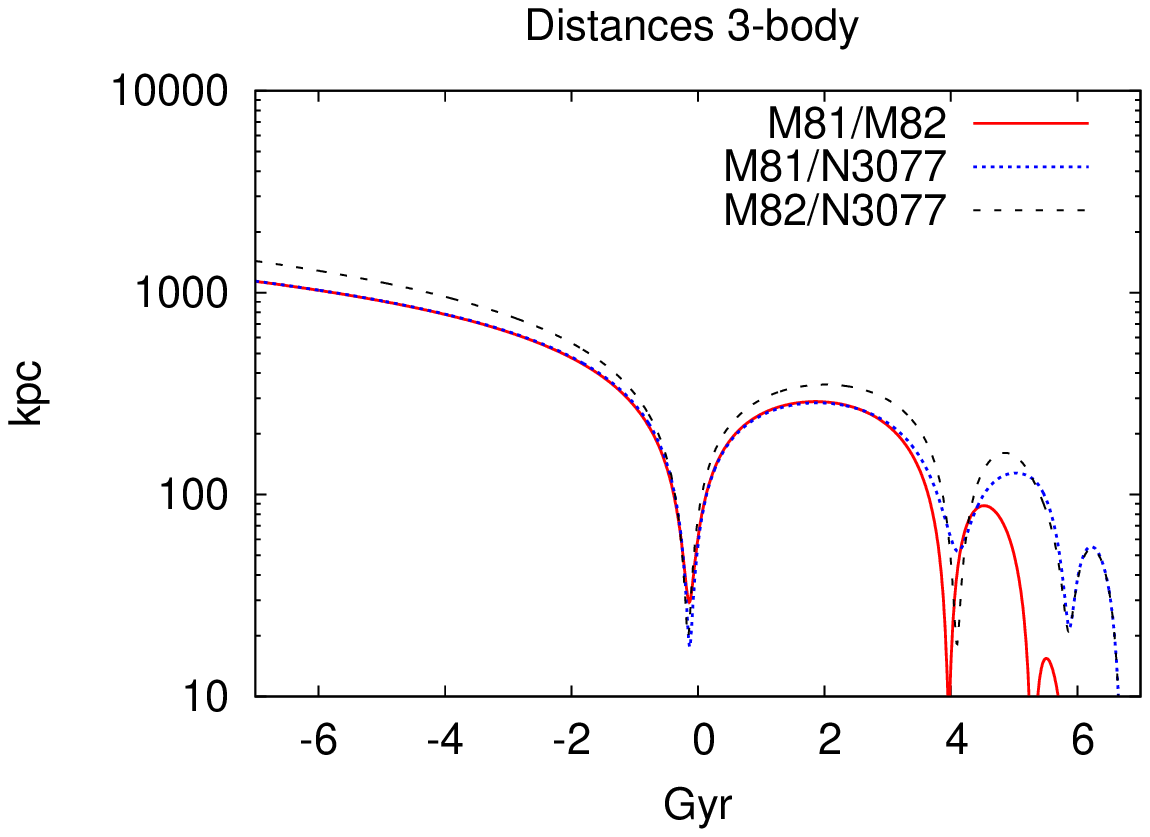} & \includegraphics[width=8.0cm]{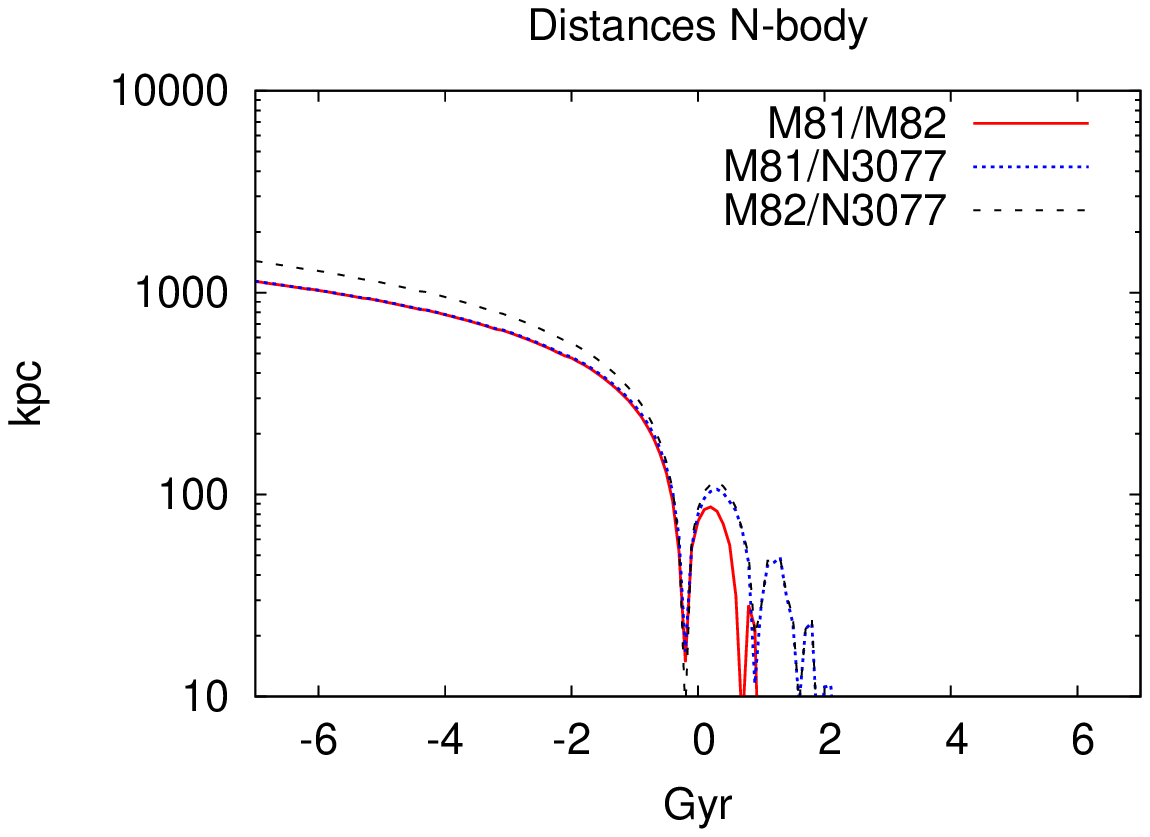} \\
\end{tabular}
\caption{Same as Figure~\ref{solutions}, but four solutions generated by means of 
  the extended fitness function Eq.~\ref{eq:fit-E}. 
  First row (solution 1750-314): most rapid merger ($0.3$~Gyr).
  Second row (1750-224): merger after $1.5$~Gyr.
  Third row (1750-423): merger after $3.5$~Gyr.
  Fourth row (1750-445): latest merger after $5.6$~Gyr.}
\label{solutions2}
\end{figure*}

The models studied in this section correspond to the 3-body genetic
algorithm solutions No. 1728-2, 1728-6, 1728-16, and 1729-633 
for fitness function Eq.~\ref{eq:fit-I}, and to the solutions No. 
1750-314, 1750-224, 1750-423, and 1750-445 for the extended
fitness function Eq.~\ref{eq:fit-II}.
They are shown on the left side of Figure~\ref{solutions} and
\ref{solutions2}, respectively, and were chosen randomly from the 
full set of GA solutions.

\subsection{Results}\label{sec:results:ramses}

Figure~\ref{snapshots} shows snapshots of the encounter leading to the
merger of all halos for model 1729-633. As can be seen the dynamical
friction causes the in-falling halos of NGC~3077 (red) and M82 (green)
to decelerate quickly and merge with the halo of M81 (blue) after a
few oscillations. Figure~\ref{solutions} and Figure~\ref{solutions2} 
show the mutual distances for the galaxies. 
The left-hand frames depict the results from the 3-body
GA calculations while the right columns contain the results of the
RAMSES $N$-body simulations. 
The solid line refers to the
distances of M82 to M81, the dotted line those of NGC3077 to M81, and
the distances between the minor components M82 and NGC3077 are shown
by the dashed line. All mutual distances follow a time evolution very
similar to the semi-analytic solution. The RAMSES $N$-body
computations lead to slightly to substantially lower post-encounter
velocities compared to the semi-analytical 3-body calculations.  The
reason for this is probably the large mass loss along two dense
tidal streams which causes a larger energy loss during the
encounter. This seems to contradict \citet{Boy-Kol+al:2008} who
predict longer merger times especially for large satellite-to-host
mass ratios compared to the results of semi-analytic calculations
using Chandrasekhar's formula.  In the Coulomb logarithm model used in
the previous sections, on the other hand, the merger timescale for the
semi-analytic calculations are slightly longer than in the RAMSES
simulation rendering that Coulomb logarithm more conservative.  The
halos merge as fast as in GA and even faster in the right panel, thus
validating the GA solutions.  There are some additional effects to be
noted:
\begin{itemize}
\item The in-falling halos of M82 and NGC~3077 suffer substantial mass
  loss during the merging process. The ejected DM forms tidal streams
  in almost opposite directions (NGC~3077, see Fig. \ref{snapshots})
  as well as a wide spray (M82), depending on the encounter
  parameters.
\item The density centres of M82 and NGC~3077 first oscillate around
  M81 and then merge shortly after the present (marked by time~$=0$).
\item All halos merge faster or lose more kinetic energy in the full
  $N$-body computations than in the semi-analytic model (right panels
  of Fig. \ref{solutions}). This emphasises the importance of
  dynamical friction.
\end{itemize}

\section{Conclusions}
\label{sec:conc}

We summarize the results obtained as follows:

\begin{itemize}

\item Solutions which fulfill the broad criterion COND and where the three
  galaxies have not merged at the present do exist within the LCDM
  model.

\item The treatment of the inner M81 group by high-resolution
  simulations of live self-consistent systems presented in
  Section~\ref{sec:N-body} confirms that the simplified three-body
  model of rigid NFW halos serves as a physical basis for our
  statistical evaluations. The semi-analytical approach for treating
  dynamical friction turns out to be a cautious estimate as the
  comparison with orbits obtained in Section~\ref{sec:N-body} show
  that in full N-body calculations the galaxies merge even more
  efficiently due to dynamical friction and as a result of energy
  removal through expulsion of dark matter particles as tidal
  material.

\item Two independent statistical methods (Section~\ref{sec:MCMC} and
  Section~\ref{sec:GA}) yield results in good agreement with each other.

\item According to these results, only $7\%$ of all initial conditions
  of either statistical population reflect pre-infall three-body bound
  states at $-7$~Gyr fulfilling condition COND. Such three-body bound
  states merge rapidly in the near cosmological future, the average future
  lifetimes being given by $1.7$~Gyr (according to the MCMC search)
  and $1.3$~Gyr (according to the GA search). The longest lifetimes
  are given by $2.7$~Gyr for MCMC and $2.8$~Gyr for the GA search (see
  Table \ref{tab:merger}).

\item All solutions (which fulfill the condition COND by the requirement
  to be a solution) of both statistical populations not merging within
  the next 7~Gyr from today comprise orbits where both companions initially are
  not bound to M81 and arrive from a far distance.

\item For the sake of avoiding those constellations  
  where both companions arrive nearly simeltaneously from a far distance, 
  an extended fitness function searching for states of negative
  pre-infall three-body energy (see Section~\ref{sec:eff}) delivers
  mostly results where one companion is bound to M81 while the other
  one arrives from a far distance. 
The longest future lifetime for this population is $5.6$~Gyr.

\end{itemize}

Thus, using three-body restricted simulations that model dynamical
friction on dark matter halos, which are in fact confirmed by
full N-body models, we find that the three inner galaxies in the M81
group are likely to merge within the next 1-2~Gyr.  If we assume that
the constraint COND is right then it is very likely that both or at
least one of the two dwarfs started out from far away and was not
bound at $-7\,$Gyr.  For a bound system which survives for a longer
time-span (into the past and also into the future) we either have to
relax the condition COND, or if COND is true, then the amount of DM of
the three galaxies assumed in this study needs to be reduced
significantly. Additional constraints will be available with relative
proper motion measurements within the M81 group of galaxies which will
significantly reduce the allowed range of solutions and by taking 
into account the tidally displaced gas distribution in the M81 system. 

Other evidence for dynamical friction through dark matter halos or the
lack thereof has been discussed, and it is worth to briefly mention
it here. The observed compact groups of galaxies would also
require highly tuned initial conditions such that its members are all
together about 1--2~Gyr from merging (see
e.g. \citealt{Sohn2015}), thus confirming our results regarding the
inner M81 group of galaxies.
Measured proper motions of the satellite galaxies of the Milky Way
provide additional constraints \citep{Angus2011}.

\section*{Acknowledgements}

W. Oehm would like to express his gratitude for the support of
{scdsoft~AG} in providing a SAP system enviroment for the numerical
calculations. Without the support of {scdsoft's} executives
\emph{P. Pfeifer} and \emph{U. Temmer} the innovative approach of
programming the numerical tasks in SAP's language ABAP wold not have
been possible. \\
We are grateful to the reviewer for constructive 
support and we thank \emph{Ch. Theis} for a fruitful general
discussion on appropriate approaches for an implementation of the
genetic algorithm in the context of galaxy encounters.


\appendix

\section{Program Development}
\label{app:pd}

Apart from Section~\ref{sec:N-body} (Numerical Simulations with
$\textnormal{RAMSES}$), all numerical computations were realized
within the frame of SAP's Development Workbench, utilizing the
programming language ABAP and the related debugging facilities, the
Repository (SE11) for easy creation of the requested statistical
databases, and their evaluation (SE16, REUSE\_ALV\_GRID\_DISPLAY). In
comparison to development environments like C++ or FORTRAN, the ABAP
Development Workbench facilitated a program development time at least
two or three times faster. However aspects like the performance of
floating point calculations, or easy linkage to mathematical standard
libraries need to be improved for the sake of full acceptance in the
world of numerical programming.

\section{Abbreviations}
\label{app:abr}

\begin{flushleft}
\begin{tabular}{ll}

{ABAP}:  & SAP's programming language \\

{COND}: & Condition specified in Section~\ref{sec:approach} \\

{DM}:      & Dark matter \\

{GA}:       & Genetic algorithm \\  

{MCMC}:  & Markov chain Monte Carlo \\

{MW}:     & Our galaxy (Milky Way)  \\

{NED}:     & NASA/IPAC extragalactic database \\

{NFW}:    & Navarro, Frenk \& White profile \\

{NTB}:     & North tidal bridge \\  

{POS}:     & Plane of sky \\

{LCDM}:   & Standard dark-energy plus cold-dark-matter\\ 
                 & model of cosmology \\

{STB}:      & South tidal bridge \\  

\end{tabular}
\end{flushleft}
    
\section{Numerical Solution of the three-body orbits}
\label{app:num}

The Hamilton equations, extended by the non-conservative forces $\vec{F}^{DF}_{ij}$ due to dynamical friction\footnote{Actually Newton's equations, transformed into a coupled set of differential equations of $1^{st}$ order.}, have been integrated numerically in the centre of mass system of the three galaxies M81, M82 and NGC~3077, represented by the indices $i,j$~=~1,2,3, respectively. Using Cartesian coordinates (index~$k$~=~1,2,3), they read in the 18-dimensional phase space (the advantage of reducing this dimension to 12 by the use of Jacobi variables is being cancelled by the disadvantage of complicating the structure of the equations):

\begin{displaymath}
\frac{d}{dt}{(\vec{x}_i)}_k = \frac{{(\vec{p}_i)}_k}{m_i} \ , 
\end{displaymath}

\begin{displaymath}
\frac{d}{dt}{(\vec{p}_i)}_k
= \sum_{j \neq i} \left[ {- \ \frac{\partial}{{\partial {(x_i)}_k}}} V_{ij} 
\ + \ (\vec{F}^{DF}_{ij})_k \ - \  (\vec{F}^{DF}_{ji})_k \right]  \ .
\end{displaymath}
The DF force on galaxy~$i$ moving in the DM halo of galaxy~$j$ is given by $\vec{F}^{DF}_{ij}$. Vice versa, the DF force on galaxy~$j$ moving in the DM halo of galaxy~$i$ appears in the equation of motion for galaxy~$i$ via \emph{actio est reactio} with reversed sign.
\\

The potential energies $V_{ij}$ betweeen the galaxies are detemined by the NFW density profiles given by Eq.~\ref{eq:NFW} and Table~\ref{NFW}. At distances where the DM halos of the galaxies are completely separated the gravitational force is given by Netwon's law for point masses. In case of overlapping DM halos the potential energy between two halos~$i,j$ is numerically calculated by reducing halo~$j$ to volume elements $h^3$ with $h~=~1~kpc$ and summation of those volume elements in the potential~$\Phi_i$ of halo~$i$: 

\begin{displaymath}
\Phi_i(s_i) = \left\{
\begin{array}{ll}
- \ G \ \int\limits_{0}^{s_i} \ \frac{m_i(r_i)}{{r_i}^2} \  dr_i \ \ + \ const , &  s_i \le R_{200_i} \ ,\\
\\
- \ G \ \frac{m_i}{s_i} \ , &  s_i > R_{200_i} \ ,\\
\end{array} \right. 
\end{displaymath}
$const$ being determined by continous transition at $R_{200_i}$,

\begin{displaymath}
V_{ij}(r_{ij}) = \left\{
\begin{array}{ll}
\sum_{n}\Phi_i(s_{i_n}) \ \rho_j(s_{j_n}) \  h^3 , &  r_{ij} \le R_{200_i} + R_{200_j}\ ,\\
\\
- \ G \ \frac{m_i m_j}{r_{ij}} \ , &  r_{ij} > R_{200_i} + R_{200_j}\ ,\\
\end{array} \right. 
\end{displaymath}
with $s_i$ and $s_j$ denoting the distances of volume element $h^3$ to the centres of halos~$i$ and~$j$, respectively. For overlapping halos the derivative of the potential energies are then obtained numerically. The first row of Figure~\ref{forces} visually proves the stability of our numerical computations and their accuracy, which of course additionally was checked by comparing the results for varying seizes of the volume elements. 
\\

Furthermore, incorporating the dynamical friction between the DM halos of the three galaxies according to Eq.~\ref{eq:chandra}, first of all one has to replace $\vec{\textbf{v}}_M$ by the relative velocities

\begin{displaymath}
\vec{\textbf{v}}_{ij} = \vec{\textbf{v}}_i - \vec{\textbf{v}}_j \ .
\end{displaymath}
 Employing \emph{actio est reactio} we find

\begin{displaymath}
\frac{d}{dt} \vec{\textbf{v}}_{ij} = 
\frac{d}{dt} \left( \vec{\textbf{v}}_i - \vec{\textbf{v}}_j \right) = \frac{1}{\mu_{ij}} \frac{d\vec{p}_i}{dt} 
\ \ \ \rm{with} \ \ \  \frac{1}{\mu_{ij}} =  \frac{1}{m_i} + \frac{1}{m_j} \ ,
\end{displaymath}
with the result for Chandrasekhar's formula:

\begin{displaymath}
\vec{F}^{DF}_{ij} = - \mu_{ij}\frac{4{\pi}G^2 {m_i}{\rho_j}}{\textnormal{v}_{ij}^3}\ \textnormal{ln}{\Lambda}_{ij}
\left[\textnormal{erf}(X_j) - \frac{2X_j}{\sqrt{{\pi}}}\textnormal{e}^{-X_j^2} \right] {\vec{\textbf{v}}_{ij}} \ .
\end{displaymath}
The dependence of the Coulomb logarithm on the masses (see Eq.~\ref{eq:cl}) is indicated by $\Lambda_{ij}$ and the entity $X_j$ is determined by the relative velocity $\textnormal{v}_{ij}$ and the dispersion~$\sigma$ of galaxy~$j$.

However, this equation is based on the assumption of Maxwellian type velocity distributions which is not the case for NFW profiles. Since most of the contribution to dynamical friction is certainly caused by the immediate vicinity of an intruder we replace the dispersion~$\sigma$ by the distance dependent circular speed $v_c$

\begin{displaymath}
X_j(s_j) = \textnormal{v}_{ij} / v_{c_j}(s_j) \ ,
\end{displaymath}
at a distance $s_j$ from the halo centre of galaxy~$j$. Similarly to the potential energies, here the total DF force between two overlapping halos is numerically obtained by reducing the DM halo~$i$ to volume elements $h^3$ with distance~$s_i$ to its centre, and computing the sum over them:

\begin{displaymath}
\vec{F}^{DF}_{ij} = - \mu_{ij}\frac{4{\pi}G^2 }{\textnormal{v}_{ij}^3}  f_{ij} \ {\vec{\textbf{v}}_{ij}} \ , 
\end{displaymath}
with

\begin{displaymath}
f_{ij} = \sum_{n} \rho_i(s_{i_n}) h^3 \ \rho_j(s_{j_n}) \ \textnormal{ln}{\Lambda}_{ij} 
\left[\textnormal{erf}(X_j) - \frac{2X_j}{\sqrt{{\pi}}}\textnormal{e}^{-X_j^2} \right]  
\end{displaymath}
and

\begin{displaymath}
\textnormal{ln}{\Lambda}_{ij}(s_j) = \textnormal{ln} \left[1 + m_j(s_j) / m_i \right] \ .
\end{displaymath}
For the sake of a better performance, the potential energies $V_{ij}$ and the functions $f_{ij}$ where calculated once on a grid of appropriate base points and interpolated when numerically solving the equations of motion.
\\

For the computation of the differential equations we implemented methods for the Runge-Kutta and leapfrog integration schemes. In order to save performance, a practical variable step size method for the solution of the Hamilton equations was created by adapting the step size according to the minimum of the, at the corresponding point of time, given distances between M81, M82 and NGC~3077, denoted as $d_{min}$. For the leapfrog integration scheme it reads:

\begin{displaymath}
\Delta t = \left\{
\begin{array}{ll}
0.1 \ \rm{Myr}, &  d_{min} \le 1 \ \rm{kpc} \\
0.1 \ (d_{min}/\rm{kpc}) \ \rm{Myr},  &  1 \ \rm{kpc} < d_{min} \le 1000 \ \rm{kpc} \\
100 \ \rm{Myr}, &  d_{min} > 1000 \ \rm{kpc} \ .
\end{array} \right. 
\end{displaymath}
This method was thoroughly tested by comparing solutions achieved in that way with integration runs using fixed small stepsizes employing the leapfrog and the Runge-Kutta integration schemes.

\section{The Markov Chain Monte Carlo Method}
\label{app:MCMC}

A Markov chain ensemble consists of $N_L$ walkers represented by vectors $\vec{X}_l\in {\textrm{R}}^{N_P}$ with $l\in\{1,...,N_L\}$ and $N_P$ denoting the dimension of the parameter space. In our case we have $\vec{X}_l=\left(P_1 ... P_6\right)$ (see Table~\ref{parameters}). For the evolution of an affine-invariant Markov chain of ensembles by means of the Metropolis-Hastings algorithm we implemented the stretch move as explained in \citet{Foreman2013}, which is based on the determination of the posterior probability density for one walker

\begin{equation}
\label{eq:posterior}
\pi(\vec{X}\mid D) = const \cdot p(\vec{X}) \cdot P(\vec{X}\mid D)\ .
\end{equation}
In our regard the normalization factor $const$ is not relevant because only the ratio of the posterior probability densities for two walkers have to be determined when performing the stretch move. $p(\vec{X})$ denotes the prior distribution, and the likelihood function $P(\vec{X}\mid D)$ may not only depend on the parameters $P_1$ to $P_6$, but on further entities symbolized by $D$, too.

The mean value of parameter~$P_j$ for sample~$t$ is given by

\begin{equation}
\label{eq:mv1}
f_j(\vec{X}(t)) = \frac{1}{N_L} \sum_{l=1}^{N_L} {\left[ \vec{X}_l(t) \right]}_j \ ,
\end{equation}
and dealing with a total number of $N_T$ samples we have for the overall mean value:

\begin{equation}
\label{eq:mv2}
<f_j>
= \frac{1}{N_T} \sum_{t=1}^{N_T} f_j(\vec{X}(t)) \ .
\end{equation}
For sufficiently large values of $N_T$ the autocovariance function for parameter~$P_j$ can be approximated by

\begin{eqnarray}          
\label{eq:ac1}
&& C_j(t) \approx \frac{1}{{N_T} - t} \ \times \\
&&\times \sum_{t'=1}^{{N_T} - t} 
{\left( f_j(\vec{X}(t'+t)) - <f_j> \right)} {\left( f_j(\vec{X}(t')) - <f_j> \right)}, \nonumber
\end{eqnarray}
and we finally arrive at estimates for the autocorrelation functions,

\begin{equation}
\label{eq:ac2}
\rho_j(t) = \frac{C_j(t)}{C_j(0)} \ ,
\end{equation}
and the integrated autocorrelation times

\begin{equation}
\label{eq:ac3}
\tau_j \approx 1 + 2 \sum_{t=1}^{N_T} \rho_j(t) \ .
\end{equation}

\end{document}